\documentclass[journal=jacsat,manuscript=article]{achemso}
\setkeys{acs}{keywords = true}

\usepackage[version=3]{mhchem} 
\usepackage{amsmath} 
\usepackage{amsthm} 
\usepackage{amssymb}	
\usepackage{graphicx} 

\usepackage{amsmath, amsfonts, amssymb}
\usepackage{amsthm}
\usepackage{graphicx,color}
\usepackage[colorlinks=true,citecolor=blue,linkcolor=blue,pdfhighlight =/O]{hyperref}
\hypersetup{urlcolor=blue}
\usepackage{gensymb}
\usepackage{xspace}
\usepackage{upgreek}
\usepackage{xcolor}
\usepackage{upgreek}
\usepackage{blindtext}
\usepackage{titlesec}
\usepackage[utf8]{inputenc}
\usepackage{microtype}
\usepackage{longtable}
\usepackage{soul}
\usepackage{minitoc}

\newcommand{\beginsupplement}{%
	\setcounter{table}{0}
	\renewcommand{\thetable}{S\arabic{table}}%
	\setcounter{figure}{0}
	\renewcommand{\thefigure}{S\arabic{figure}}%
	\setcounter{equation}{0}
	\renewcommand{\theequation}{S.\arabic{equation}}
	\setcounter{section}{0}
	\renewcommand{\thesection}{S\arabic{section}}%
}

\author{Sayanti Samaddar}
\affiliation{2nd Institute of Physics B and JARA-FIT, RWTH Aachen University, Otto-Blumenthal-Stra{\ss}e, 52074 Aachen, Germany}
\alsoaffiliation{National Physical Laboratory, Hampton Road, Teddington TW11 0LW, United Kingdom}

\author{Jeff Strasdas}
\affiliation{2nd Institute of Physics B and JARA-FIT, RWTH Aachen University, Otto-Blumenthal-Stra{\ss}e, 52074 Aachen, Germany}

\author{Kevin Janßen}
\affiliation{2nd Institute of Physics B and JARA-FIT, RWTH Aachen University, Otto-Blumenthal-Stra{\ss}e, 52074 Aachen, Germany}
\alsoaffiliation{Peter Gr\"unberg Institute 6 \& 9, Forschungszentrum Jülich GmbH, 52425 Jülich, Germany}

\author{Sven Just}
\affiliation{2nd Institute of Physics B and JARA-FIT, RWTH Aachen University, Otto-Blumenthal-Stra{\ss}e, 52074 Aachen, Germany}
\alsoaffiliation{Leibniz Institute for Solid State and Materials Research Dresden (IFW), 01171 Dresden, Germany}

\author{Tjorven Johnsen}
\affiliation[RWTH Aachen University]
{2nd Institute of Physics B and JARA-FIT, RWTH Aachen University, Otto-Blumenthal-Stra{\ss}e, 52074 Aachen, Germany}

\author{Zhenxing Wang}
\affiliation[AMO Gmbh]
{Advanced Microelectronic Center Aachen (AMICA), AMO GmbH, Otto-Blumenthal-Str. 25, 52074 Aachen, Germany}

\author{Burkay Uzlu}
\affiliation[AMO Gmbh]
{Advanced Microelectronic Center Aachen (AMICA), AMO GmbH, Otto-Blumenthal-Str. 25, 52074 Aachen, Germany}
\alsoaffiliation{Chair of Electronic Devices, RWTH Aachen University, 52074 Aachen, Germany}

\author{Sha Li}
\affiliation[AMO Gmbh]
{Advanced Microelectronic Center Aachen (AMICA), AMO GmbH, Otto-Blumenthal-Str. 25, 52074 Aachen, Germany}

\author{Daniel Neumaier}
\affiliation[University of Wuppertal]
{University of Wuppertal, 42285 Wuppertal, Germany}
\alsoaffiliation[AMO Gmbh]
{Advanced Microelectronic Center Aachen (AMICA), AMO GmbH, Otto-Blumenthal-Str. 25, 52074 Aachen, Germany}

\author{Marcus Liebmann}
\affiliation[RWTH Aachen University]
{2nd Institute of Physics B and JARA-FIT, RWTH Aachen University, Otto-Blumenthal-Stra{\ss}e, 52074 Aachen, Germany}

\author{Markus Morgenstern}
\email{mmorgens@physik.rwth-aachen.de}
\affiliation{2nd Institute of Physics B and JARA-FIT, RWTH Aachen University, Otto-Blumenthal-Stra{\ss}e, 52074 Aachen, Germany}

\title{Evidence for local spots of viscous electron flow in graphene at moderate mobility} 


\begin{document}
\date{\today}

\makeatletter
\setlength\acs@tocentry@height{8.25 cm}
\setlength\acs@tocentry@width{4.45 cm}
\makeatother


\begin{abstract}
Dominating electron-electron scattering enables viscous electron flow exhibiting hydrodynamic current density patterns such as  Poiseuille profiles or vortices. The viscous regime has recently been observed in graphene by non-local transport experiments and mapping of the Poiseuille profile. Here, we probe the current-induced surface potential maps of graphene field effect transistors with moderate mobility using scanning probe microscopy at room temperature. We discover micron-sized large areas appearing close to charge neutrality that show current induced electric fields opposing the externally applied field.  By estimating the local scattering lengths from the gate dependence of local in-plane electric fields, we find that electron-electron scattering dominates in these areas as expected for viscous flow. Moreover, we suppress the inverted fields by artificially decreasing the electron-disorder scattering length via mild ion bombardment.
 These results imply that viscous electron flow is omnipresent in graphene devices, even at moderate mobility.    
\end{abstract}



\noindent


\maketitle

\chapter{}
\section*{Introduction}
Since electron-electron scattering is momentum conserving, if Umklapp scattering is absent, the corresponding electric resistance is not related to momentum relaxation, but to viscous properties of the electron liquid.\cite{Polini2020,Mayzel2019,Schaefer2009} Indeed, Navier-Stokes type equations have been employed  to calculate resistance and charge flow patterns when electron-electron scattering\cite{Narozhny2017,Narozhny2019,Torre2015,Polini2020} dominates.
This revealed that the resistance of  a constriction can drop below its ballistic Landauer-B\"uttiker-type value via lateral drag (as dubbed the Gurzhi effect \cite{Gurzhi1968, Govorov2004}), that a Poisseuile flow implying an inverted parabolic velocity profile appears across a ribbon \cite{Torre2015,Guo2017,Moessner2019,Holder2019} and that vortices of current flow can develop \cite{Mohseni2005} sideways from a  current injection point \cite{Torre2015,Levitov2016,Danz2020,Chandra2019,Lent1990, Mendoza2011,Pellegrino2016} or within a disorder potential \cite{Li2020}. 

The first experimental evidence of Gurzhi effect was found for GaAs constrictions.\cite{deJong1995} 
More recently, indications of a dominant viscous electron flow were observed in other 2D materials such as graphene \cite{Bandurin2016,Crossno2016,Ghahari2016,Gallagher2019,Kumar2017,Berdyugin2019,Bandurin2018,Geurs2020,Lucas2018} or PdCoO$_2$ \cite{Moll2016} as well as in the 3D Dirac- and Weyl-type materials PtSn$_4$\cite{Fu2020} and WP$_2$\cite{Gooth2018}, respectively. These experiments proved viscous flow indirectly via electric or heat transport experiments, partly at optical frequencies, \cite{Gallagher2019,Block2020} using the detailed parameter dependence. A real-space visualization has been accomplished for graphene displaying Poiseuille charge flow profile \cite{Sulpizio2019, Ku2020} and its transition to ohmic \cite{Jenkins2020} or ballistic \cite{Sulpizio2019} transport profiles . Additionally, artificial constrictions in the viscous regime have been probed for GaAs by scanning gate microscopy \cite{Braem2018} and for graphene by scanning tunneling potentiometry \cite{krebs2021}.\\   

Here, we employ Kelvin probe force microscopy (KPFM)\cite{Melitz2011} and electrostatic force microscopy (EFM)\cite{Xu2018} on graphene field effect transistors under current flow at moderate mobility. We screen the gate electrode by large contact pads such that its influence on the cantilever is minimized. The resulting current-induced potential maps feature local textures that we attribute to viscous electron flow.\cite{Torre2015,Pellegrino2016,Levitov2016} 
 In these areas, the local potential drop opposes the externally applied source-drain voltage $V_{\rm SD}$ implying an inverted electric field. Such areas appear with increasing frequency, if the sample is tuned towards charge neutrality, i.e., the Dirac point voltage $V_{\rm D}$. The inverted fields could be partly attributed to source-drain voltage induced local doping (SDILD) \cite{Geurs2020}, i.e. we could reproduce them as a consequence of SDILD using the previously measured electron concentration maps. For the areas of  inverted fields which could not be fully explained by SDILD, we use the gate dependence of the measured current-induced potentials to estimate the local electron-disorder scattering length $l_{\rm dis}$ that, for low charge carrier densities, turns out to  be larger than the local electron-electron scattering length $l_{\rm ee}$. 
Consistently, the areas of relatively short $l_{\rm ee}$ exhibit inverted electric fields relating these fields to the hydrodynamic regime.\cite{Bandurin2016,Falkovich2017,Pellegrino2016,Lucas2018,Polini2020} 
Reducing $l_{\rm dis}$ artificially by low-energy ion bombardment, to establish $l_{\rm dis} < l_{\rm ee}$ for all $V_{\rm gate}$, consistently removed the areas of of current-induced inverted fields.
Since the devices exhibit moderate mobilities $\mu=1000-4000$\,cm$^2$/Vs and the effects are observed at $300$\,K, our results imply that viscous electron flow is ubiquitous in graphene devices.

\section*{Results and discussion}
\begin{figure}
  \centering
    \includegraphics[width=165 mm]{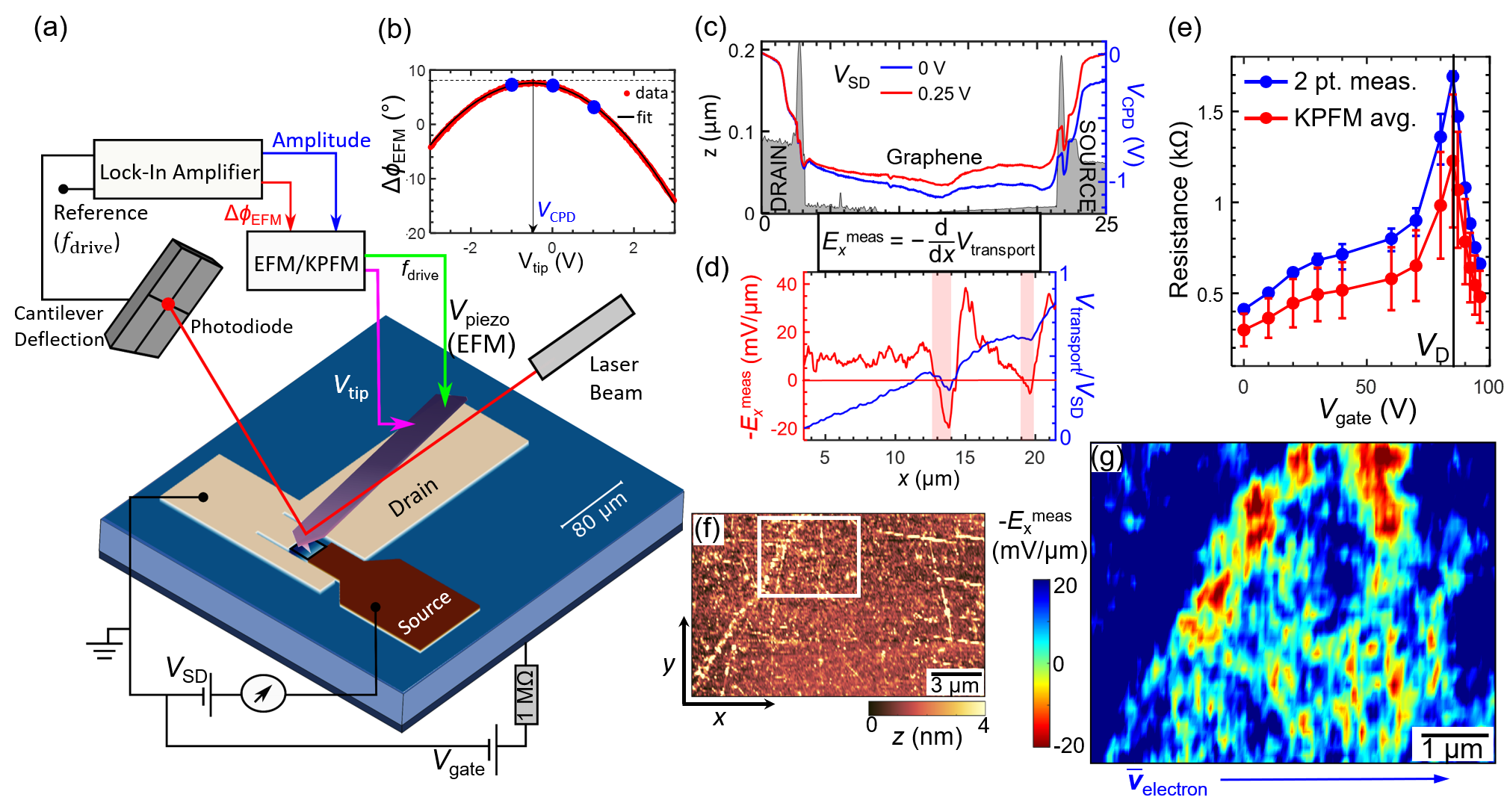}
    \caption{{\bf Measurement scheme, sample characterization and inverted electric fields.} (a) Setup for mapping the lateral electric fields induced by charge transport $E_x^{\rm meas}(x,y)$, $V_{\rm gate}$: gate voltage, $V_{\rm SD}$: source-drain voltage, $V_{\rm piezo}$: piezo excitation for EFM, $V_{\rm tip}$: tip voltage excitation for KPFM, $f_{\rm drive}$: drive frequency for both methods. The large, grounded drain pad (beige) protects the cantilever (violet) from influences of $V_{\rm gate}$. Black rectangle between source and drain marks graphene. (b) Red: $\Delta \phi_{\rm EFM}(V_{\rm tip})$ measured by EFM. Black: parabolic fit with deduced contact potential difference $V_{\rm CPD}$ ($V_{\rm tip}$ at maximum). Blue dots: $\Delta \phi_{\rm EFM}(V_{\rm tip})$ at three $V_{\rm tip}$ probed at the same location as the red data points. These three values are recorded for each position $(x,y)$ to reconstruct the full parabola and, hence, $V_{\rm CPD}(x,y)$  (Supplementary Section \ref{sec3:EFM}). (c) Blue, red: $V_{\rm CPD}(x)$\ along the same line at different $V_{\rm SD}$, KPFM, $V_{\rm gate}= 85$\,V (charge neutrality). Grey areas: topography along the same line. Graphene, source and drain electrode are marked. (d) Blue: normalized current-induced voltage drop across graphene $\frac{V_{\rm transport}}{V_{\rm SD}}(x) := \frac{V_{\rm CPD}(x,V_{\rm SD}) - V_{\rm CPD}(x,V_{\rm SD} = 0\hspace{0.5mm} {\rm V})}{V_{\rm SD}}$\ using the two $V_{\rm CPD}(x, V_{\rm SD})$  from c.  Red: deduced lateral electric field $E_x^{\rm meas}(x)$ via the formula displayed between c and d. Inverted $E_x^{\rm meas}(x)$ areas are  highlighted in pink. (e) Blue: two point resistance of the graphene device. Red: resistance of the graphene area only deduced by the spatially averaged $R=\langle E_{ x}^{\rm meas}(x,y)\rangle \cdot L/I_{\rm SD}$ with $I_{\rm SD}$: source-drain current, $L$: graphene length,  $V_{\rm D}$: deduced Dirac point. (f) Topography of graphene area (tapping mode AFM). (g) $E_x^{\rm meas}(x,y)$ in the area of the white rectangle in f, KPFM, $V_{\rm gate}= 80$\,V, $V_{\rm SD}=500$\,mV. Preferential electron flow direction $\overline{v}_{\rm electron}$ is marked.
    }
    \label{fig1:ExptProcedure}
\end{figure}

We use a graphene monolayer (Graphenea SE) deposited on a Si(100)/SiN(150\,nm) backgate with edge-contacted source and drain electrodes (Ni/Al (12/50 nm)) structured via optical lithography (Supplementary Section \ref{sec:prep}).\cite{Shaygan2017}  The large drain contact (Fig.~\ref{fig1:ExptProcedure}a), set to ground during all measurements, protects the cantilever from direct influences of the gate voltage $V_{\rm gate}$.
The remaining influence of penetration fields through the graphene is adequately described  by the quantum capacitance model (Supplementary Section \ref{sec:compressibility}).
A commercial atomic force microscope (Bruker Dimensions Icon PT) enables EFM and KPFM with lateral resolution down to 20~nm, while applying
$V_{\rm gate}$ and $V_{\rm SD}$. Both methods map the surface potential of graphene as the contact potential difference $V_{\rm CPD}(x,y)$ between the tip and the surface below the tip, with a resolution of $\sim 10$\,mV for KPFM and $\sim 2$\,mV for EFM (Supplementary Sections \ref{sec2:AM KPFM}, \ref{sec3:EFM}). At $V_{\rm SD}=0$\,V, $V_{\rm CPD}(x,y)$ is related to the charge carrier density $n_0(x,y)$ (Supplementary Section \ref{Sec5_DopingDistr_CPD}). Comparing the $V_{\rm CPD}(x,y)$ maps at $V_{\rm SD}=0$\,V and $V_{\rm SD}\neq 0$\,V enables us to produce the current-induced potentials and, via derivatives, the current induced electric fields.

Of the two techniques, KPFM has faster acquisition times, so is generally more appropriate for measurements at ambient conditions that are prone to temporal changes of the potential landscape. In contrast, EFM is slower, provides better $V_{\rm CPD}$ resolution and is less sensitive to any remaining influences of undesired stray fields that penetrate to the cantilever.\cite{Panchal2013,Xu2018}  Hence, we mostly use KPFM, which requires us to subtract a smooth background from the $V_{\rm CPD}(x,y)$ images (Supplementary Section \ref{subsec_BG}), and employ EFM only if quantitative potential values matter, using smaller areas that are recorded more rapidly. For the latter, we use three different tip voltages $V_{\rm tip}$ to deduce the maximum of the inverted parabola of the phase lag $\Delta \phi_{\rm EFM} (V_{\rm tip})$ between applied voltage oscillation and resulting cantilever oscillation. $V_{\rm tip}$ at this maximum is directly $V_{\rm CPD}$
(Fig.~\ref{fig1:ExptProcedure}b, Supplementary Section \ref{sec3:EFM}).

Figure~\ref{fig1:ExptProcedure}c shows two measured $V_{\rm CPD}(x)$ profiles ($x$: lateral position)  along the same line of graphene at $V_{\rm SD}=0$\,V and $V_{\rm SD}=0.25$\,V for $V_{\rm gate}=V_{\rm D}$. 
The corresponding topography (grey areas) reveals the positions of the Ni/Al electrodes, where $V_{\rm CPD}(x)$ exhibits steps due to a work function difference between Al and graphene. Both $V_{\rm CPD}(x)$ lines show fluctuations across the graphene, attributed to charge puddles \cite{Martin2007}. Application of $V_{\rm SD}$ changes the slope of $V_{\rm CPD}(x)$ indicating the current induced potential drop. To disentangle charge puddles and current induced potential $V_{\rm transport}(x)$, we subtract the two curves. Dividing $V_{\rm transport}(x)$ by the applied $V_{\rm SD}$ reveals that about 80\,\% of $V_{\rm SD}$ drops across graphene (Fig.~\ref{fig1:ExptProcedure}d). Notably, $V_{\rm transport}/V_{\rm SD}$ exhibits negative slopes (highlighted areas in Fig.~\ref{fig1:ExptProcedure}d), i.e. an inverted voltage drop with respect to the applied $V_{\rm SD}$. The resulting in-plane electric field $E_x^{\rm meas}(x)=-dV_{\rm transport} (x)/dx$ is therefore also inverted with respect to the electric field direction caused by $V_{\rm SD}$ (red line, Fig.~\ref{fig1:ExptProcedure}d). In this work, we consistently plot $-E_x^{\rm meas}(x,y)$ such that inverted electric fields always appear negative in maps and curves.
The electric field inversion  is ubiquitous in $E_x^{\rm meas}(x,y)$ maps, if recorded close to charge neutrality at $V_{\rm gate}\in \left[V_{\rm D}- 15\,{\rm V},V_{\rm D}+ 15\,{\rm V} \right]$ (Fig.~\ref{fig1:ExptProcedure}g). Areas of inverted $E_x^{\rm meas}(x,y)$ partly correlate with topographic features. For example the diagonal fold starting at the lower left in the topography map (Fig.~\ref{fig1:ExptProcedure}f) has multiple inverted $E_x^{\rm meas}(x,y)$ areas to its right (Fig.~\ref{fig1:ExptProcedure}g, discussion in Supplementary Section \ref{sec:origin}). We determined the gate-dependent resistance of graphene from KPFM by spatially averaging $E_x^{\rm meas}(x,y)$, multiplying by the sample length $L=18.5$\,$\mu$m (Supplementary Section \ref{sec:prep}) and dividing by the applied source-drain current $I_{\rm SD}$. We then crosschecked that this agreed with the simultaneously recorded two-point resistance $R_{\rm transport}$ (Fig.~\ref{fig1:ExptProcedure}e). The two data sets match except of an offset of 80-160 $\Omega$. The offset is attributed to the metal-graphene contact resistance as corroborated by 4-point measurements of identically prepared samples (Supplementary Section \ref{sec:transportmobility}) \cite{Shaygan2017}.  

The inverted electric fields with respect to $V_{\rm SD}$ (Fig.~\ref{fig1:ExptProcedure}g) imply a complex charge redistribution by the current flow.
Such charge redistribution is known to appear in the hydrodynamic regime, e.g., via current induced vorticity (Supplementary Section \ref{sec:origin}) \cite{Polini2020,Danz2020,Falkovich2017,Torre2015}. Hence, it is tempting to assume that the inverted
$E_x^{\rm meas}(x,y)$ is due to viscous electron flow.\cite{Polini2020,Mayzel2019,Schaefer2009} 

\begin{figure}
  \centering
    \includegraphics[width=160 mm]{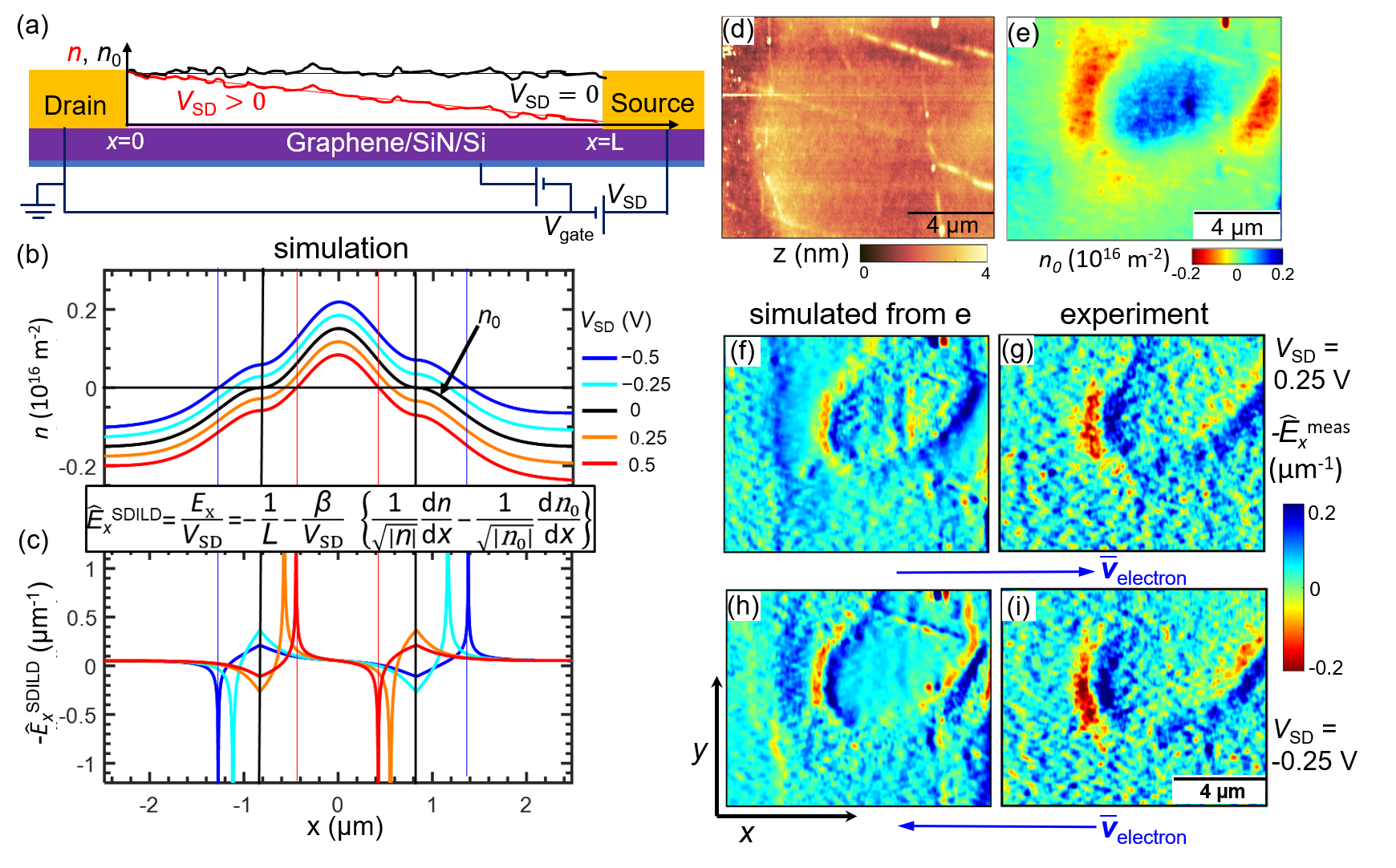}
    \caption{{\bf Inverted $\widehat{E}_{ x}^{\rm meas}(x)$ by source-drain voltage induced local doping (SDILD).} (a) Sample sketch with exemplary charge carrier density  $n_0(x)$ (black line, $V_{\rm SD}=0$\,V) and resulting $n(x)$  (red line, $V_{\rm SD}>0$\,V). Thin lines: $n_0(x)$, $n(x)$ without charge puddles. The drain remains grounded. 
   (b) Calculated example of $n_0(x)$ and $n(x)$ for a small scale structure at multiple $V_{\rm SD}$ with some zero crossings marked by vertical lines of identical color. The $n_0(x)$ profile corresponds to a Gaussian shaped potential that crosses $E_{\rm F}$ twice. 
    (c) $\widehat{E}_{x}^{\rm SDILD}(x)$ deduced from b according to eq.~(\ref{eq:2}) and (\ref{eq:3}) as displayed between b and c ($\beta = \hbar v_\mathrm{F} \sqrt{\pi}/2|e|$). Note the extrema at the vertical lines corresponding to zeroes of charge carrier density in b. 
    (d) Graphene topography (tapping mode AFM).
    (e) Equilibrium doping $n_0(x,y)$ of the same area as d, deduced from measured $V_{\rm CPD} (x,y,V_{\rm SD}=0)$ (Supplementary Section \ref{Sec5_DopingDistr_CPD}), $V_{\rm gate} = V_{\rm D} = 85\,$V, KPFM.
    (f), (h) $\widehat{E}_{ x}^{\rm SDILD}(x,y)$  deduced from e by eq.~(\ref{eq:2}) and (\ref{eq:3}) (displayed between b and c) after tilting $V_{\rm CPD}(x,y,V_{\rm SD}=0)$ to get $n(x,y)$ according to eq.\ref{eq:SDILD} (Supplementary Section \ref{subsec_SDILD}), $V_{\rm SD}$ is marked on the right of g, i. (g), (i) Measured $\widehat{E}_{x}^{\rm meas}(x,y)$ at the same $V_{\rm SD}$ as f, h. The similarity of inverted $\widehat{E}_{ x}(x,y)$ areas (yellow/red) implies that the inverted fields in the experimental maps are largely caused by SDILD. Remaining discrepancies are discussed in Supplementary Section \ref{subsec_SDILD}}
    \label{fig2:SDind_doping}
\end{figure}

However, there is a known artifact leading to an apparent inverted $E_x^{\rm meas}(x,y)$ at $V_{\rm gate}\simeq V_{\rm D}$.\cite{Geurs2020} It results from the local doping by the applied $V_{\rm SD}$ acting as a gate (Fig.~\ref{fig2:SDind_doping}a) and has to be carefully distinguished from a current induced inverted $E_x^{\rm meas}(x,y)$. To understand the artifact, we recall that $E_x^{\rm meas}(x,y)$ at given $V_{\rm SD}$ is determined from $V_{\rm CPD}(x,y)$ maps via
\begin{equation}
\label{eq:1}
E_x^{\rm meas}(x,y)=-\frac{d(V_{\rm CPD} (x,y,V_{\rm SD})-V_{\rm CPD} (x,y,V_{\rm SD}=0))}{dx}.
\end{equation}

Assuming that the applied $V_{\rm SD}$ drops linearly across the graphene, one straightforwardly obtains (Supplementary Section \ref{Sec:EField_SDILD}, eq.\ref{eq:E_SDLD})
\begin{equation}
\label{eq:2}
  E_x^\mathrm{meas}(x,y)  = -\frac{V_{\rm SD}}{L} -  \frac{\hbar v_\mathrm{F} \sqrt{\pi}}{2|e|} \left( \frac{1}{\sqrt{|n(x,y)|}}  \frac{dn(x,y)}{dx} - \frac{1}{\sqrt{|n_0(x,y)|}} \frac{dn_0(x,y)}{dx} \right):=E_x^\mathrm{SDILD}(x,y)  
\end{equation}
  with graphene's Fermi velocity $v_{\rm F}\simeq 10^6$\,m/s and the charge carrier densities $n(x,y)$ ($n_0(x,y)$) at applied $V_{\rm SD}$ 
(without $V_{\rm SD}$). We include quantum capacitance to calculate $n(x,y)$ and $n_0(x,y)$, but not negative compressibility, which is usually irrelevant for graphene \cite{Li2011,Sheehy2007}  (Supplementary Section \ref{sec:compressibility}). The resulting field from source-drain voltage induced local doping (SDILD) is dubbed $E_x^\mathrm{SDILD}(x,y)$. $E_x^\mathrm{SDILD}(x,y)$ diverges at $n(x,y)=0$ and $n_0(x,y)=0$ with sign depending on the spatial derivative of the corresponding charge carrier density, and being opposite for $n(x,y)$ and $n_0(x,y)$. 
After normalizing $E_x^\mathrm{SDILD}(x,y)$ to $V_{\rm SD}$ for easier comparison (analogously to $E_x^\mathrm{meas}(x,y)$):  
\begin{equation}
\label{eq:3}
\widehat{E}_x^\mathrm{SDILD}(x,y)= \frac{E_x^\mathrm{SDILD}(x,y)}{V_{\rm SD}}, \hspace{0.7cm}
\widehat{E}_x^\mathrm{meas}(x,y)= \frac{E_x^\mathrm{meas}(x,y)}{V_{\rm SD}},
\end{equation}
the sign of the divergence also depends on the sign of $V_{\rm SD}$. 

Since $V_{\rm SD}$ shifts each zero crossing of charge carrier density along $x$, parallel to $V_{\rm SD}$, (Fig.~\ref{fig2:SDind_doping}b), one mostly obtains doublets of divergences with opposite sign in $\widehat{E}_x^\mathrm{SDILD}(x)$  
(Fig.~\ref{fig2:SDind_doping}c), one at $n_0(x)=0$ ($V_{\rm SD}=0$) and one at the shifted $n(x)=0$ ($V_{\rm SD} \neq 0$).
Figure~~\ref{fig2:SDind_doping}b and c display a calculated example with two zero crossing $n_0(x)=0$ featuring an electron puddle with diameter 1.4\,$\mu$m embedded into a hole density. Tilting the $n_0(x)$ profile by $V_{\rm SD}$ shifts the zero crossings laterally (Fig.~\ref{fig2:SDind_doping}b). Consequently, the resulting
$\widehat{E}_x^\mathrm{SDILD}(x)$ curves (Fig.~\ref{fig2:SDind_doping}c) show two doublets with negative dip and positive peak each. It is also clear that the slope at $n_0(x)=0$ and $n(x)=0$
determines the strength of dips and peaks, respectively.

\begin{figure}
  \centering
    \includegraphics[width=160 mm]{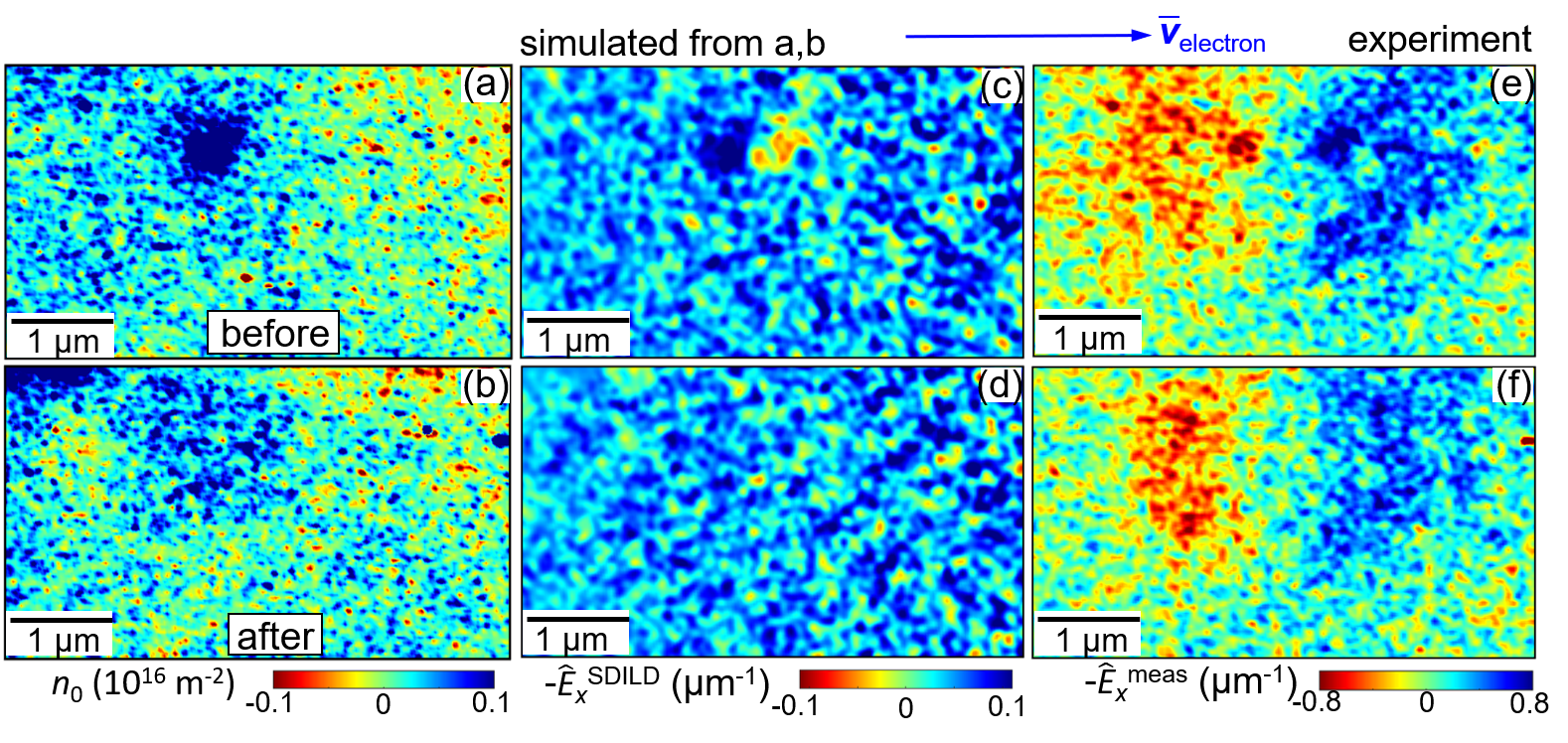}
    \caption{{\bf Current induced inverted electric fields not caused by $\widehat{E}_{x}^{\rm SDILD}$.} (a), (b) Graphene doping maps $n_0(x,y)$ as deduced from measured $V_{\rm CPD}(x,y,V_{\rm SD} = 0 V)$, EFM. Both images display the same sample area recorded prior (a) and after (b) $V_{\rm CPD}(x,y,V_{\rm SD}=0.1\,{\rm V})$ as necessary to determine $\widehat{E}_{x}^{\rm meas}(x,y)$ in e, f (eq.~(\ref{eq:1})), $V_{\rm gate}= V_{\rm D}=\,$85\,V.
    (c), (d) $\widehat{E}_{x}^{\rm SDILD}(x,y)$\ deduced from a, b using eqs.~(\ref{eq:2}) and (\ref{eq:3}), $V_{\rm SD}=0.1$\,V. (e), (f) Experimental $\widehat{E}_{x}^{\rm meas}(x,y)$ at $V_{\rm SD} = 0.1$\,V deduced from a measured $V_{\rm CPD}(x,y,V_{\rm SD}=0.1\,{\rm V})$ map  and the $V_{\rm CPD}(x,y,V_{\rm SD}=0$\,V) leading to a, b, respectively (eq.~(\ref{eq:1})). The strong discrepancy between c, d compared to e, f rules out that the partially inverted field pattern in e, f results from SDILD. Note the different color scales in c, d and e, f.
    }
    \label{fig3:NegFields_ViscousFlow}
\end{figure}

The SDILD effect is indeed identified in the experiments. Figure~\ref{fig2:SDind_doping}d, e display the topography of a graphene area and the corresponding $n_0(x,y)$ for $V_{\rm gate}\approx V_{\rm D}$. They are deduced from a recorded $V_{\rm CPD}(x,y, V_{\rm SD}=0\,{\rm V})$ map (Supplementary Section \ref{Sec5_DopingDistr_CPD}, eq.\ref{DopingCPDrelation}). An electron puddle of size 5\,$\mu$m is apparent (blue area in e) surrounded by hole doped areas. Figure~\ref{fig2:SDind_doping}f and h display  $\widehat{E}_x^\mathrm{SDILD}(x,y)$ as calculated from Fig.~\ref{fig2:SDind_doping}e for two opposite $V_{\rm SD}$ via 
eq.~(\ref{eq:2}) and (\ref{eq:3}) after deducing $n(x,y)$ by tilting the potential and including quantum capacitance (Supplementary Section \ref{subsec_SDILD}, eq.\ref{eq:SDILD}). $\widehat{E}_x^\mathrm{meas}(x,y)$ resulting from two recorded $V_{\rm CPD}(x,y,V_{\rm SD})$ maps (eq.~(\ref{eq:1})) are displayed in Fig.~\ref{fig2:SDind_doping}g, i. These include both, current induced electric fields and SDILD effects. As is evident, most details of the experiment are reproduced by the simulation.
Smaller discrepancies can be attributed to experimental noise, slightly varying tip potentials during recording of the two $V_{\rm CPD}(x,y)$ images and temporal fluctuations in the doping distribution (Supplementary Section \ref{sec:fig2_discrepancies}). Importantly, these discrepancies ($\sim 10$\,\% with respect to the strongest signal) are more than an order of magnitude lower than the discrepancy of $\sim 800$\,\% between inverted fields attributed to viscous flow and the expected inverted fields from SDILD in the same area as discussed below (Fig.~\ref{fig3:NegFields_ViscousFlow}). 

Since SDILD leads to inverted electric fields,
it is important to distinguish SDILD artifacts from real current induced inverted $\widehat{E}_x^\mathrm{meas}(x,y)$.
Figure~\ref{fig3:NegFields_ViscousFlow} shows an example, where this has been accomplished. Figure \ref{fig3:NegFields_ViscousFlow}a displays the charge carrier density $n_0(x,y)$ deduced from $V_{\rm CPD}(x,y,V_{\rm SD}= 0\,{\rm V})$. Figure~\ref{fig3:NegFields_ViscousFlow}c displays the resulting $\widehat{E}_x^\mathrm{SDILD}(x,y)$ according to eq.~(\ref{eq:2}) and (\ref{eq:3}) and Fig.~\ref{fig3:NegFields_ViscousFlow}e shows the measured $\widehat{E}_x^\mathrm{meas}(x,y)$, both at $V_{\rm SD}=0.1$\,V. The $\widehat{E}_x^\mathrm{meas}(x,y)$ map features an extended doublet structure (total width: $\sim 4$\,$\mu$m) consisting of two lobes with opposite fields showing the inverted electric field on the left (Fig.~\ref{fig3:NegFields_ViscousFlow}e). In contrast, the weak doublet structure of $\widehat{E}_x^\mathrm{SDILD}(x,y)$ is a factor of ten smaller in amplitude (note the different extents of the color bars), a factor of  four smaller in $x$ extension and has the inverted electric field on the right (Fig.~\ref{fig3:NegFields_ViscousFlow}c). Since the doping profile $n_0(x,y)$ temporarily fluctuates at ambient conditions, we firstly minimized these fluctuations by adequate waiting times before recording the $V_{\rm CPD}(x,y)$ and by optimizing the sequences to change $V_{\rm SD}$ and $V_{\rm gate}$ (Supplementary Section \ref{subsec_DopingStability}, Fig. \ref{fig8:ElecField_ImagingConditions}). Moreover, we recorded $V_{\rm CPD}(x,y,V_{\rm SD}=0\,{\rm V})$
before and after $V_{\rm CPD}(x,y, V_{\rm SD}\neq 0\,{\rm V})$ (eq.~(\ref{eq:1})). The resulting two image sets consisting of
$n_0(x,y)$, $\widehat{E}_x^\mathrm{SDILD}(x,y)$, and $\widehat{E}_x^\mathrm{meas}(x,y)$ are compared in Fig.~\ref{fig3:NegFields_ViscousFlow} revealing that the small changes in $n_0(x,y)$ (Fig.~\ref{fig3:NegFields_ViscousFlow}a, b) barely change the measured $\widehat{E}_x^\mathrm{meas}(x,y)$ doublet (Fig.~\ref{fig3:NegFields_ViscousFlow}e, f) that in both cases strongly deviates from $\widehat{E}_x^\mathrm{SDILD}(x,y)$ (Fig.~\ref{fig3:NegFields_ViscousFlow}c, d).

Consequently, we attribute the  $\widehat{E}_x^\mathrm{meas}(x,y)$ doublet, including a large area of inverted $\widehat{E}_x^\mathrm{meas}(x,y)$ (Fig.~\ref{fig3:NegFields_ViscousFlow}e--f), to a spatially inhomogeneous current flow. Such inverted  $\widehat{E}_x^\mathrm{meas}(x,y)$ can be rationalized by a strong, passing current that reduces the charge carrier density in a nearby area with reduced current density by viscous friction, i.e. charge puddles are sucked from the quieter area by the passing current without being compensated by the forward electron flow (Supplementary Section \ref{sec:origin}). This has been observed in simulations for viscous electron flow within a disorder potential\cite{Mendoza2011,Falkovich2017,Lucas2018,Polini2020} where a lateral viscous force pulls electrons out of a relatively quiet area protected from electron flow, e.g., by an upstream obstacle.\cite{Polini2020,Danz2020,Pellegrino2016}
However, inhomogeneous current induced potentials also appear in the ohmic and the ballistic regime.\cite{Chandra2019,Landauer1957,Morr2017}
Most prominently, the Landauer resistivity dipole around an obstacle produces an enhanced-inverted-enhanced triplet-like electric field structure along the current path\cite{Landauer1957,Morr2017} that is, however, not observed in our experiments. Ballistic patterns, which are unlikely in our low mobility samples, can also produce field inversions depending on boundary conditions (discussion in Supplementary Section \ref{sec:origin}).\cite{Chandra2019, Pellegrino2016}  

\begin{figure}
  \centering
    \includegraphics[width=160 mm]{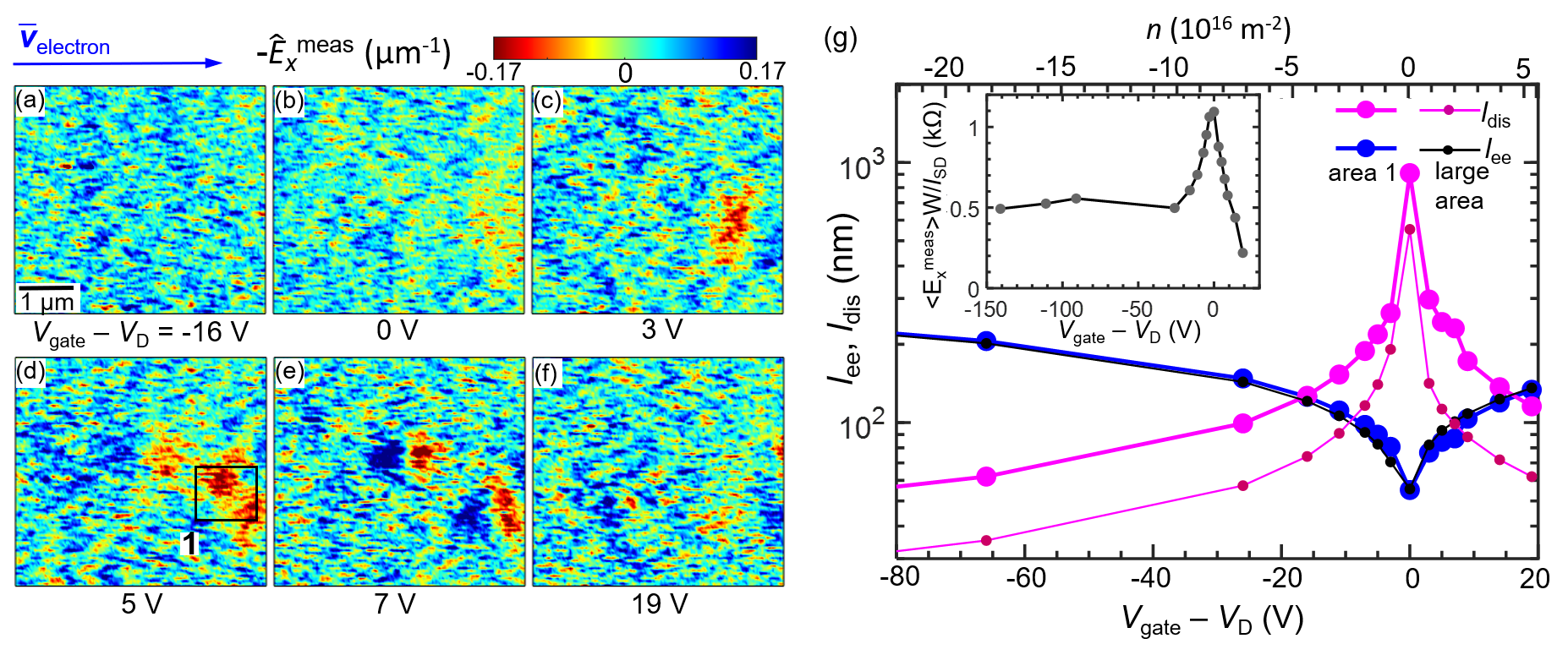}
    \caption{{\bf Relation between inverted ${E}_x^\mathrm{meas}(x,y)$ and scattering lengths $l_{\rm ee}$, $l_{\rm dis}$.} (a)-(f) $\widehat{E}_x^\mathrm{meas}(x,y)$ of the same area at different $V_{\rm gate}-V_{\rm D}$, $V_{\rm D} = 81$\ V, KPFM. Black rectangle in d highlights an area of inverted $\widehat{E}_x^\mathrm{meas}(x,y)$ as analyzed in g. Multiple areas of inverted field  appear  (red). (g) Gate dependence of electron-electron scattering length  $l_{\rm ee}$ (blue, black, eq.~({\ref{eq:4}})) and electron-disorder scattering length $l_{\rm dis}$ (pink, eq.~(\ref{eq:5})) averaged across the black rectangle of d (large symbols) and across a larger area ($12.5$\,$\mu{\rm m}$\ $\times$\ $8.33$\, $\mu {\rm m}$, small symbols). At $V_{\rm gate} = V_{\rm D}$,  $n_0(x,y)=n_{\rm Th}$ is used to avoid an unphysical divergent $l_{\rm dis}$ (eq.~(\ref{eq:5})). A large gate voltage region features $l_{\rm ee}< l_{\rm dis}$, in particular, for area 1 showcasing inverted $\widehat{E}_x^\mathrm{meas}(x,y)$ in c$-$e. Inset: Spatially averaged $\left<E_x^\mathrm{meas}(x,y)\right>$ scaled to represent the local resistivity $\left<\rho_{\rm local}(x,y)\right>=\left<{E}_x^\mathrm{meas}(x,y)\right>\cdot W/I_{\rm SD}$ ($W$: sample width). Constant resistivity appears at large hole doping and, thus, $\left<\rho_{\rm local}(x,y)\right>$ is attributed to short-range disorder scattering.\cite{DasSarma2011}}
    \label{fig4:Relevance_Viscous}
\end{figure}

To corroborate our claim that electron viscosity is responsible for the field inversion, we demonstrate that the puddles of inverted $\widehat{E}_x^\mathrm{meas}(x,y)$ exhibit conditions favoring viscous flow, namely $l_{\rm ee} < l_{\rm dis}$.\cite{Polini2020,Lucas2018} 
Figures~\ref{fig4:Relevance_Viscous}a$-$f show $\widehat{E}_x^\mathrm{meas}(x,y)$ of graphene at various $V_{\rm gate}$ with areas of inverted $\widehat{E}_x^\mathrm{meas}(x,y)$ at $V_{\rm gate}\approx V_{\rm D}$, here at the electron side ($V_{\rm gate}>V_{\rm D}$). The measured $n_0(x,y)$ of the area does not exhibit any zeroes at $V_{\rm gate}-V_{\rm D}\ge 3$\,V, hence, SDILD is negligible.\\
We estimate  $l_{\rm ee}(x,y)$ as local property via $n(x,y)$ deduced from the measured $n_0(x,y)$ (Supplementary Section \ref{subsec_SDILD}, eq.\ref{eq:SDILD}) by \cite{Giuliani_Vignale_2005, Polini2016_NoNonsensePhysicist}
\begin{equation}
 \label{eq:4}
 l_{\rm ee}(x,y) = \frac{4}{\pi}\left(\frac{\hbar v_{\rm F}}{k_{\rm B} T}\right)^2 \left<\sqrt{\pi |n (x,y)|} \frac{1}{\ln{ \frac{2\hbar v_\mathrm{F} \sqrt{\pi |n(x,y)|}}{k_\mathrm{B} T}}}\right>  
\end{equation}
($T=298$\,K: temperature, $k_{\rm B}$: Boltzmann constant, $\hbar$: reduced Planck's constant, $\left<...\right> $: spatial average).
This formula is valid except very close to $V_{\rm D}$ in the so-called quantum critical or Dirac liquid regime below $|n_{\rm QC}|\approx 2\cdot 10^{14}$/\,m$^2$.\cite{Kim2020,Sheehy2007} Even then, graphene at 0\,T does not exhibit negative compressibility \cite{Li2011,Sheehy2007} in line with experiments.\cite{Martin2007} Outside the quantum critical regime, only small deviations from eq.~(\ref{eq:4}) by less than a factor of 1.5 are expected. \cite{Li2013, Kim2020} 

To determine $l_{\rm dis}(x,y)$, we firstly deduce whether short-range or long-range impurity scattering dominates.\cite{DasSarma2011}
We employ the local resistivity $\rho_{\mathrm{local}}(x,y)=E_x^\mathrm{meas}(x,y)\cdot W/I_{\rm SD}$ ($W=28$\,$\mu$m: width of sample)  shown in
the inset of Fig.~\ref{fig4:Relevance_Viscous}g after averaging across $ 100$\,$\mu$m$^2$. 
The maximum of $\left<\rho_{\mathrm{local}}(x,y)\right>(V_{\rm gate}-V_{\rm D})$ is close to $V_{\rm D}$ as expected. More importantly, a rather constant $\left<\rho_{\mathrm{local}}(x,y)\right>$ appears at large hole doping implying a dominant short-range impurity scattering.\cite{DasSarma2011}
At these large densities, electron-electron scattering is irrelevant (eq.~(\ref{eq:4})) and the current flow is unidirectional. Moreover, electron-phonon contributions can be neglected (Supplementary Section \ref{subsec_disorderLength})\cite{DasSarma2011, Li2013} such that deducing
$l_{\rm dis}(x,y)$ is straightforward using:\cite{DasSarma2011,Shon1998}  
\begin{equation}
 \label{eq:5}
 l_{\rm dis}(x,y) = \left(\frac{h}{2e^2} \right) \left< \frac{1}{\sqrt{\pi |n (x,y)|}\cdot \rho_{\mathrm{local}}(x,y)} \: \right>.
\end{equation} 
Equation~(\ref{eq:5}) is derived from a graphene model with two Dirac cone valleys\cite{Shon1998} and is therefore, applicable up to a Fermi energy $E_{\rm F} \sim 1.0$\,eV away from charge neutrality\cite{Ohta2007,Plochocka2008}. At low densities, it is valid down to the largest of $|n_{\rm QC}|$, the thermal limit $n_{\rm Th} = 4 \cdot10^{14}$\, m$^{-2}$\ \cite{Polini2016_NoNonsensePhysicist}, the lateral $n_0(x,y)$ fluctuations {$\Delta n_0(x,y)\approx 2\cdot10^{15}$/m$^2$}, and $E_{\rm F}$ equaling the imaginary part of the self energy\cite{Shon1998} ($\approx 30$\,meV $\Rightarrow |n|\approx 7\cdot 10^{14}$/m$^2$). 
Hence, we can extrapolate eq.~(\ref{eq:5}) down to $|n|\approx 2\cdot
10^{15}$/m$^2$.\cite{DasSarma2011}

For consistency, we always spatially average $l_{\rm ee}(x,y)$ and $l_{\rm dis}(x,y)$ across areas larger than the scattering lengths (Supplementary Section \ref{sec:ScatteringLength}). Figure~\ref{fig4:Relevance_Viscous}g displays $\left<l_{\rm dis}(x,y)\right>$ and $\left<l_{\rm ee}(x,y)\right>$ within the black rectangle of Fig.~\ref{fig4:Relevance_Viscous}d (area 1) and for a 100~$\mu$m$^2$ area  (large area).
As is evident, the $V_{\rm gate}$ range with $l_{\rm ee} < l_{\rm dis}$ is significantly larger for area 1, featuring inverted $\widehat{E}_x^\mathrm{meas}(x,y)$, than for the large area. Moreover, the $V_{\rm gate}$ range with $l_{\rm ee} < l_{\rm dis}$ extends well into the valid regions for eqs.~(\ref{eq:4}) and (\ref{eq:5}).
Such an extended $V_{\rm gate}$ range with $l_{\rm ee} < l_{\rm dis}$ is consistently observed in most regions of
inverted $\widehat{E}_x^\mathrm{meas}(x,y)$ (Supplementary Section \ref{subsec_CompLocalLengths}, Fig.~\ref{fig:Relevance_Hydrodynamics_II}f).
Hence, the conditions for viscous electron flow are generally realized around 
$V_{\rm gate}\simeq V_{\rm D}$ in our sample, most pronounced
in areas of inverted $\widehat{E}_x^\mathrm{meas}(x,y)$. This is strong evidence that the inverted $\widehat{E}_x^\mathrm{meas}(x,y)$ areas (except if attributed to SDILD) are caused by hydrodynamic electron flow. 

%
\begin{figure}
    \centering
    \includegraphics[width=\textwidth]{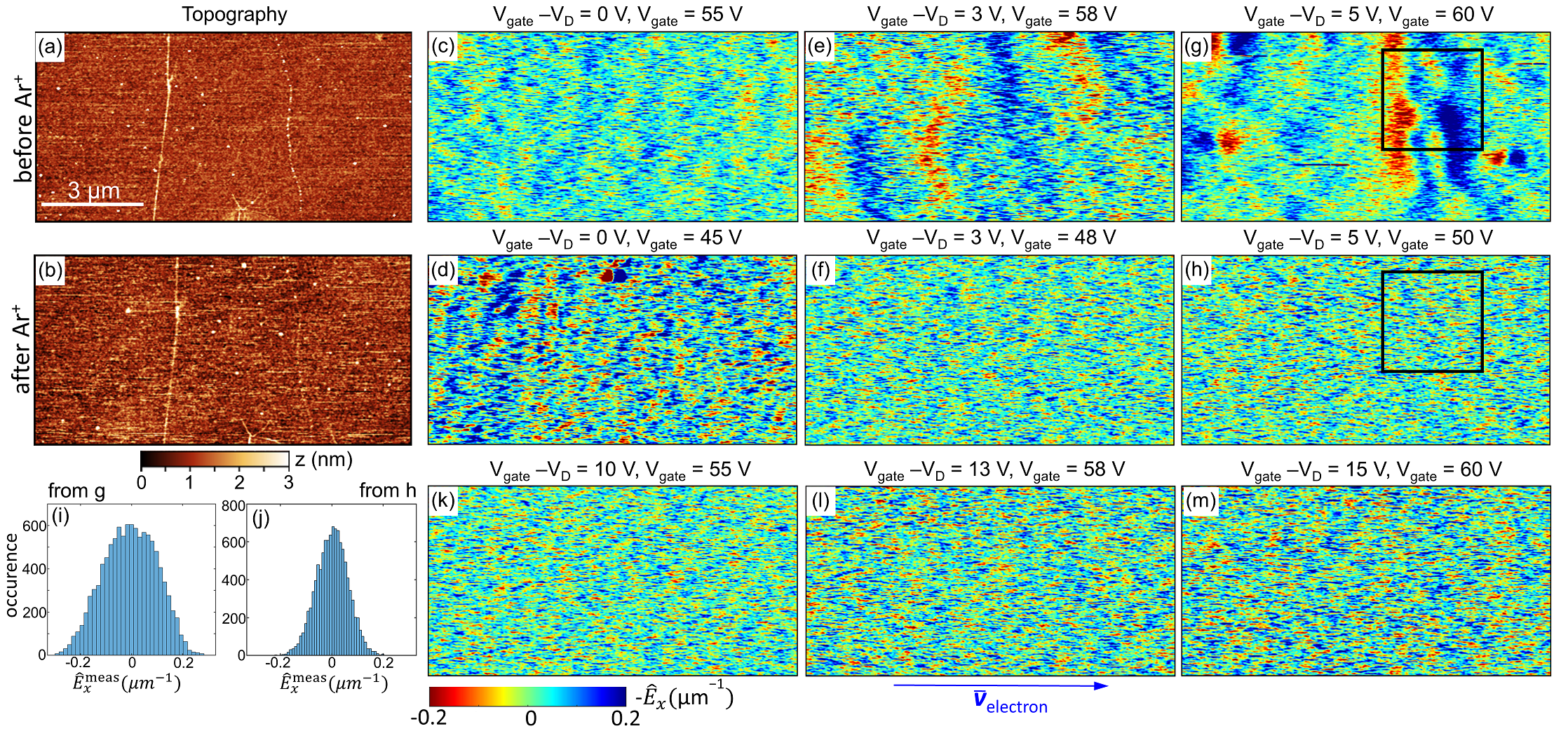}
    \caption{{\bf Removing inverted electric fields by increasing disorder scattering.} (a), (b) Topography (tapping mode AFM) of the same graphene area prior and after Ar$^+$ bombardment  ($E_{\rm kin}=50$\,eV, fluence: $2\cdot 10^{16}$\,m$^{-2}$) leading to a vacancy density $n_{\rm vac}=7\cdot 10^{15}$\,m$^{-2}$.
    (c)--(h) $\widehat{E}_{x}^{\rm meas}(x,y)$ of the same area as a--b prior (c, e, g) and after (d, f, h) Ar$^+$ bombardment at the marked $V_{\rm gate}-V_{\rm D}$ and $V_{\rm gate}$, $V_{\rm SD}=\,0.2$\,V. Inverted electric fields disappear after bombardment due to the reduced $l_{\rm dis}$ (eq.~(\ref{eq:ldis_nvac})). (i), (j) Histograms of the marked areas in (g), (h) showing the suppression of fluctuations by the reduced $l_{\rm dis}$.
    (k)--(m) $\widehat{E}_{x}^{\rm meas}(x,y)$ after ion bombardment at $V_{\rm gate}$ values that are identical to c, e , g, respectively.}
    \label{fig5:ion_bombardment}
\end{figure}

To further corroborate this evidence, we reduced $l_{\rm dis}$ artificially by inserting vacancies into the graphene using Ar$^+$ bombardment at kinetic energy $E_{\rm kin}=50$\,eV. This low energy restricts the ion induced damage to single vacancies with density $n_{\rm vac}$ as calibrated by scanning tunneling microscopy images  (Supplementary Section \ref{sec:ionbomb}).\cite{Just2014} For charge carrier densities $|n_0|$ larger than $n_{\rm vac}$ and larger than $\Delta n_0$, the resulting $l_{\rm dis}$ is given by \cite{Stauber2007, Giannazzo2011}
\begin{equation}
 l_{\rm dis}=\frac{E_{\rm F}}{\hbar \pi^2 v_{\rm F}n_{\rm vac}}\ln{\left(\frac{E_{\rm F}R_0}{\hbar v_{\rm F}}\right)^2}  \label{eq:ldis_nvac} 
\end{equation}
with vacancy radius  $R_0\simeq 0.14$\,nm (including the influence of midgap states) \cite{Stauber2007}.
Using $n_{\rm vac}=7\cdot 10^{15}$\,m$^{-2}$, we get $l_{\rm dis}=40$\,(50)\,nm for $|n_0|=2\cdot10^{16}$\,$(1\cdot10^{17})$\,m$^{-2}$. Hence, $l_{\rm dis} < l_{\rm ee}$ for all gate voltages with $n_0 > \Delta n_0$ (Fig.~\ref{fig4:Relevance_Viscous}g). Figure~\ref{fig5:ion_bombardment} compares topography and
$\widehat{E}_{x}^{\rm meas}(x,y)$ maps at various $V_{\rm gate}$ of the identical graphene area prior and after ion bombardment.
This graphene device was swept by contact-mode AFM \cite{Lindvall2012,Goossens2012} prior to the experiment to remove remaining resists from the surface (Supplementary Section \ref{sec:sweep}, \ref{sec:dirt}). The bombardment reduced the mobility of the device from $\mu=1000$\,cm$^2$/Vs to $\mu=700$\,cm$^2$/Vs and the Dirac point voltage from {$V_{\rm D}=55$\,V} to $V_{\rm D}=45$\,V. Hence, Fig.~\ref{fig5:ion_bombardment} displays both, the comparison at the same $V_{\rm gate}-V_{\rm D}$ and at the same $V_{\rm gate}$.
The reduction of $l_{\rm dis}$ removed most of the inverted field regions except for a few remainders at charge neutrality that can be attributed to SDILD. Thus, $l_{\rm dis} > l_{\rm ee}$ is indeed the central requirement for observing patches with inverted $\widehat{E}_{x}^{\rm meas}(x,y)$. This substantiates our central claim that electron viscosity is of prime importance for observing current-induced inverted fields. 
This result is remarkable, since we operate at moderate mobility, $\mu = 1000-4000$\,cm$^2$/Vs, and ambient conditions similar to typical graphene devices.\cite{Neumaier2019}  It implies that hydrodynamic electron flow is also relevant to corresponding graphene applications.
Currently, we cannot pinpoint a unique trigger of the inverted electric fields in either the topography or the equilibrium potential maps $V_{\rm CPD}(x,y,V_{\rm SD}=0)$, however some correlations are observed as discussed in Supplementary Section \ref{sec:origin}.

Our results establish EFM and KPFM as commercially available methods to probe consequences of hydrodynamic electron flow with high spatial resolution.\cite{Falkovich2017}
This is helpful for regimes where  relatively short length scales of viscous patterns prohibit their detection by negative vicinity resistance. For example, one could map inverted fields very close to a constriction, where it appears that field inversion is distinctive between the viscous and the ballistic regime,\cite{Guo2017,Pellegrino2016,Levitov2016,Li2021} while vicinity resistance is not.\cite{Shytov2018,Wang2019, Bandurin2016}
EFM, since less sensitive to background electric fields, could also map other signatures of viscous flow  such as Poiseuille profiles within less disordered samples, previously probed only by more elaborate scanning probe methods.\cite{Ella2019,Sulpizio2019,Ku2020,Jenkins2020,Sinterhauf2020}
This also works at mK temperatures and high magnetic field\cite{McCormick1999,Hedberg2010}, but then looses its advantage of simplicity. 

In conclusion, we have discovered areas of inverted electric field with respect to the applied source-drain voltage in graphene field-effect transistors at room temperature and moderate mobility. Via carefully analyzing artifacts such as SDILD, we attribute several of these features to local viscous electron flow, in particular, by correlating its appearance with strongly dominating electron-electron scattering compared to electron-disorder scattering and by removing them via reducing $l_{\rm dis}< l_{\rm ee}$ artificially with the help of ion bombardment. This indicates that viscous electron flow is relevant for material parameters used for applications and provides a new method to study these intriguing electron transport phenomena with high spatial resolution.\\

\section*{Acknowledgement}
The authors thank for helpful discussions with K.~Sotthewes, C. B.~Winkelmann, O.~Kolosov, V.~Falko, and A. Tzalenchuk, as well as T. Vincent for cross-reading the final manuscript. This project has received funding from the European Union's Horizon 2020 research and innovation programme under grant agreement number 881603 (Graphene Core3), the German Research Foundation (DFG) via Mo 858/14-1 as well as the Humboldt foundation via a grant of S.S.

\newpage

\section*{Supplementary Information}
\beginsupplement
\tableofcontents
\newpage

\section{Sample Preparation and Transport Characterization}
\subsection{Preparation of Graphene Field Effect Devices}
\label{sec:prep}
\begin{figure}
	\centering
	\includegraphics[width = 160 mm]{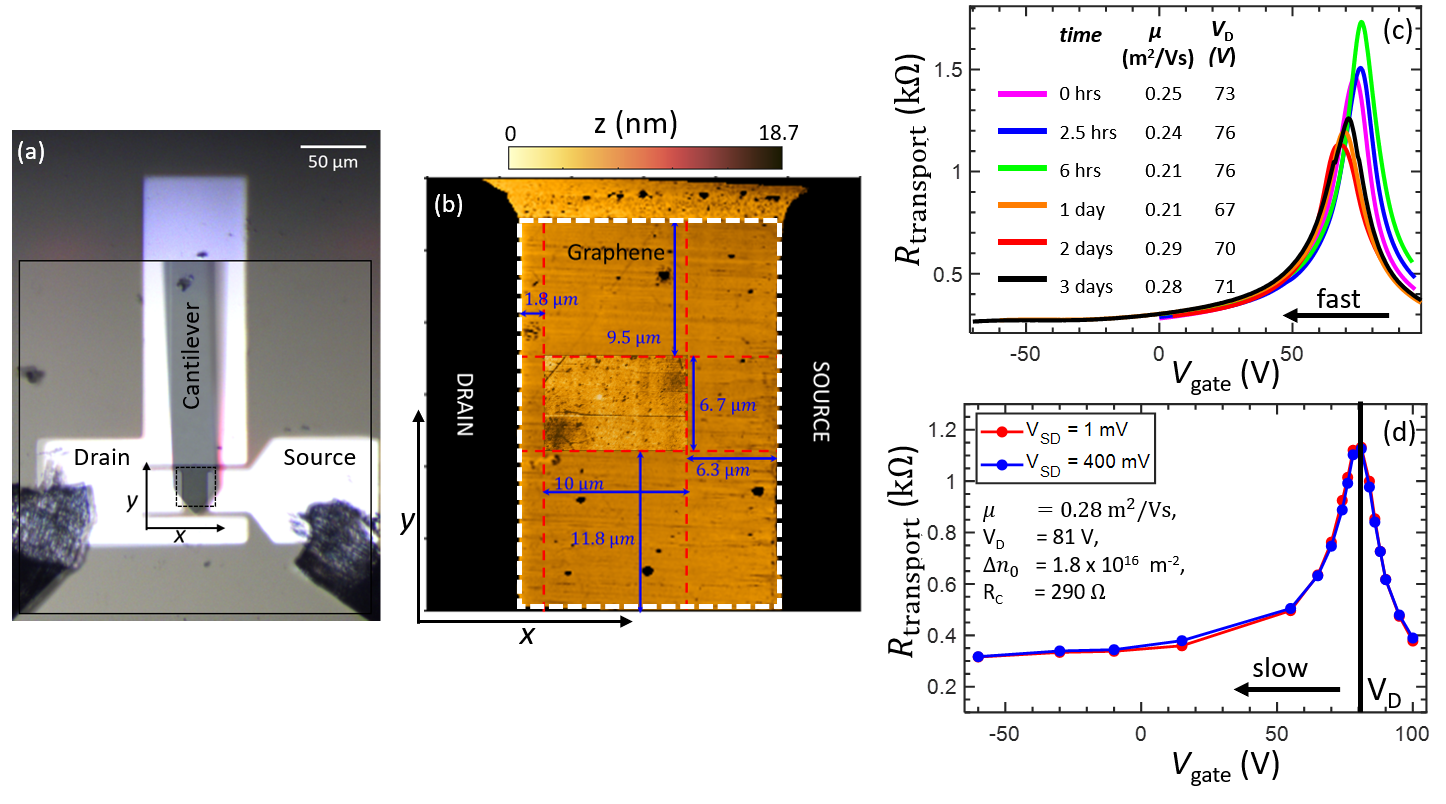}
	\caption{{\bf Experimental Setup.} (a) Optical image of the device  in top view with the cantilever above it. The graphene is the darker region below the cantilever. The dashed rectangle marks the area of b. The $x$ and $y$ coordinates depict the scan directions. (b) Topography of the area marked in a as recorded by tapping mode AFM. The white dashed rectangle marks the graphene. The central, red rectangle marks the region analyzed in Fig.~\ref{fig7:Efield_filteProcedure}a--f. It is displayed at increased contrast with respect to the color bar on top that is valid for the rest of the image. (c) Gate dependent two terminal resistance for the device analyzed in Fig.~\ref{fig4:Relevance_Viscous}, main text. The curves are recorded after the time delays as marked, recording time per curve: 8/40/40/1.5/13/2.8 min (from top (0 hrs) to bottom (3 days)), 0.5 s/measurement point. Mobilities $\mu$ and Dirac point voltages $V_{\rm D}$ are deduced from fits by eq.~(\ref{eq1:RVg})
		revealing, additionally, the residual doping    $\Delta n_0=(1.7\pm 1) \times 10^{16}\,{\rm m^{-2}}$ 
		and the saturation resistance at high doping $R_{\rm C} = (230 \pm 70)\,{\rm \Omega}$. 
		The arrow indicates the direction of $V_{\rm gate}$ sweep for all curves. (d) Two-point resistance of the same device as c, but acquired during imaging the maps of Fig.~\ref{fig4:Relevance_Viscous}a-f, main text . Two  $V_{\rm SD}$ (blue, red) are applied subsequently at each $V_{\rm gate}$, recording time: 20 min/point. The arrow marks the direction of $V_{\rm gate}$ sweep. Parameters $\mu$, $V_{\rm D}$, $\Delta n_0$, and $R_{\rm C}$ result from a fit by eq.~(\ref{eq1:RVg}).
	}
	\label{fig1_Transport_SampleOrientation}
\end{figure}

The field effect devices were made from commercially obtained CVD graphene grown on copper (Graphenia SE) that has been transferred to a SiN (150 nm)/Si(100) substrate after wet chemical etching of copper with FeCl$_3$, using PMMA as a supporting layer for the transfer \cite{Li2009}. This partially leaves PMMA residues on the surface. It turned out that residues with heights up to 2\,nm do not leave any fingerprints in the surface potential maps (section~\ref{sec:dirt}). Moreover, the $V_{\rm CPD}(V_{\rm gate})$ curves at residues with height up to 5\,nm can well be described by the quantum capacitance model (section~\ref{sec:compressibility}) indicating that they only cause local doping. 
Larger clusters are partly appearing, but are carefully excluded for the quantitative analysis of the images (section~\ref{sec:dirt}). In addition, sweeping the graphene by contact mode AFM prior to mapping the surface potential removes all larger clusters and most of the small height residues (section~\ref{sec:sweep}), but barely changes the surface potential maps as well as the presence of current induced inverted electric fields (section~\ref{sec:dirt}). 
The thickness of the SiN $t_{\rm SiN}=151 \pm 1$\,nm has been determined via ellipsometry and its dielectric constant  $\epsilon_{\rm SiN} = 7.6 \pm 0.3$ has been measured capacitively. The source and drain electrodes are connected to the graphene at its edges according to a procedure described elsewhere \cite{Shaygan2017}. In short, the contacts are defined by a polymer resist AZ5214E via optical lithography followed by removal of graphene from the exposed areas by oxygen plasma in a reactive ion etching chamber operated at 100 W for 30 mins.  The contact metals are deposited directly afterwards in order to contact the open bonds of graphene.
This implies a small contact resistance at negligible overlap of the contacts and the graphene and, hence, negligible contact doping. 
For the contacts, firstly, a 12\,nm thick film of Ni is deposited by sputtering, which facilitates edge contact with low contact resistance. Then, a 50 nm thick film of Al is deposited via e-beam evaporation followed by lift-off of the resist. Subsequently, optical lithography is used to remove undesired graphene areas in the transversal direction of the current flow by etching in oxygen plasma. This leads to a size of the remaining graphene with width $W=28$\,$\mu$m and length $L=18.5$\,$\mu$m (Fig.~\ref{fig1_Transport_SampleOrientation}a--b). Finally, the samples are carefully rinsed in acetone to remove remaining AZ5214E from the surface.

Figure~\ref{fig1_Transport_SampleOrientation}a shows an optical image of a typical device with the AFM cantilever on top. The device design ensures minimum exposure of the cantilever body to the Si/SiN substrate and, hence, to the electric fields of the gate voltage $V_{\rm gate}$. Only $2.7\,$\,\% of the cantilever are in line of sight of the gate dielectric SiN as barely changing during scanning. Hence, gate voltages as high as $\pm 100\,V$ can be applied, while requiring less than 3\,V of compensation voltage in KPFM mode as enabled by the Bruker instrument. 
Furthermore, to reduce the influences of the source and drain voltages, we restrict the measurements to graphene areas away from source and drain as marked, e.g., by the central red rectangle in Fig.\ref{fig1_Transport_SampleOrientation}b exhibiting an increased contrast.

The graphene flake is marked as a  white rectangle in Fig.~\ref{fig1_Transport_SampleOrientation}b barely reaching below the contacts. 

\subsection{Transport Characterization}
\label{sec:transportmobility}
The contact resistivity has been determined for identically prepared devices with four transport contacts as $\rho_{\rm C}\sim 1\,{\rm k \Omega \mu m}$, slightly varying between samples and depending on the graphene doping. Hence, we deduce a contact resistance  $R_{\rm C}\sim 100 \,{\rm \Omega}$  for our samples.
This matches to the difference between the 2-point resistance and the deduced resistance of the graphene from $V_{\rm CPD}(x,y)$ maps (Fig,~\ref{fig1:ExptProcedure}e, main text) as well as to the two-point saturation resistance $R_{\rm C}=200\,{\rm \Omega}$ (Fig.~\ref{fig1_Transport_SampleOrientation}c). Note that the samples probed by EFM and KPFM are restricted to two contacts in order to reduce the influence of the backgate onto the tip.

The device mobility $\mu$, the gate voltage at charge neutrality $V_{\rm D}$, the residual doping $\Delta n_0$ at charge neutrality, and the contact resistance  $R_{\rm C}$ are deduced from  fitting the two-terminal $R_{\rm transport} (V_{\rm gate})$ using a relation developed for long range Coulomb type disorder \cite{Adam2007, DasSarma2011} (constant mobility):

\begin{equation}
R_{\rm transport} = R_{\rm C} + \frac{1}{\mu} \frac{L/W} {\sqrt{\left( C_{\rm gate}(V_{\rm gate}-V_{\rm D})\right)^2 +(\Delta n_0e)^2}}
\label{eq1:RVg}
\end{equation}

\noindent
with the length $L= 18.5$\,$\mu$m and the width $W = 28 $\,$\mu$m of the graphene area. The gate capacitance per unit area reads $C_{\rm gate} = \epsilon_{\rm SiN} \epsilon_0 /t_{\rm SiN}$ with $\epsilon_{\rm SiN} = 7.6$ and $t_{\rm SiN} = 150\,$nm as dielectric constant and thickness of the dielectric, respectively, and $\epsilon_0$ as the vacuum dielectric constant. The five  devices featured $\mu=1000 - 4000$\,${\rm cm^2/Vs}$, $\Delta n_0=(0.8 - 3.8) \times 10^{16}\,{\rm m^{-2}}$, and  a saturation resistance at high doping $R_{\rm C} = (160 - 300)\,{\rm \Omega}$ that we identify with the contact resistance. 

\subsection{Sequences of KPFM and EFM Images}

Five devices from two different chips are probed by KPFM and EFM, each for about one week and, at least, twice.  Since the measurements were performed at ambient conditions, the two terminal resistance $R_{\rm transport}$ of the devices varied continuously with time, namely the voltage at charge neutrality $V_{\rm D}$ and the mobility $\mu$. Moreover, a gate voltage hysteresis appeared, where the sweeps with decreasing $V_{\rm gate}$ (reverse sweeps) were more reproducible. A series of such sweeps for one device is shown in Fig.~\ref{fig1_Transport_SampleOrientation}c. Most of the presented data are recorded during sequences changing $V_{\rm gate}$ in reverse sweep direction with the only exception of Fig.~\ref{fig1:ExptProcedure}g. Moreover, we took care that $V_{\rm SD}$ was changed in the same direction prior to recording images that are compared directly. In addition, a waiting time was established after setting $V_{\rm gate}$ and $V_{\rm SD}$ for each EFM/KPFM image (section~\ref{subsec_DopingStability}, Fig.~\ref{fig8:ElecField_ImagingConditions}) as given in  table~\ref{table_Record_Parameters}. Finally, the two-point resistance $R_{\rm transport}(V_{\rm gate})$ was continuously monitored during imaging revealing minor variations as depicted in Fig.~\ref{fig1:ExptProcedure}e, main text and 
Figure~\ref{fig1_Transport_SampleOrientation}d. The latter  nicely reproduces the faster $R_{\rm transport}(V_{\rm gate})$ sweeps of the same device recorded without imaging (Fig.~\ref{fig1_Transport_SampleOrientation}c).

\subsection{Sweeping Graphene for Removal of Polymers}
\label{sec:sweep}
As discussed in section~\ref{sec:prep} and in more detail in section~\ref{sec:dirt}, there are remaining polymers on the graphene with larger clusters being detrimental for KPFM and EFM studies. To remove these polymers, we used contact mode AFM with a hard Si cantilever ($\mu$masch QQ-NSC15/AlBS, 40\,N/m) serving as a broom\cite{Lindvall2012,Goossens2012}. The broom was scanned in contact-mode from drain to source with speed $4\,\mu$m/s. The setpoint has been optimized manually, being large enough to remove the polymers and low enough to avoid rupture of the graphene. After cleaning an area of ($\sim 10$\,$\mu$m)$^2$, ridges and hills of residues appear at the surroundings of the swept area with heights of 50--100\,nm and widths of about 500\,nm, while the swept interior is free of larger clusters (Fig.\ref{fig:S12c}g). Subseqeuntly, we changed the cantilever back to the PtIr covered SCM-PIT-V2 for KPFM or EFM. 

\subsection{Ion Bombardment}
\label{sec:ionbomb}
For ion bombardment, the chip with multiple graphene field effect devices was firstly placed into a contacted Cu box ($8\times 8 \times 5$\,mm$^3$) shielding the ion flux that features a mm sized hole for targeting the investigated device with ions. This device was contacted to the Cu box via its bond wire such that charging effects on the chip by the ion flux are minimized and the ion current could be comfortably measured across the relatively large area of the Cu box. The Ar$^+$ ions are produced by a plasma source equipped with a Wien filter for energy selection. We checked that the ion current is homogeneous across the size of the Cu box on the 10\% level. Moreover, the vacancy production yield per ion has been carefully calibrated by repetitively counting the produced defects per area with scanning tunneling microscopy in ultra-high vacuum.\cite{Just2014} This revealed a yield of 0.35 vacancies/ion at $E_{\rm kin}=50$\,eV. It is well known that only single vacancies can be produced at such low ion energy \cite{ElBarbary2003}. We used a low ion flux ($3.5\cdot10^{13}$/m$^2$s) for moderate time (580\,s) leading to a vacancy density $n_{\rm vac}=7\cdot10^{15}$/m$^2$, i.e. the vacancies are on average 12\,nm apart.

\section{Kelvin Probe Force Microscopy}
\label{sec2:AM KPFM}

\begin{figure}
	\centering
	\includegraphics[width=160 mm]{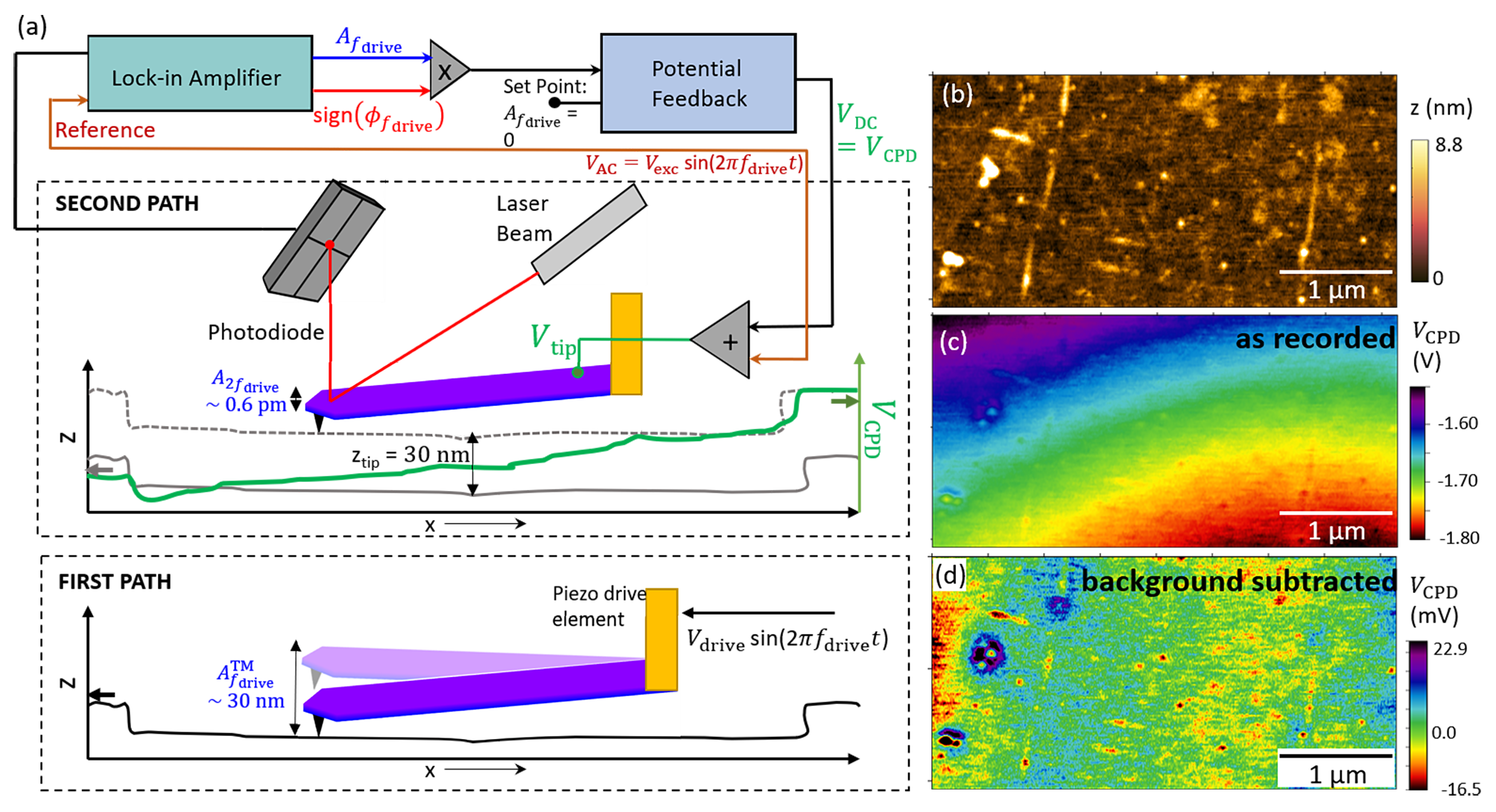}
	\caption{\textbf{Kelvin Probe Force Microscopy}. (a) Schematic of amplitude modulation KPFM as implemented in the Dimension Icon AFM setup from Bruker \cite{DimensionIcon}. The lower image shows a profile line $z(x)$ (black) recorded by the cantilever (violet) in tapping mode AFM with amplitude $A^{\rm TM}_{f_{\rm drive}}$
		excited via the voltage $V_{\rm drive}\sin{(2\pi f_{\rm drive} t)}$ applied to the piezo drive of the cantilever. The middle image shows the profile line from the lower image (full grey line) and the same line lifted by $z_{\rm tip}$ (dashed grey line) as traced during recording $V_{\rm CPD}(x)$  (green line). 
		$V_{\rm CPD}$ is determined at each point by the feedback loop depicted on top. It minimizes the cantilever  amplitude $A_{f_{\rm drive}}$ at frequency $f_{\rm drive}$, while the AC voltage $V_{\rm AC}$ is applied to the tip with constant amplitude $V_{\rm exc}$   simultaneously with a DC voltage $V_{\rm DC}$, that itself is optimized by the feedback. A remaining amplitude $A_{ 2f_{\rm drive}}$ appears at frequency $2f_{\rm drive}$  (eq.~(\ref{eq345_Fes_2w})). $\phi_{f_{\rm drive}}$ measures the relative phase between $V_{\rm AC}$ excitation and detected cantilever oscillation.
		(b) Topography of graphene area for one of the measured devices recorded by tapping mode AFM. (c) Corresponding $V_{\rm CPD}(x,y)$ map, acquired by KPFM at $V_{\rm gate} = V_{\rm D} = \,$87\,V,  $V_{\rm SD} = 0$\,V. (d) Same map as in c after subtracting a second order polynomial background for horizontal and vertical direction \cite{Necas2012}.}
	\label{fig2_KPFM}
\end{figure}

One of the two scanning probe microscopy methods employed to image the current induced electric fields is Kelvin Probe Force Microscopy with amplitude modulation (AM-KPFM). Figure~\ref{fig2_KPFM}a illustrates its  implementation in the Dimension Icon AFM setup from Bruker as used here. It employs a two-path process conducted  for each scan line. During the first path, the topography of the sample is acquired by tapping mode AFM with cantilever amplitude of 30--33\,nm. At the second path, called the lift mode, the tip retraces the measured topography from the first path with an additional tip-sample distance $z_{\rm tip}\approx 30$\,nm, while the contact potential difference of the tip with respect to the sample, $V_{\rm CPD}(x,y)$, is determined at each tip position $(x,y)$. $V_{\rm CPD}(x,y)$ measures the difference between the work function of the tip $W_{\rm tip}$ and the the work function of the sample area below the tip $W_{\rm sample}(x,y)$, i.e. $V_{\rm CPD}(x,y)=W_{\rm tip}-W_{\rm sample}(x,y)$. During tapping mode, the cantilever oscillation is mechanically driven by a piezo-electric element while the cantilever scans the sample surface employing a $z$-feedback that regulates the tip-sample distance $z$ to maintain a constant amplitude of the cantilever oscillation $A_{f_{\rm drive}}^{\rm TM}\approx\,$30 nm. The drive amplitude of the piezoelectric element is $V_{\rm drive}=2-3$\,V at a drive frequency $f_{\rm drive}$ that is chosen at 50\,Hz below the resonance frequency of the free cantilever $f_{\rm res}$ and given in table~\ref{table_Record_Parameters}. 

During the second path, both the $z$-feedback and the piezo drive voltage are switched off, while a potential feedback regulates the DC voltage applied to the cantilever $V_{\rm DC}$. An AC modulation $V_{\rm AC}(t)=V_{\rm exc}{\rm sin}(2\pi f_{\rm drive}t)$ of constant amplitude $V_{\rm exc}$ is superposed to $V_{\rm DC}$ such that the total tip voltage reads $V_{\rm tip}(t) = V_{\rm DC} + V_{\rm exc}{\rm sin}(2\pi f_{\rm drive}t)$. This causes the electrostatic tip-sample interaction force $F_{\rm es}$ to be modulated as \cite{Zerweck2005}

\begin{equation}
F_\mathrm{es} (t)= \frac{1}{2}\, \frac{d C_{\rm ts}(z)}{dz} (V_\mathrm{tip}(t) - V_\mathrm{CPD})^2 = F_\mathrm{es}^0 + (F_\mathrm{es})_{f_{\rm drive}}(t) + (F_\mathrm{es})_{2f_{\rm drive}}(t) 
\label{eq2:TipSampleElecForce}
\end{equation}

\noindent 
where $C_{\rm ts}(z)$ is the distance dependent tip-sample capacitance and

\begin{align}
F_\mathrm{es}^0 & = \frac{1}{2} \frac{d C_{\rm ts}}{dz} \left[(V_\mathrm{DC} - V_\mathrm{CPD})^2 + \frac{V_{\rm exc}^2}{2} \right] \\
(F_\mathrm{es})_{f_{\rm drive}} (t)& = \frac{d C_{\rm ts}}{dz} (V_\mathrm{DC} - V_\mathrm{CPD}) V_{\rm exc} \, {\rm sin}(2\pi f_{\rm drive} t) \\
(F_\mathrm{es})_{2f_{\rm drive}} (t)& = \frac{1}{4} \frac{dC_{\rm ts}}{dz} \, V_{\rm exc}^2 \, {\rm cos}(4\pi f_{\rm drive} t)
\label{eq345_Fes_2w}
\end{align}

\noindent

It can be shown that the oscillating  $F_{\rm es}(t)$ causes the cantilever to vibrate with $z_{f_{\rm drive}}(t) \propto Q\cdot(F_{\rm es})_{f_{\rm drive}}(t)/k$ at excitation frequency $f_{\rm drive}$, if $f_{\rm drive}$ is close to $f_{\rm res}$.\cite{Zerweck2005} Here, $k\approx \,$3 N/m is the stiffness constant of the cantilever and $Q \approx 250$  its quality factor  (SCM-PIT-V2, Bruker \cite{SCMPITV2}). An additional oscillation $z_{2f_{\rm drive}}(t) \propto (F_{\rm es})_{2f_{\rm drive}}(t)/k$ appears at frequency $2f_{\rm drive}$. The amplitude and phase lag of the cantilever   oscillation at $z_{f_{\rm drive}}(t)$ with respect to the driving $V_{\rm AC}(t)$ is detected by a lock-in amplifier and passed to the potential feedback that nullifies $z_{f_{\rm drive}}(t)$ by adjusting $V_{\rm DC}$. Using $V_{\rm CPD}$ as the adjusted $V_{\rm DC}$ for each position $(x,y)$, a map $V_{\rm CPD}(x,y)$ results as displayed in Fig.~\ref{fig2_KPFM}c. 

Since $z_{f_{\rm drive}}(t)$ is rendered zero by the feedback, the cantilever oscillates only with $z_{2f_{\rm drive}}(t)$. The amplitude can be estimated  by eq.~(\ref{eq345_Fes_2w}) using $V_{\rm exc} = 1.066$\,V, tip-sample distance $d = A_{f_{\rm drive}}^{\rm TM} + z_{\rm tip} = 60$\,nm  and assuming the simple model of a charged sphere above an infinite metal surface for the tip-sample capacitance leading to \cite{Hudlet1998} $\frac{dC_{\rm ts}}{dz} = 2\pi \epsilon_0 \frac{R^2}{d(d+R)}$ with tip radius $R = 25$\,nm. One obtains an amplitude $A_{2f_{\rm drive}}  \approx 0.6$\,pm $\ll d$.

\subsection{Cantilever Contribution}
\label{Sec2_subsec1_CantContr}

\begin{figure}
	\centering
	\includegraphics[width=160 mm]{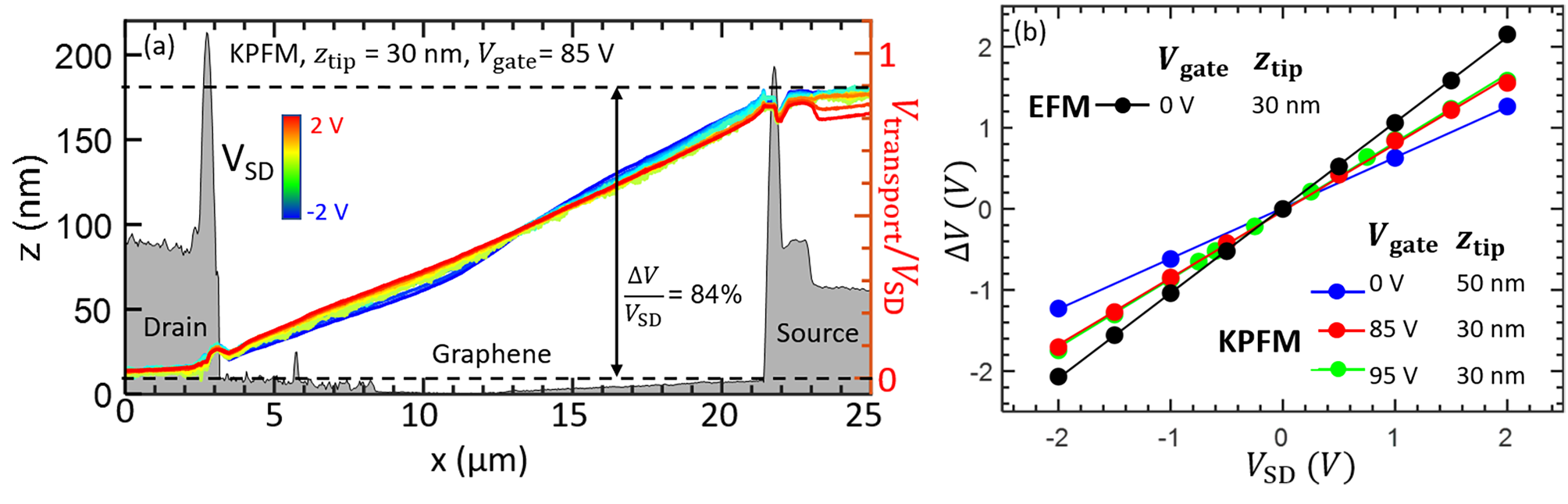}
	\caption{\textbf{Transport Potential Drop in KPFM.} (a) Transport potential drops across graphene (colored lines), each averaged from 50 distinct profile lines at the same $V_{\rm SD}$ and normalised by $V_{\rm SD}$ (right vertical scale). Various $V_{\rm SD}$ differing by 0.5\,V are  color coded, $z_{\rm tip} = 30\,$nm, $V_{\rm gate} = 85\,$V, $\Delta V := V_{\rm transport}(x=25\,{\rm \mu m}) - V_{\rm transport}(x=0\,{\rm \mu m})$. A topography profile probed by tapping mode AFM is added in grey (left vertical scale). (b) $\Delta V (V_{\rm SD})$ for different $V_{\rm gate}$, $z_{\rm tip}$ as marked (slopes: blue: 0.62, red: 0.84, green: 0.82, black: 1.0). For the EFM data, $\Delta V$ is deduced from shifts of the maxima of the EFM parabolas measured on the source ($x=\,41\,{\rm \mu m}$) while the drain is grounded (section \ref{sec3:EFM}).}
	\label{fig3:CantContr}
\end{figure}

The above description is simplified by assuming a homogeneous surface potential and by neglecting the influence of the electrostatic interaction of the cantilever body with the device. Both are relevant due to the long range nature of electrostatic forces. More precisely, the recorded $V_{\rm CPD}(x,y)$ is described at each point $(x,y)$ by a convolution of the surface potential map with a  point spread function (PSF) deduced from the tip geometry at distance  $z_{\rm tip}$ \cite{Strassburg2005, Xu2018}. Since AM-KPFM nullifies the $f_{\rm drive}$ component of the tip sample force $F_{\rm es}(t)$ (eq.~(\ref{eq:2})) and not the force gradient $dF_{\rm es}/dz$ as EFM (section~\ref{sec3:EFM}), the cantilever body as well as tip areas more distant from the apex influence the measured $V_{\rm CPD}$ \cite{Xu2018}. This impact is known to reduce the signal intensity, but barely the spatial resolution \cite{Xu2018}. For the cantilever  of our measurements (SCM-PIT-V2) and $z_{\rm tip} = 30\,$nm, the approximate PSF has been given  by Xu {\it et al.}\cite{Xu2018} revealing a spatial resolution of 60\,nm and a reduction of signal intensity by $20\,\%$ assuming a homogeneous surface potential.
This is compatible with our results. 
Indeed, the relative drop of transport potential  across the graphene as measured by AM-KPFM is only $\sim 80$\,\% of the applied $V_{\rm SD}$ (Fig.~\ref{fig3:CantContr}), but it is 100\,\%  using EFM  (Fig.~\ref{fig3:CantContr}b, section \ref{sec3:EFM}).\cite{Xu2018} The relation for KPFM turns out to be independent on the applied $V_{\rm SD}$ (Fig.~\ref{fig3:CantContr}a) and $V_{\rm gate}$ (Fig.~\ref{fig3:CantContr}b), but changes with $z_{\rm tip}$ (Fig.~\ref{fig3:CantContr}b) as predicted via its PSF.\cite{Xu2018} 
Note that the potential drop across graphene was determined 3\,$\mu$m away from the lateral metal-graphene interface avoiding influences of the potential inhomogeneities there. Hence, our results confirm the considerations by PSF\cite{Xu2018} implying a residual long range influence.

\subsection{Background Subtraction}
\label{subsec_BG}

Figure~\ref{fig2_KPFM}c shows a KPFM map of a graphene area recorded at $V_{\rm SD} =$0\,V and $V_{\rm gate} =$87\,V, i.e., at charge neutrality. The displayed area is, at least,  $3$\,${\rm \mu}$m away from the source and drain electrodes and from exposed SiN areas. Thus, residual backgrounds cannot result from the tip apex rendering them long range.  Indeed, a dominating, long range contrast developing from the  lower right to the upper left appears. It is likely caused by remaining interactions of the cantilever body with exposed SiN areas. Such long-range background is only found in AM-KPFM maps but not in EFM maps corroborating that it is not caused by the tip apex. The increase of $V_{\rm CPD}(x,y)$ towards the upper left meets the expectation from the geometry of the exposed SiN areas (Fig.~\ref{fig1_Transport_SampleOrientation}a) and the positive $V_{\rm gate}$. Due to the inclination of $10\degree$ of the cantilever towards the tip, the distance of the cantilever body part above SiN decreases relative to the SiN by moving the cantilever into this direction and, hence, the interaction force to the SiN increases.  

To get rid of the long-range background, we apply a 2$^{\rm nd}$ order polynomial background subtraction using Gwyddion \cite{Necas2012}. The resulting map after subtraction is displayed in Fig.~\ref{fig2_KPFM}d featuring an obvious correlation with the simultaneously recorded topography (Fig.~\ref{fig2_KPFM}b),  e.g., at the two nearly vertically propagating wrinkles or at the two large bumps on the left that obviously change the local doping. This justifies the subtraction method for background removal featuring a spatial resolution of about 20\,nm and a $V_{\rm CPD}$ resolution  below 10\,mV. 

Of course, the background subtraction also removes the spatial average of $V_{\rm CPD}$ and the linear potential drop induced by $V_{\rm SD}$. They are, however, required for the doping maps (section~\ref{Sec5_DopingDistr_CPD}) and the electric field maps(section~\ref{sec4_EField}), respectively.
For the electric field maps, we do not apply the background subtraction, but rely on the subtraction of two $V_{\rm CPD}(x,y)$ maps
recorded at the same $V_{\rm gate}$ (eq.~(\ref{eq10:Vtransport})) and, hence, removing the background, that is dominated by the gate voltage that penetrates to the cantilever, automatically.  For the doping maps (eq.~(\ref{DopingCPDrelation})), we add an averaged value $\overline{V}_{\rm CPD}$ to the background subtracted $V_{\rm CPD}(x,y)$ as deduced from a straightforward capacitive charging model of the graphene by $V_{\rm gate}$ (section \ref{Sec5_DopingDistr_CPD}). This procedure is corroborated by the excellent agreement between the simulated electric field maps (Fig.~\ref{fig2:SDind_doping}f,h, main text) as directly deduced from doping maps (section~\ref{Sec5_DopingDistr_CPD}) with the experimentally measured electric field maps (Fig.~\ref{fig2:SDind_doping}g,i, main text).

\subsection{Noise and Sensitivity}
\label{subsec_Noise_Sensitivity}

Two major sources of noise are known for KPFM.\cite{Li2012} Firstly, thermal fluctuations of the cantilever oscillation are present due to Brownian motion. Secondly, sensor noise from detecting the optical beam deflection (OBD)  contributes. Using established formulas,\cite{Li2012} we find the thermal noise to be $n_{\rm thermal} = 1.9\,{\rm pm}/\sqrt{{\rm Hz}}$ using our parameters $Q\,=\,250, \, f_{\rm res} = 62.34\,{\rm kHz}\,$ and $T = 300\,$K. A typical value for the OBD sensor noise is $n_{\rm OBD} = 100\,{\rm fm}/\sqrt{{\rm Hz}}$,\cite{Li2012} i.e., negligible. For the typical bandwidth of the feedback loop $B\approx 5 \,$Hz, we get an amplitude noise at $f_{\rm drive}$ of $n_{\rm thermal}\sqrt{B}\approx 4$\,pm. This imposes a lower limit on the $V_{\rm CPD}$ precision. At $z_{\rm tip} = 30\,$nm and AC voltage amplitude $V_{\rm AC} = 1.066\,$V, we find this limit to be $\delta V_{\rm CPD}=10\,$mV.\cite{Li2012} 
This is roughly consistent with the noise in $V_{\rm CPD}(x,y)$ maps, e.g., in Fig.~\ref{fig2_KPFM}d, i.e., we find RMS fluctuations within areas of $0.5\,\mu$m $\times 0.5\,\mu$m of 5--20 mV. It also roughly agrees with the data sheet from Bruker promising a noise level of $\sim 10\,$mV and with other experiments in the literature.\cite{Yu2009, Wilke2016, Zerweck2005}

\section{Electrostatic Force Microscopy}
\label{sec3:EFM}

\begin{figure}
	\centering
	\includegraphics[width = 160 mm]{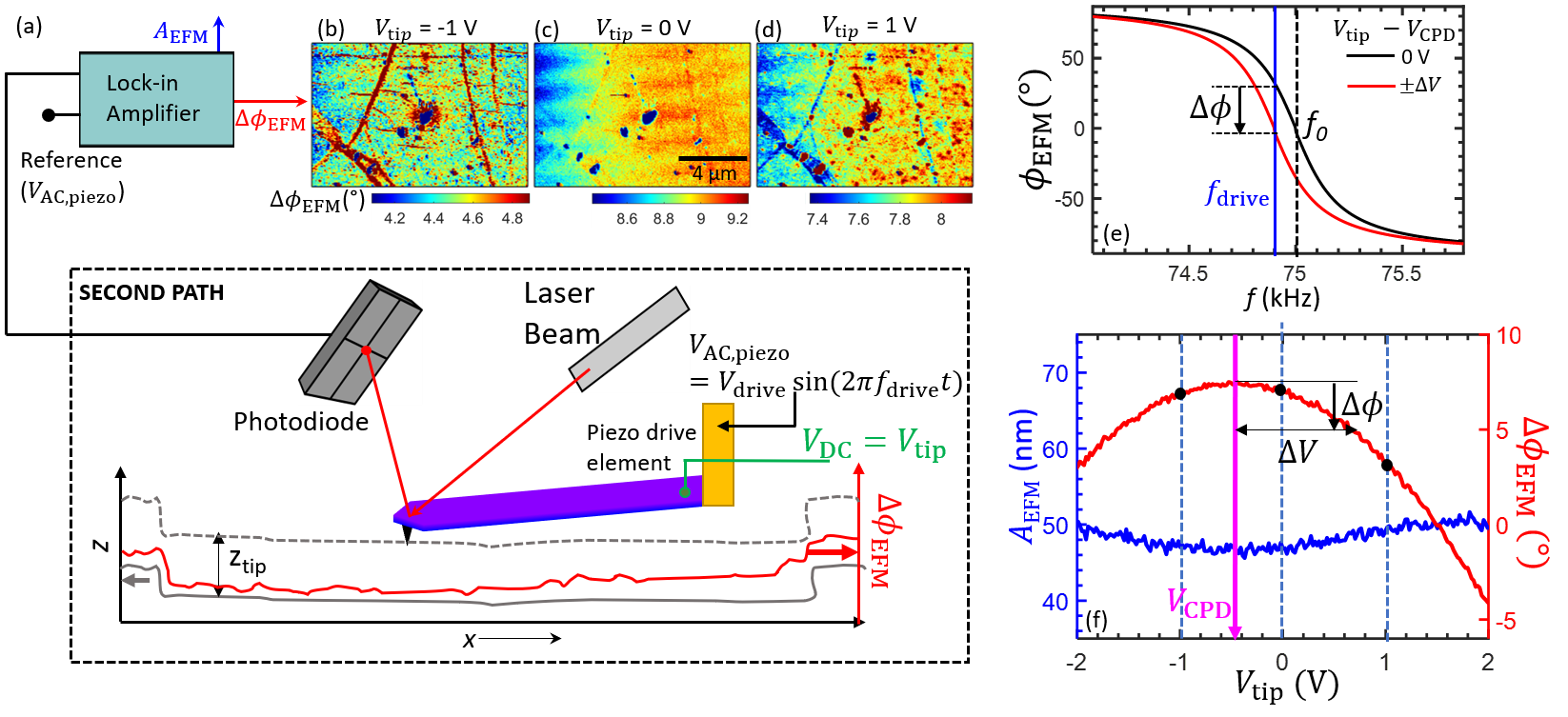}
	\caption{{\bf Electrostatic Force Microscopy}. (a) Sketch of the cantilever (violet) during the second path (lift mode) including the detection scheme (laser, photodiode). Full grey line: Topography $z(x)$ recorded during the first path. Dashed grey line: Topography line after lifting the cantilever by $z_{\rm tip}$ as tracked during the second path. Red line: phase lag of cantilever oscillation $\Delta \phi_{\rm EFM}(x)$ as recorded during the second path. The cantilever is continuously excited by $V_{\rm AC, piezo}$, applied to the piezo element at frequency $f_{\rm drive}$ slightly below the resonance frequency of the free cantilever $f_{\rm res}$. A DC tip voltage $V_{\rm tip}$ changes the amplitude $A_{\rm EFM}$ and phase lag $\Delta \phi_{\rm EFM}$ of the cantilever oscillation at $f_{\rm drive}$ as recorded by lock-in amplifier. (b--d) Three exemplary  $\Delta \phi_{\rm EFM}(x,y)$ maps of {\it one} graphene area (topography in Fig.~\ref{fig5:Gallery_FieldInversion}h) recorded at three different $V_{\rm tip}$ as marked. (e) Simulation of $\Delta \phi_{\rm EFM}(f)$ for two distinct $ V_{\rm tip}-V_{\rm CPD}$,  $f_0 = 75\,$kHz, $Q= 240$ as present for the used  SCM-PIT-V2 cantilever. The downward arrow marks the decrease in $\Delta \phi_{\rm EFM}$ at $f_{\rm drive}$ due to the electrostatic force induced by $V_{\rm tip}$. (f) Measured  $A_{\rm EFM}$ (blue, left axis) and $\Delta \phi_{\rm EFM}$ (red, right axis) as function of $V_{\rm tip}$ for a centrally located point on graphene, $V_{\rm gate}=65\,$V, $V_{\rm SD} = 0\,$V. $V_{\rm CPD}$ marks the voltage at the $\Delta \phi_{\rm EFM}$ maximum. $\Delta \Phi$ marks the phase lag shift with respect to $V_{\rm CPD}$ induced by $V_{\rm tip}=0.8$\,V. The vertical dashed lines with black dots indicate the three $V_{\rm tip}$ as used for recording $\Delta \phi_{\rm EFM}(x,y)$ maps and subsequently deducing $V_{\rm CPD}(x,y)$ maps.}
	\label{fig4_EFM}
\end{figure}

As second method, we employed Electrostatic Force Microscopy (EFM),  well established to map surface potentials with high resolution \cite{Girard2001, Altvater2019}. Like KPFM (section \ref{sec2:AM KPFM}), it uses a two step process acquiring the topography during a first path in tapping mode.  After lifting the tip for the second path  by $z_{\rm tip}$, the cantilever is oscillated by an AC voltage applied to a piezo-electric element using a frequency $f_{\rm drive}\sim62$\,kHz slightly below the resonance frequency of the free cantilever. A DC voltage $V_{\rm tip}$ is applied to the tip  (Fig.~\ref{fig4_EFM}a) inducing an attractive electrostatic force between tip and sample due to image charges in the sample (eq.~(\ref{eq2:TipSampleElecForce})). This causes a decrease in resonance frequency of the cantilever $f_0$ with resulting shift of the phase vs. frequency curve $\phi_{\rm EFM}(f)$ (Fig.~\ref{fig4_EFM}e). Probing at the excitation frequency $f_{\rm drive}< f_0$, the electrostatic force changes the phase lag $\Delta \phi_{\rm EFM}$ between cantilever oscillation and voltage oscillation at the piezo  (Fig.~\ref{fig4_EFM}e) as well as the amplitude of the cantilever oscillation $A_{\rm EFM}$. The measured $\Delta \phi_{\rm EFM}$ depicts the additional phase lag of the exciting oscillation with respect to the cantilever oscillation after being nullified prior to each measurement at the resonance frequency of the free cantilever oscillation. Both, $A_{\rm EFM}$ and $\Delta \phi_{\rm EFM}$ are related to the electrostatic tip-sample force gradient that must be nullified to detect the local $V_{\rm CPD}$. 

The resulting $\Delta \phi_{\rm EFM} $ as function of tip voltage $V_{\rm tip}$ features an inverted parabola $\Delta \phi_{\rm EFM} (V_{\rm tip})$ (Fig.~1b, main text). This is due to the cantilever softening that is caused by the attractive electrostatic interaction between the charged tip and the induced charge within the sample. The maximum of the parabola is, hence, at vanishing electrostatic force, i.e., at $V_{\rm tip}=V_{\rm CPD}$.\cite{Xu2018} 

We repeat the derivation of $\Delta \phi_{\rm EFM} (V_{\rm tip})$ in the following.\cite{Girard2001, Altvater2019} Assuming the cantilever as a damped, driven harmonic oscillator disturbed by a force that is small compared to the restoring force of the cantilever and that varies slowly on the scale of the oscillation amplitude, the following expression applies for the induced resonance frequency shift $\Delta f_0$
\cite{Giessibl2003} ($k$: cantilever stiffness constant, $F_{\rm es}$: electrostatic force according to eq.~(\ref{eq2:TipSampleElecForce})).

\begin{equation}
\Delta f_0 = -\frac{f_0}{2k}\, \frac{\partial F_{\rm es}}{\partial z}.
\label{eq6:FrequencyShift}
\end{equation}

For $f_{\rm drive}\approx f_0$ and $Q\cdot \Delta f_0/f_0 \ll 1$, the corresponding phase shift reads:\cite{Altvater2019}

\begin{equation}
\Delta \phi_{\rm EFM}(f_{\rm drive}) = -\arcsin\left({\frac{2Q}{f_0}\,\Delta f_0}\right) \approx -\frac{Q}{k}\,\frac{\partial F_{\rm es}}{\partial z}.
\label{eq7:Phase_force}
\end{equation}

\begin{figure}
	\centering 
	\includegraphics[width = 160 mm]{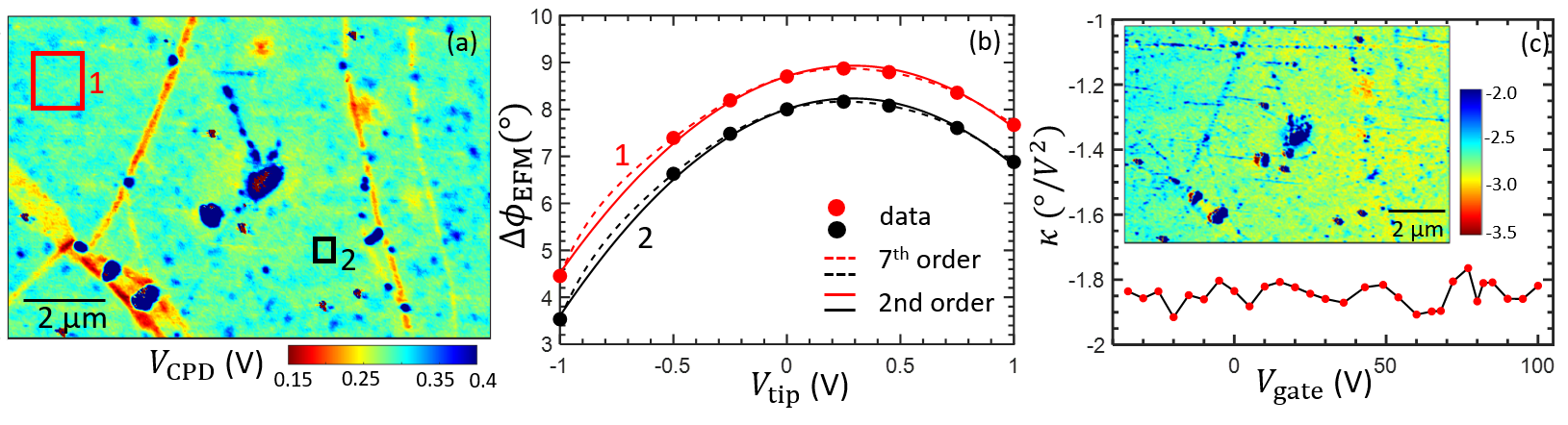}
	\caption{{\bf Parabolic fits of $\Delta \phi_{\rm EFM}(V_{\rm tip})$}. (a) $V_{\rm CPD}(x,y)$ map of graphene measured by EFM via parabolic fits of eight recorded $\Delta \phi_{\rm EFM}(V_{\rm tip})$ values  for each location $(x,y)$ (eq.~(\ref{eq9:Phi_Vtip_fit})), $V_{\rm gate} = 100\,V,\, V_{\rm SD} = 0\,V, \, z_{\rm tip} = 30\,$nm. Corresponding topography: Fig.~\ref{fig5:Gallery_FieldInversion}h. (b) Variation of $\Delta \phi_{\rm EFM}$ with $V_{\rm tip}$ for the two marked regions in a. The solid/dashed lines are fits to the measured data points with the specified polynomial order. The black curves are shifted downwards for better visibility. (c) Curvature $\kappa(V_{\rm gate})$ resulting from parabolic fits as in b (eq.~\ref{eq9:Phi_Vtip_fit}) and displayed for a single location. Inset: $\kappa(x,y)$ of the same area as in a, $V_{\rm gate} = 100\,V,\, V_{\rm SD} = 0\,V, \, z_{\rm tip} = 30\,$nm.}
	\label{fig5:EFM_3ParabolaPmts}
\end{figure}

\noindent 
Hence, the change of $\Delta \phi_{\rm EFM}(f_{\rm drive})$ is proportional to the force gradient. This renders EFM more sensitive to local surface potentials than AM KPFM, that minimizes the force instead (section~\ref{Sec2_subsec1_CantContr}). Long range forces acting on  the cantilever body contribute much less to EFM via their less steep gradient \cite{Xu2018}. Indeed, $\Delta \phi_{\rm EFM}(x,y)$ maps are found devoid from any significant long range background (Fig.~\ref{fig4_EFM}b--d, Fig.~\ref{fig5:EFM_3ParabolaPmts}a). The spatial resolution of EFM improves with decreasing $z_{\rm tip}$. For the selected SCM-PIT-V2 cantilever and $z_{\rm tip}=20$\,nm  (Fig.~\ref{fig3:NegFields_ViscousFlow}, main text), the resolution is approximately $1.1 \cdot z_{\rm tip}+ 11$\,nm $= 33\,$nm.\cite{Xu2018}

Substituting $F_{\rm es}$ from eq.~(\ref{eq2:TipSampleElecForce}) into eq.~(\ref{eq7:Phase_force}) implies a quadratic dependence of $\Delta \phi_{\rm EFM}$ on $V_{\rm tip}$ reading

\begin{equation}
\Delta \phi_{\rm EFM}(x,y) = -\frac{Q(x,y)}{2k}\,\frac{d^2 C_{\rm ts}(x,y)}{dz^2}\,\left(V_{\rm tip} - V_{\rm CPD}(x,y)\right)^2,
\label{eq8:Phi_Vtip_th}
\end{equation}

\noindent
where the quality factor $Q$ and the tip sample capacitance $C_{\rm ts}$ are position dependent as affected by local variations of dissipation and screening, respectively, due to, e.g., adsorbates or doping fluctuations. 
Nevertheless, the maximum of $\Delta \phi_{\rm EFM}(V_{\rm tip})$ is a direct measure of $V_{\rm CPD}$.
Indeed, the measured $\Delta \phi_{\rm EFM}(V_{\rm tip})$ at a single point features a parabola as shown by the fit in Fig.~\ref{fig1:ExptProcedure}b, main text. Minor deviations at larger voltages are most likely caused by the influence of quantum capacitance that is not captured by eq.~(\ref{eq2:TipSampleElecForce}).
To determine $V_{\rm CPD}$ for each position $(x,y)$, we probe $\Delta \phi_{\rm EFM}(x,y)$  at only three $V_{\rm tip}= -1, 0, 1$\,V (Fig.~\ref{fig4_EFM}f) and use the three points for fitting a parabolic function 
\begin{equation}
\Delta \phi_{\rm EFM}(x,y) = -\kappa(x,y) \left(V_{\rm tip} - V_{\rm CPD}(x,y)\right)^2 + \Delta \phi_0(x,y)
\label{eq9:Phi_Vtip_fit}
\end{equation}
via three parameters, namely the curvature $\kappa (x,y)$ and the maximum at $(\Delta \phi_0$, $V_{\rm CPD})(x,y)$. 
Hence, we take the local nature of all three parameters into account with $\Delta \phi_0(x,y)$ caused by local forces that are not dependent on $V_{\rm tip}$.
We choose the extrema of $V_{\rm tip}=\pm 1\,$V large enough to obtain high precision of the fit, but still small enough to avoid deviations from the parabola due to quantum capacitance.  

To illustrate the principle, Fig.~\ref{fig5:EFM_3ParabolaPmts}a displays a $V_{\rm CPD}(x,y)$ map of an electron doped graphene area as obtained by a parabolic fit using eight $V_{\rm tip}$ for each position $(x,y)$ instead of three, hence, enabling better precision. Features are discernable at several graphene folds, wrinkles and point defects  (compare with the topography in Fig.~\ref{fig5:Gallery_FieldInversion}h). These topographic features obviously exhibit distinct surface potentials. Importantly, no long range background had to be subtracted. Figure~\ref{fig5:EFM_3ParabolaPmts}b shows two selected sets of measured $\Delta \phi_{\rm EFM}(V_{\rm tip})$ at the eight $V_{\rm tip}$ together with a parabolic fit and a 7$^{\rm th}$ order polynomial fit. The two fits reveal
nearly identical maxima. 

Figure~\ref{fig5:EFM_3ParabolaPmts}c displays the variation of the curvature of the fitted parabolas with $V_{\rm gate}$ (main) and with position (inset).
While the $V_{\rm gate}$ dependence is rather irregular exhibiting fluctuations of about 5 \% only, the position dependence exhibits features at folds and wrinkles that likely exhibit different dissipation strengths leading to different local $Q(x,y)$.
Variations in $\Delta \Phi_0$, related to the forces that do not depend on $V_{\rm tip}$, are about 0.6$^\circ$ for, both, $V_{\rm gate}$ dependence and position dependence, but without obvious correlations to topographic features (not shown).

\subsection{Noise and Sensitivity}
\label{sec:EFMnoise}

As described above, for $V_{\rm CPD}(x,y)$ mapping by EFM, we recorded only three $\Delta  \phi_{\rm EFM}(V_{\rm tip})$ at each location
that are fitted by a parabola (eq.~(\ref{eq9:Phi_Vtip_fit})).
The resulting noise is estimated in the following. Continuous sweeps of $\Delta \phi_{\rm EFM} (V_{\rm tip})$ consisting of 512 points and recorded at different $V_{\rm gate}$ revealed an average RMS deviation from the parabola of $\delta \phi_{\rm EFM} =0.09\degree$ at a recording time per point of 1.5\,ms. The RMS deviation with respect to the parabola dropped continuously, if one employs Gaussian averaging of $\Delta \phi_{\rm EFM} (V_{\rm tip})$ prior to fitting, as expected for uncorrelated deviations. Moreover, fits of higher polynomial order only slightly decreased $\delta \phi_{\rm EFM}$, e.g., by less than 5\% up to  a fit order of ten. This corroborates that the parabola is a very good approximation of  $\Delta \phi_{\rm EFM} (V_{\rm tip})$.
Since recording times of $3.4\,{\rm ms}$ are employed for the three $\Delta  \phi_{\rm EFM}(V_{\rm tip})$ used for $V_{\rm CPD}(x,y)$ maps, the RMS error becomes $\delta \phi_{\rm EFM} = 0.09 \degree \times \sqrt{1.5/3.4} = 0.06 \degree$.
We used this $\delta \phi_{\rm EFM}$ to deduce the resulting error in $V_{\rm CPD}$ by numerical simulations revealing $\delta V_{\rm CPD}\approx 2$\,mV at $V_{\rm CPD}\approx 0.2$\,V such as in Fig.~3, main text. This largely fits with the uncorrelated noise that we observe in maps experimentally (section~\ref{Efield_noise_filter}).   

\section{Parameters for Recording KPFM and EFM Maps}

The following table~\ref{table_Record_Parameters} describes all key parameters that are used to record $V_{\rm CPD}(x,y)$ maps. These maps are the base of all figures in the main text and in the supplement. Maps that are used in multiple images are only described once. For the other images, a reference is given in the caption to the image where the map is first used. 

\begin{center}
	\setlength{\tabcolsep}{0.25pt}
	\begin{longtable}{|c|c|c|c|c|c|c|c|c|c|c|}
		
		\caption{Parameters for recording $V_{\rm CPD}(x,y)$ maps. Most figures in the main text and the supplement are based on several $V_{\rm CPD}(x,y)$ maps numbered consecutively by $V_{\rm CPD}$ No., $z_{\rm tip}$: lifting height after tapping mode AFM, $V_{\rm tip, ac}$: amplitude of tip voltage for KPFM, $A_{\rm ac}$: cantilever oscillation amplitude for EFM,  $f_{\rm drive}$: excitation frequency, $\Delta t$: recording time of $V_{\rm CPD}(x,y)$ image,
			$\delta t$: waiting time without recording after last $V_{\rm CPD}(x,y)$ image.} \\
		
		\hline
		\multicolumn{1}{|c|}{Figure} & 
		\multicolumn{1}{c|}{Method} & 
		\multicolumn{1}{c|}{$V_{\rm SD}$} & 
		\multicolumn{1}{c|}{$V_{\rm CPD}$} & \multicolumn{1}{c|}{$z_{\rm tip}$} & \multicolumn{1}{c|}{$V_{\rm tip, ac}$} & \multicolumn{1}{c|}{$A_{\rm ac}$} & 
		\multicolumn{1}{c|}{$f_{\rm drive}$} & \multicolumn{1}{c|}{$\Delta t$} & 
		\multicolumn{1}{c|}{$\delta t$} & 
		\multicolumn{1}{c|}{previous} \\ 
		
		\multicolumn{1}{|c|}{} & 
		\multicolumn{1}{c|}{} & 
		\multicolumn{1}{c|}{(V)} & 
		\multicolumn{1}{c|}{No.} & 
		\multicolumn{1}{c|}{(nm)} & 
		\multicolumn{1}{c|}{(V)} & 
		\multicolumn{1}{c|}{(nm)} & 
		\multicolumn{1}{c|}{(kHz)} & 
		\multicolumn{1}{c|}{(mins)} & 
		\multicolumn{1}{c|}{(mins)} & 
		\multicolumn{1}{c|}{$V_{\rm CPD}$}\\
		\hline 
		\endfirsthead
		
		\multicolumn{3}{c}%
		{{\bfseries \tablename\ \thetable{} -- continued from previous page}} \\
		\hline 
		\multicolumn{1}{|c|}{Figure} & \multicolumn{1}{c|}{Method} & \multicolumn{1}{c|}{$V_{\rm SD}$} & \multicolumn{1}{c|}{$V_{\rm CPD}$} & \multicolumn{1}{c|}{$z_{\rm tip}$} & \multicolumn{1}{c|}{$V_{\rm tip, ac}$} & \multicolumn{1}{c|}{$A_{\rm ac}$} & \multicolumn{1}{c|}{$f_{\rm drive}$} & \multicolumn{1}{c|}{$\Delta t$} & \multicolumn{1}{c|}{$\delta t$} & \multicolumn{1}{c|}{previous} \\ 
		
		\multicolumn{1}{|c|}{} & \multicolumn{1}{c|}{} & \multicolumn{1}{c|}{(V)} & \multicolumn{1}{c|}{No.} & \multicolumn{1}{c|}{(nm)} & \multicolumn{1}{c|}{(V)} & \multicolumn{1}{c|}{(nm)} & \multicolumn{1}{c|}{(kHz)} & \multicolumn{1}{c|}{(mins)} & \multicolumn{1}{c|}{(mins)} & \multicolumn{1}{c|}{$V_{\rm CPD}$}\\
		\hline 
		\endhead
		
		
		\hline
		\endfoot
		
		\hline \hline
		\endlastfoot
		
		\ref{fig1:ExptProcedure}c & KPFM & 0.0 & 1 &30 & 1.1 & - & 61.34 & 22 & - & - \\
		\ref{fig1:ExptProcedure}c & KPFM & 0.25 & 2 &30 & 1.1 & - & 61.34 & 22 & 5 & 1\\
		
		\hline
		\ref{fig1:ExptProcedure}g & KPFM & 0.0 & 1 &30 & 1.1 & - & 61.34 & 11 & - & - \\
		\ref{fig1:ExptProcedure}g & KPFM & 0.5 & 2 &30 & 1.1 & - & 61.34 & 11 & 5 & 1\\
		
		\hline
		\ref{fig2:SDind_doping}i & KPFM & -0.25 & 1 &30 & 1.1 & - & 61.355 & 22 & - & - \\
		\ref{fig2:SDind_doping}e & KPFM & 0.0 & 2 &30 & 1.1 & - & 61.355 & 22 & 5 & 1 \\
		\ref{fig2:SDind_doping}g & KPFM & 0.25 & 3 &30 & 1.1 & - & 61.355 & 22 & 5 & 2\\
		
		\hline
		\ref{fig3:NegFields_ViscousFlow}a & EFM & 0.0 & 1 & 20 & - & 31 & 61.72 & 34 & - & - \\ 
		\ref{fig3:NegFields_ViscousFlow}e, f & EFM & 0.1 & 2 & 20 & - & 31 & 61.72 & 34 & 83 & 1 \\ 
		\ref{fig3:NegFields_ViscousFlow}b & EFM & 0.0 & 3 & 20 & - & 31 & 61.72 & 34 & 5 & 2\\ 
		
		\hline
		\ref{fig4:Relevance_Viscous}a-f & KPFM & 0.0 & 1 & 30 & 1.1 & - & 61.344 & 20 & 5  & - \\
		\ref{fig4:Relevance_Viscous}a-f & KPFM & 0.4 & 2 & 30 & 1.1 & - & 61.344 & 20 & 5  & 1\\
		
		\hline
		\ref{fig5:Gallery_FieldInversion}a--f & KPFM & 0.0 & 1 & 30 & 1.1 & - & 61.34 & 11 & 5 & -\\ 
		\ref{fig5:Gallery_FieldInversion}a--f & KPFM & 0.5 & 2 & 30 & 1.1 & - & 61.34 & 11 & 5 & 1\\ 
		
		\hline
		\ref{fig5:Gallery_FieldInversion}k & EFM & 0.1 & 1 & 30 & - & 31 & 62.34 & 21 & - & -\\ 
		\ref{fig5:Gallery_FieldInversion}l & EFM & 0.2 & 2 & 30 & - & 31 & 62.34 & 21 & 5 & 1\\ 
		\ref{fig5:Gallery_FieldInversion}m & EFM & 0.1 & 3 & 30 & - & 31 & 62.34 & 21 & 5 & 2\\ 
		\ref{fig5:Gallery_FieldInversion}i-m & EFM & 0.0 & 4 & 30 &  & 31 & 62.34 & 21 & 83 & 3\\ 
		
		\hline
		\ref{fig5:Gallery_FieldInversion}p & KPFM & -1.0 & 1 & 30 & 1.1 & - & 62.34 & 20 & - & -\\
		\ref{fig5:Gallery_FieldInversion}o & KPFM & -0.5 & 2 & 30 & 1.1 & - & 62.34 & 20 & 5 & 1\\ 
		\ref{fig5:Gallery_FieldInversion}n & KPFM & -0.25 & 3 & 30 & 1.1 & - & 62.34 & 20 & 5 & 2\\ 
		\ref{fig5:Gallery_FieldInversion}n-s & KPFM & 0.0 & 4 & 30 & 1.1 & - & 62.34 & 20 & 5 & 3\\ 
		\ref{fig5:Gallery_FieldInversion}q & KPFM & 0.25 & 5 & 30 & 1.1 & - & 62.34 & 20 & 5 & 4\\ 
		\ref{fig5:Gallery_FieldInversion}r & KPFM & 0.5 & 6 & 30 & 1.1 & - & 62.34 & 20 & 5 & 5\\ 
		\ref{fig5:Gallery_FieldInversion}s & KPFM & 1.0 & 7 & 30 & 1.1 & - & 62.34 & 20 & 5 & 6\\ 
		
		\hline
		\ref{fig2_KPFM}c, d & KPFM & 0.0 & 1 & 25 & 1.1 & - & 61.35 & 11 & 5 & - \\ 
		
		\hline
		\ref{fig6:ElecField_BG_direction}a-h & KPFM & 0.0 & 1 & 30 & 1.1 & - & 61.35 & 11 & 5 & - \\ 
		\ref{fig6:ElecField_BG_direction}a-h & KPFM & 0.5 & 2 & 30 & 1.1 & - & 61.35 & 11 & 5 & 1 \\
		
		\hline
		\ref{fig7:Efield_filteProcedure}a-f & EFM & 0.0 & 1 & 20 & - & 31 & 61.72 & 34 & 5 & - \\ 
		\ref{fig7:Efield_filteProcedure}a-f & EFM & 0.1 & 2 & 20 & - & 31 & 61.72 & 34 & 83 & 1 \\ 
		
		\hline
		\ref{fig8:ElecField_ImagingConditions}a-f & EFM & 0.0 & 1 & 20 & - & 31 & 61.72 & 34 & 5 & - \\
		
		\hline
		\ref{DopingFromCPD_fig}d-f & KPFM & 0.0 & 1 & 30 & 1.1 & - & 62.34 & 20 & 5 & - \\
		
		\hline
		\ref{fig:NegEfields_SDILD}c-d & KPFM & 0.0  & 1 & 30 & 1.1 & - & 62.34 & 21 & 5   & - \\
		\ref{fig:NegEfields_SDILD}d   & KPFM & -1.0 & 2 & 30 & 1.1 & - & 62.34 & 21 & 120 & 1 \\
		
		\hline
		\ref{fig:S12c}b-c & EFM & 0.0 & 1 & 30 & - & 31 & 62.34 & 21 & 5   & - \\
		\ref{fig:S12c}c   & EFM & 0.1  & 2 & 30 & - & 31 & 62.34 & 21 & 177 & 1 \\
		
		\hline
		\ref{fig:S12c}e-f & EFM & 0.0 & 1 & 20 & - & 31 & 61.72 & 27 & 5   & - \\
		\ref{fig:S12c}f   & EFM & 0.1  & 2 & 20 & - & 31 & 61.72 & 27 & 133 & 1 \\
		
		\hline
		\ref{fig:S12c}h-i & KPFM & 0.0 & 1 & 25 & 1.1 & - & 60.39 & 11 & 5   & - \\
		\ref{fig:S12c}i   & KPFM & 0.2  & 2 & 25 & 1.1 & - & 60.39 & 11 & 5   & 1 \\
		
		\hline
		\ref{fig:S12b}b-c & KPFM & 0.0 & 1 & 20 & 1.1 & - & 62.34 & 20 & 5   & - \\
		\ref{fig:S12b}c   & KPFM & 0.4 & 2 & 20 & 1.1 & - & 62.34 & 20 & 5   & 1 \\
		
		\hline
		\ref{fig:Relevance_Hydrodynamics_II}a-d & KPFM & 0.0 & 1 & 30 & 1.1 & - & 61.34 & 11 & 5 & - \\
		\ref{fig:Relevance_Hydrodynamics_II}a-d & KPFM & 0.5 & 2 & 30 & 1.1 & - & 61.34 & 11 & 5 & 1 \\
		
		\hline
		\ref{fig:S15}b-c & KPFM & 0.0 & 1 & 30 & 1.1 & - & 61.34 & 11 & 5 & 5 \\
		\ref{fig:S15}c   & KPFM & 0.5 & 2 & 30 & 1.1 & - & 61.34 & 11 & 5 & 1 \\
		
		\hline
		\ref{fig:S15}e-f & KPFM & 0.0 & 1 & 20 & 1.1 & - & 62.34 & 20 & 5 & 5 \\
		\ref{fig:S15}f   & KPFM & 0.4  & 2 & 20 & 1.1 & - & 62.34 & 20 & 5 & 1 
		
		
		
		\label{table_Record_Parameters}
	\end{longtable}
\end{center}

\section{Electric Fields due to Current Flow}
\label{sec4_EField}

To map the current induced electric fields, we measure the change of the contact potential difference $V_{\rm CPD}(x,y)$ after applying a source-drain voltage $V_{\rm SD}$. 
This is dubbed the transport voltage $V_{\rm transport}(x,y,V_{\rm SD})$ defined as.

\begin{equation}
V_{\rm transport}(x,y,V_{\rm SD}) = V_{\rm CPD}(x,y,V_{\rm SD}) - V_{\rm CPD}(x,y,V_{\rm SD} = 0\,V)
\label{eq10:Vtransport}
\end{equation}

\noindent
The negative spatial gradient of $V_{\rm transport}(x,y,V_{\rm SD})$ is the transport induced electric field reading

\begin{equation}
\mathbf{E^{\rm meas}}(x,y) = - \nabla V_{\rm transport}(x,y).
\label{eq11:EField_BG_direction}
\end{equation}



The derivation implicitly assumes that the static work-function fluctuations of the two subtracted $V_{\rm CPD}(x,y)$ maps remain unchanged and therefore cancel (section~\ref{Sec:EField_SDILD}). 
\noindent

\begin{figure}
	\centering
	\includegraphics[width = 165 mm]{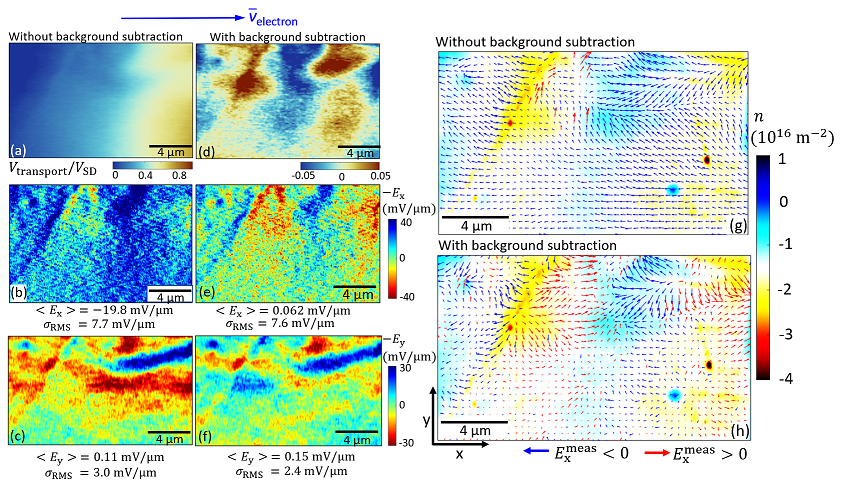}
	\caption{{\bf Current Induced Electric Fields after Background Subtraction.} (a) $V_{\rm transport}(x,y)/V_{\rm SD}$ map using KPFM for recording $V_{\rm CPD}(x,y)$ at $V_{\rm SD}=0.5$\,V and $V_{\rm SD}=0$\,V as input for  eq.~(\ref{eq10:Vtransport}), $V_{\rm gate} - V_{\rm D} = -5\,$V, corresponding topography:  Fig.~\ref{fig1:ExptProcedure}f, main text. (b), (c) Maps of $x$ and $y$ component of the electric field, respectively,  as deduced by eq.~(\ref{eq11:EField_BG_direction}) using a. Average electric fields $\left<E_{x,y} \right>$ and the RMS standard deviations $\sigma_{\rm RMS}$ are marked below the images. The latter is determined within boxes of ($1\,\mu$m)$^2$ and averaged over all boxes afterwards. (d) Same as a, but after removal of a 2$^{\rm nd}$ order polynomial background from the two constituting $V_{\rm CPD}(x,y)$ maps (section~\ref{subsec_BG}). (e), (f) Maps of $x$ and $y$ component of the electric field 
		derived from d via eq.~(\ref{eq11:EField_BG_direction}). (g) Vector plot of in-plane electric field $\mathbf{E^{\rm meas}}(x,y)$ deduced via eq.~(\ref{eq11:EField_BG_direction}) from a--c. The arrows are colored blue (red), if pointing forward (backward) with respect to the applied $V_{\rm SD}$. The background color shows the doping distribution $n(x,y)$ at $V_{\rm SD} = 0.5\,$V (eq.~(\ref{eq:SDILD})). (h) Same as g, but deduced from d--f.}  
	\label{fig6:ElecField_BG_direction}
\end{figure}

Figure~\ref{fig6:ElecField_BG_direction}a
shows a $V_{\rm transport}(x,y)$ map derived from $V_{\rm CPD}(x,y)$ maps recorded by KPFM according to eq.~(\ref{eq10:Vtransport}) and after normalizing to $V_{\rm SD}$. The image employs $V_{\rm SD}=0.5$\,V at charge neutrality.
The $x$ and $y$ components of the resulting in-plane electric field $\mathbf{E^{\rm meas}}(x,y)$ (eq.~(\ref{eq11:EField_BG_direction})) are displayed in Fig.~\ref{fig6:ElecField_BG_direction}b and c, respectively.  Figure~\ref{fig6:ElecField_BG_direction}g shows a vector representation of $\mathbf{E^{\rm meas}}(x,y)$ represented by arrows on the colored background of the deduced charge carrier density $n(x,y)$ as present at $V_{\rm SD}=0.5$\,V (eq.~(\ref{eq:SDILD})). 

Figure~\ref{fig6:ElecField_BG_direction}d-f and h show the same data, but deduced from the $V_{\rm CPD}(x,y)$ maps after 2$^{\rm nd}$ order polynomial background subtraction (section~\ref{subsec_BG}).
This naturally removes the constant electric field along the applied $V_{\rm SD}$ and, thus, highlights deviations from this average electric field, but it prohibits the direct identification of inverted electric fields.
Correlations between doping and electric field get more apparent, e.g., hole (electron) doped regions in this particular area tend to reduce (increase) the electric field in $x$ direction with respect to the average field. 
Moreover, it gets more easy within the vector maps with background subtraction to identify electric fields pointing in different $y$ directions 
(Fig.~\ref{fig6:ElecField_BG_direction}g). Finally, one can identify source like areas (e.g. upper fold area) or sink like areas (e.g. upper left area) of the current induced electric fields.

\subsection{Noise Filtering}
\label{Efield_noise_filter}

\begin{figure}
	\centering
	\includegraphics[width = 165 mm]{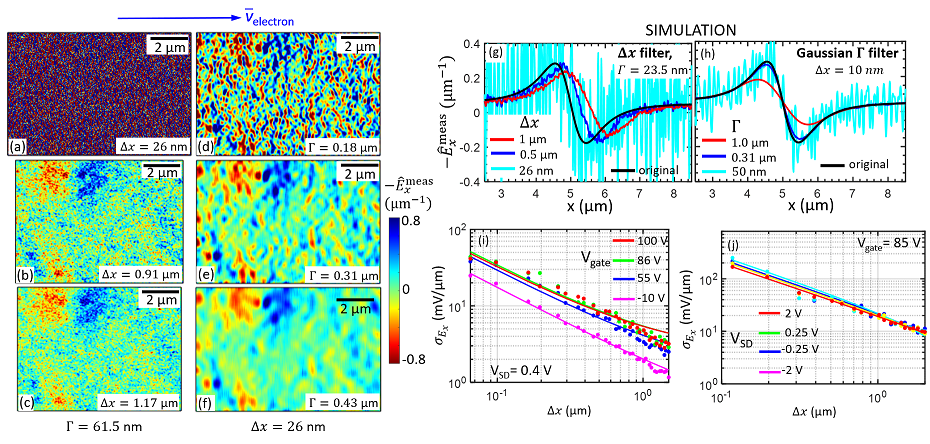}
	\caption{{\bf Filtering procedure  for electric field maps.} (a)--(c) Experimentally measured $\widehat{E}_{x}^{\rm meas}(x,y)$ (EFM) using eq.~(\ref{eq:Ex_VtrDer_filter}) with different values of $\Delta x$ as indicated, after mild Gaussian averaging of $V_{\rm transport}(x,y)$ with $\Gamma=61.5$\,nm. The upper left area of b corresponds to Fig.~\ref{fig3:NegFields_ViscousFlow}e, main text. (d)--(f) $\widehat{E}_{x}^{\rm meas}(x,y)$ using the same $V_{\rm transport}(x,y)$ map as in a--c, but using Gaussian averaging with different $\Gamma$ before employing $\Delta x = 26$\,nm for differentiation (eq.~(\ref{eq:Ex_VtrDer_filter})). (g) Simulated dipolar feature of $\widehat{E}_x (x)$ (black line)  and the same feature after superposition with white noise of strength $\delta V_{\rm transport} = 16\,$mV in the constituting $V_{\rm transport}(x)$ at $V_{\rm SD} = 0.1\,$V and subsequent application of eq.~(\ref{eq:Ex_VtrDer_filter}) with  the indicated $\Delta x$ (colored lines). (h) Same as g concerning simulated dipole (black line) and noise superposition, but using subsequently   Gaussian averaging with different $\Gamma$ as marked (colored lines). (i), (j) RMS noise of experimental $E_{x}^{\rm meas}(x,y)$ maps, $\sigma_{E_x} (\Delta x)$,  for different $V_{\rm gate}$ (i) and $V_{\rm SD}$ (j), KPFM. The  noise for each $\Delta x$ is deduced  within boxes  of $0.3\,\mu{\rm m} \times 0.3\,\mu$m covering the whole image. Subsequently, the noise of all boxes is averaged. The solid lines are linear fits revealing $\sigma_{E_x} \propto \Delta x^{-1}$.}
	\label{fig7:Efield_filteProcedure}
\end{figure}

Performing a nearest neighbor differentiation of $V_{\rm transport}(x,y)$ to obtain $\mathbf{E^{\rm meas}}(x,y)$ maps (eq.~(\ref{eq11:EField_BG_direction})) revealed  uncorrelated noise for both components with strength $\delta E^{\rm meas}_x\approx \delta E^{\rm meas}_y \approx 100\,{\rm mV}/\mu$m  (Fig.~\ref{fig7:Efield_filteProcedure}a). This uncorrelated noise obscured the observation of any feature. This is likely of electric origin at the piezoelectric actuators reducing the accuracy to determine $V_{\rm CPD}$ in EFM (section \ref{sec3:EFM}). 

To reduce the overwhelming noise, we employ two  filtering processes. On one hand, we apply a spatial Gaussian averaging with full width at half maximum $\Gamma$ to $V_{\rm transport}(x,y)$ maps. On the other hand, we use locations that are further apart from each other for the differentiation, i.e.,  

\begin{align}
E_{x}^{\rm meas}(x,y) = \frac{V_{\rm transport}(x,y) - V_{\rm transport}(x-\Delta x,y)}{\Delta x} \nonumber \\
E_{ y}^{\rm meas}(x,y) = \frac{V_{\rm transport}(x,y) - V_{\rm transport}(x,y-\Delta y)}{\Delta y} 
\label{eq:Ex_VtrDer_filter}
\end{align}

\noindent
with $\Delta x$ and $\Delta y$ being the chosen distances in the respective directions. 
The RMS fluctuations $\delta E_x^{\rm meas}$
and $\delta E_y^{\rm meas}$ scale with $1/\Delta x$ and $1/\Delta y$, respectively, according to error propagation. This naturally improves the signal to noise ratio at the expense of spatial resolution.

This second approach is visualized for an experimentally measured $V_{\rm transport}(x,y)$ map in Fig.~\ref{fig7:Efield_filteProcedure}a--c. Stable structures appear at $\Delta x \ge 0.91$\,$\mu$m, i.e., using a distance of 35 measurement points for $\Delta x$.
The remaining uncorrelated noise exhibits a standard variation $\sigma_{E_x}=5$\,mV/$\mu$m as deduced by analyzing multiple boxes of ($0.3\,\mu$m)$^2$, much smaller than the apparent features sizes. 
This can be compared with the error  $\delta V_{\rm CPD}=2$\,mV resulting from the $\Delta \Phi_{\rm EFM}(V_{\rm tip})$ noise (section~\ref{sec:EFMnoise}).
Error propagation implies $\delta V_{\rm transport}=3$\,mV (eq.~(\ref{eq10:Vtransport})) and, respectively, $\delta E_x^{\rm meas}=5$\,mV/$\mu$m for $\Delta x =0.91\,\mu$m (eq.~\ref{eq:Ex_VtrDer_filter}) in very good agreement with the measured noise of 5\,mV/$\mu$m. 

Figure~\ref{fig7:Efield_filteProcedure}g shows the same procedure of filtering via large $\Delta x$ for a fictitious electric field $E_x(x)$ featuring a dipolar structure (black line) that is mixed with uncorrelated noise in $V_{\rm transport}(x)$ of strength $\delta V_{\rm transport}=16$\,mV prior to using different $\Delta x$ to determine $E_x^{\rm meas}(x)$ according to eq.~(\ref{eq:Ex_VtrDer_filter}) (colored lines). Obviously, the feature width and height are barely changed by the  $\Delta x$ filtering, but the dipolar feature is slightly shifted to the right. 
In contrast, the Gaussian averaging makes the features wider and weaker in amplitude, while maintaining its center position (Fig.~\ref{fig7:Efield_filteProcedure}h). This is also visible in the accordingly Gaussian filtered experimental images (Fig.~\ref{fig7:Efield_filteProcedure}d-f). 

Since our main interest is the feature size and the feature strength, in particular, during the analysis of  Fig.~\ref{fig3:NegFields_ViscousFlow}, main text, we optimize $\Delta x$ (eq.~(\ref{eq:Ex_VtrDer_filter})) towards the lowest possible signal/noise ratio, where features get significantly stronger than the noise floor, but use only a mild Gaussian averaging with $\Gamma \le 100\,$nm. Table~\ref{table1_Filter_Pmts} summarizes the chosen $\Delta x$ and $\Gamma$ for all electric field maps presented in main text and supplement.


\begin{table}[h!]
	\centering
	\begin{tabular}{ |c|c|c|c|c| } 
		\hline
		Figure & Method & $\Gamma$ & $\Delta x$ & $\Delta y$ \\ 
		\hline
		\ref{fig1:ExptProcedure}d & KPFM & 92 nm & 1.17 $\mu$m & 0.12 $\mu$m \\
		\ref{fig1:ExptProcedure}g & KPFM & 81 nm & 0.43 $\mu$m & 0.12 $\mu$m\\
		\ref{fig2:SDind_doping}f-i, \ref{fig:SDILD_discrepancy} & KPFM & 93 nm & 0.39 $\mu$m & 0.08 $\mu$m\\
		\ref{fig3:NegFields_ViscousFlow}c--d, & EFM & 86 nm & 0.47 $\mu$m & 0.03 $\mu$m \\ 
		\ref{fig3:NegFields_ViscousFlow}e--f, \ref{fig8:ElecField_ImagingConditions}m--n, \ref{fig:SDILD_PatchedDoping}e & EFM & 62 nm & 0.91 $\mu$m & 0.03 $\mu$m \\ 
		\ref{fig4:Relevance_Viscous}a--f, \ref{fig:S12b}c, \ref{fig:S15}f & KPFM & 31 nm & 0.39 $\mu$m & 0.05 $\mu$m \\ 
		\ref{fig5:ion_bombardment}c--h, k--m & KPFM & 28 nm & 0.23 $\mu$m & 0.06 $\mu$m \\
		\ref{fig6:ElecField_BG_direction}b,e, \ref{fig:Relevance_Hydrodynamics_II}a--d, \ref{fig:S15}c, \ref{fig5:Gallery_FieldInversion}a--f & KPFM & 61 nm & 0.52 $\mu$m & 0.12 $\mu$m\\ 
		\ref{fig6:ElecField_BG_direction}c,f & KPFM & 81 nm & 0.52 $\mu$m  & 0.93 $\mu$m\\
		\ref{fig6:ElecField_BG_direction}g--h & KPFM & 61 nm & 0.52 $\mu$m  & 0.93 $\mu$m\\
		\ref{fig:NegEfields_SDILD}c & KPFM & 77 nm & 0.33 $\mu$m  & 0.05 $\mu$m\\
		\ref{fig:NegEfields_SDILD}d & KPFM & 99 nm & 0.03 $\mu$m  & 0.05 $\mu$m\\
		\ref{fig:S12c}c, \ref{fig5:Gallery_FieldInversion}i--m & EFM & 77 nm & 0.39 $\mu$m & 0.04 $\mu$m\\
		\ref{fig:S12c}f & EFM & 53 nm & 0.89 $\mu$m & 0.01 $\mu$m\\
		\ref{fig:S12c}i & KPFM & 46 nm & 0.29 $\mu$m & 0.06 $\mu$m\\
		\ref{fig5:Gallery_FieldInversion}n--s & KPFM & 77 nm & 0.82 $\mu$m  & 0.05 $\mu$m\\
		\hline
	\end{tabular}
	\caption{Filter parameters for all electric field maps of main text and supplement. $\Gamma$: FWHM of a Gaussian filter for $V_{\rm transport}(x,y)$, $\Delta x$, $\Delta y$: length scales for determining $E_x^{\rm meas}(x,y)$ from $V_{\rm transport}(x,y)$ according to eq.~(\ref{eq:Ex_VtrDer_filter}).}
	\label{table1_Filter_Pmts}
\end{table}

Eventually, Fig.~\ref{fig7:Efield_filteProcedure}i and j show the rms electric field noise $\sigma_{E_x}(\Delta x)$, determined within multiple boxes of size $(0.3\,\mu$m)$^2$ (average of all boxes within one image) for different $V_{\rm gate}$ and $V_{\rm SD}$, respectively. The fitted slope in the double-logarithmic plots (lines) consistently reveals $\sigma_{E_x}\propto \Delta x^{-1}$, as expected from the discussion above. This evidences uncorrelated $E_x^{\rm meas}(x,y)$ noise. In line, the correlation length of $E_x^{\rm meas}(x,y)$ at smallest possible $\Delta x=26$\,nm (eq.~\ref{eq:Ex_VtrDer_filter}) is $\xi \approx 30 \,$nm only, i.e., the image resolution.
Note that the noise depends barely on $V_{\rm SD}$ and $V_{\rm gate}$. 

\subsection{Temporal Stability of Dopant Distribution}
\label{subsec_DopingStability}

\begin{figure}
	\centering
	\includegraphics[width = 165 mm]{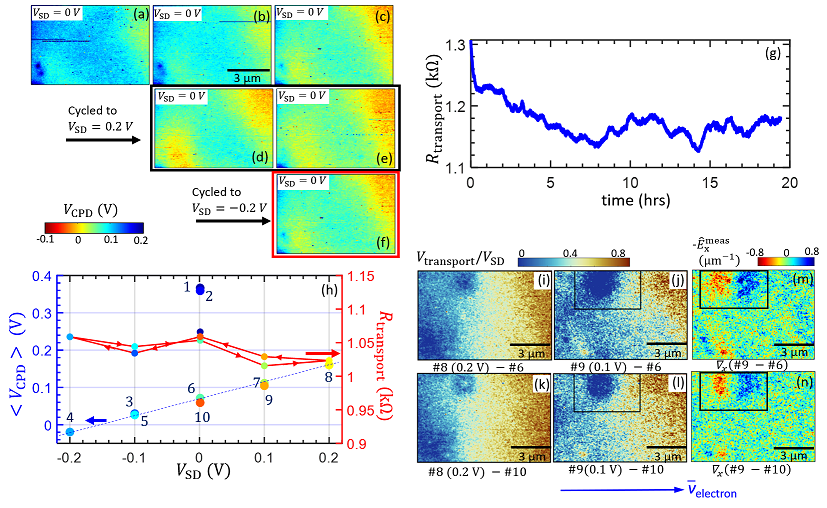}
	\caption{{\bf Temporal Stability of Doping Distribution.} (a)--(f) Subsequently recorded $V_{\rm CPD}(x,y)$ maps using EFM, $V_{\rm SD} = 0\,$V, $V_{\rm gate } = V_{\rm D} = 85\,$V, $V_{\rm tip}=2\,$V during tapping mode traces. $V_{\rm gate}$ and $V_{\rm SD}$ was changed  one full day  prior to the acquisition of a. Before recording d (f), $V_{\rm SD}$ was cycled to $+0.2$\,V ($-0.2$\,V) staying there for 10 min. (g) Two terminal resistance $R_{\rm transport}$  recorded between the setting of $V_{\rm gate}$ and the data acquisition of Fig.~\ref{fig3:NegFields_ViscousFlow}, main text, $V_{\rm gate} = V_{\rm D}=85\,$V. (h) Spatially averaged $V_{\rm CPD}$ (larger dots, left axis) and simultaneously acquired temporarily averaged device resistance (smaller dots connected by red line, right axis) for a series of $V_{\rm CPD}(x,y)$ maps measured in order as numbered, just after finishing the time trace in g. Same colors belong to the same $V_{\rm CPD}(x,y)$ map. The current induced electric field map deduced from image 9 and image 6 (10) is also shown in Fig.~\ref{fig3:NegFields_ViscousFlow}e (f), main text. (i)--(l) $V_{\rm transport} (x,y)/V_{\rm SD}$ maps as deduced from the subtraction of the $V_{\rm CPD}(x,y)$ maps labelled in h as marked below each image (eq.~(\ref{eq10:Vtransport})). Note that different reference maps at $V_{\rm SD}=0$\,V are used for i, j and k, l. (m), (n) Deduced $\widehat{E}_{ x}^{\rm meas}(x,y)$ map (eq.~(\ref{eq11:EField_BG_direction})) from j and l, respectively, using the $V_{\rm CPD}(x,y)$ maps as marked below the image. The rectangular area marked in l--n is displayed in Fig.~\ref{fig3:NegFields_ViscousFlow}e, f, main text.}
	\label{fig8:ElecField_ImagingConditions}
\end{figure}

The transport potential $V_{\rm transport}(x,y)$ according to eq.~(\ref{eq10:Vtransport}) quantifies the potential drop due to charge flow, if the work-function fluctuations in the measured region remain the same while acquiring the two $V_{\rm CPD}(x,y)$ maps, biased and unbiased. However, the ambient conditions during recording partially lead to local charging and discharging  depending on the history of $V_{\rm gate}$ and $V_{\rm SD}$. An example is shown in Fig.~\ref{fig8:ElecField_ImagingConditions}a--c, where changes in $V_{\rm CPD}(x,y)$ appear, albeit the maps are recorded subsequently without changing $V_{\rm gate}$ or $V_{\rm SD}$ and a full day after setting these voltages. However, changes after recording Fig.~\ref{fig8:ElecField_ImagingConditions}c are much less pronounced (Fig.~\ref{fig8:ElecField_ImagingConditions}d--f), albeit $V_{\rm SD}$ was cycled in between  indicating that a first imaging itself leads to equilibration of the doping pattern.

Hence, it is crucial to  reduce the uncontrolled charging processes. Therefore, we firstly monitored the two-terminal $R_{\rm transport}$ continuously. It changed minimally after ramping  $V_{\rm gate}$ to about $100\,$V and then ramp it down slowly by $\sim 0.1$\,V/s until the device resistance is maximized signalling charge neutrality. Figure~\ref{fig8:ElecField_ImagingConditions}g shows a time trace of $R_{\rm transport}$ after such stabilization revealing only small fluctuations of about 5\,\% after a waiting time of roughly 5\,hours. 
Using this procedure, stability at the local scale  has still to be ensured. This was more involved and not always successful. There is no direct way to experimentally map the doping distribution, while applying a finite $V_{\rm SD}$. Hence, we checked the doping distribution by recording $V_{\rm CPD}(x,y,V_{\rm SD}=\,0\,$V) maps before and after recording $V_{\rm CPD}(x,y,V_{\rm SD}\neq\,0\,$V) maps. Figure~\ref{fig8:ElecField_ImagingConditions}h--n display characteristic features of a successful image sequence, namely a minimum change in $V_{\rm transport}$  while cycling $V_{\rm SD}$ and very similar spatially averaged $V_{\rm CPD}$ values for images recorded at the same $V_{\rm SD}$ but after a distinct $V_{\rm SD}$ history (Fig.~\ref{fig8:ElecField_ImagingConditions}h). 
Obviously, the first two images exhibit a rather different $\left<V_{\rm CPD}(x,y)\right>$ in line with the changes in Fig.~\ref{fig8:ElecField_ImagingConditions}a--c, implying again that the first maps within a certain area change the lateral doping distribution more strongly, likely via the applied tip voltages. However, stable subsequent images can be often recorded afterwards.
Consequently,  the  $V_{\rm transport}(x,y)$ maps deduced from a stable sequence  are nearly identical, if distinct reference images recorded at $V_{\rm SD}=\,0\,$V  are employed (eq.~(\ref{eq10:Vtransport})). This is visible by comparing Fig.~\ref{fig8:ElecField_ImagingConditions}i and k as well as Fig.~\ref{fig8:ElecField_ImagingConditions}j and l, that used image 6 and image 10 as reference, respectively. The similarity naturally also applies 
for the resulting electric field maps in Fig.~\ref{fig8:ElecField_ImagingConditions}m--n. 
These kind of images, selected by adequate monitoring, are, hence, reliably attributed to consequences of the applied $V_{\rm SD}$. 

\section{Doping Distribution from $V_{\rm CPD}(x,y)$ Maps}
\label{Sec5_DopingDistr_CPD}

Spatial fluctuations of the contact potential difference between tip and graphene are related to doping fluctuations of the graphene \cite{Samaddar2016}. 
In the absence of current, the corresponding doping density $n_0(x,y)$ can be deduced via $|e|\left(V_{\rm CPD}(x,y) - V_{\rm CPD}^0\right) = E_{\rm F} -E_{\rm D}(x,y)$, where $E_{\rm F} := 0\,$eV is the Fermi energy of graphene, $E_{\rm D}(x,y)$ is the local Dirac point energy, and $V_{\rm CPD}^0$ is the contact potential difference between the tip and charge neutral graphene. The resulting doping distribution $n_0(x,y)$ at $V_{\rm SD}=0$\,V reads

\begin{equation}
n_0(x,y) =  \frac{e^2}{\pi} \, {\rm sign}{\left[V_\mathrm{CPD}(x,y,V_{\rm SD} = 0\,{\rm V}) - V_\mathrm{CPD}^\mathrm{0} \right]} \, \left(\frac{V_\mathrm{CPD}(x,y,V_{\rm SD} = 0\,{\rm V}) - V_\mathrm{CPD}^\mathrm{0} }{\hbar v_\mathrm{F}} \right)^2 ,
\label{DopingCPDrelation}
\end{equation}

\noindent
whith the Fermi velocity of graphene $v_{\rm F} = 1\times 10^6\,$m/s. Thus, one needs $V_{\rm CPD}^0$. It can be deduced  from $V_{\rm CPD}(x,y,V_{\rm SD} = 0\,$V) maps recorded by EFM at charge neutrality, i.e., at $V_{\rm gate} = V_{\rm D}$ as the spatial average of $V_{\rm CPD}(x,y)$, but not by KPFM due to the required background subtraction. To determine charge neutrality, we use the maximum in 2-point resistance (Fig.~\ref{fig1:ExptProcedure}(e), main text), that fits to the maximum in average resistance deduced from the maps of current-induced potentials.

\begin{figure}
	\centering
	\includegraphics[width=160 mm]{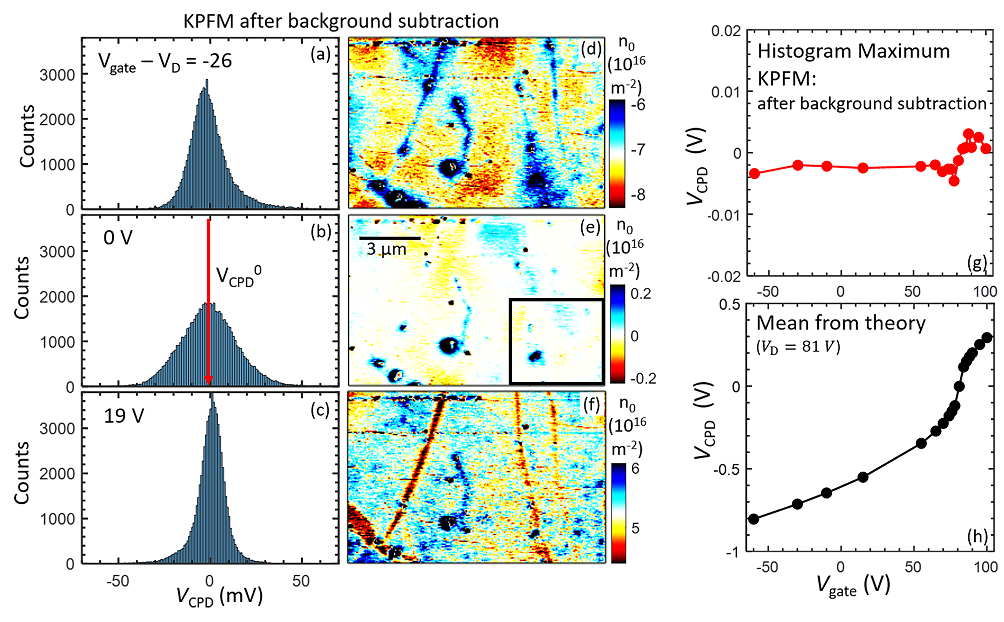}
	\caption{{\bf Doping Maps.}(a)--(c) Histograms of $V_{\rm CPD}(x,y,V_{\rm SD}=\,$0 V) as recorded by KPFM after background subtraction (section~\ref{subsec_BG}), $V_{\rm gate}-V_{\rm D}$ as marked, $V_{\rm D}= 81\,$V, red line in b: deduced $V_{\rm CPD}^0=-1.3$\,mV. (d)--(f) Doping maps $n_0(x,y)$ corresponding to the adjacent histograms and deduced from $V_{\rm CPD}(x,y,V_{\rm SD}=\,$0 V) via eq.~(\ref{DopingCPDrelation}) using the capacitively calculated background shift of (h), corresponding topography: Fig.~\ref{fig5:Gallery_FieldInversion}h, black rectangle in e: region of Fig.~\ref{fig4:Relevance_Viscous}a--f, main text. (g) Maximum of $V_{\rm CPD}$  histograms for various $V_{\rm gate}$ deduced from KPFM images after background subtraction. (h) $\left<V_{\rm CPD}\right>$ as derived from a capacitive gating model ($V_{\rm D} = 81\,$V, text) and as added to the background subtracted $V_{\rm CPD}(x,y)$ maps prior to applying eq.~(\ref{DopingCPDrelation}) for calculating $n_0(x,y)$.} 
	\label{DopingFromCPD_fig}
\end{figure}

For $V_{\rm CPD}(x,y)$ maps acquired by EFM, we then construct doping maps by applying eq.~(\ref{DopingCPDrelation}) straightforwardly. However, for KPFM, we have to consider that $V_{\rm CPD}(x,y)$ maps are obtained after subtracting a second order polynomial background  (section~\ref{subsec_BG}) that largely removes the spatial average of $V_{\rm CPD}(x,y)$ (Fig.~\ref{DopingFromCPD_fig}g). 
In order to restore the average, we calculate the average contact potential difference for a particular $V_{\rm gate}$ using $V_{\rm CPD}= V_{\rm CPD}^0 + {\rm sign}(n_{\rm gate})\,\hbar v_{\rm F}\sqrt{\pi |n_{\rm gate}|}$  with $n_{\rm gate} = C_{\rm gate}\left(V_{\rm gate} - V_{\rm D}\right)/e$ as doping density induced by the gate via a capacitive model (Fig.~\ref{DopingFromCPD_fig}h). Subsequently, the histograms of the $V_{\rm CPD}(x,y)$ maps (Fig.~\ref{DopingFromCPD_fig}a--c), i.e., each value of the map, are shifted such that the histogram maxima  (Fig.~\ref{DopingFromCPD_fig}g) are aligned  with the calculated values of Fig.~\ref{DopingFromCPD_fig}h. These adjusted $V_{\rm CPD}(x,y)$ maps are then used to apply eq.~(\ref{DopingCPDrelation}) resulting straightforwardly in the doping maps $n_0(x,y)$. Figure~\ref{DopingFromCPD_fig}d-f displays resulting doping maps of the same area at different $V_{\rm gate}$. The whole map changes from  hole doping to electron doping due to the added $V_{\rm CPD}^0$. Moreover, folds and wrinkles (topography in Fig.~\ref{fig5:Gallery_FieldInversion}h) charge less via gating, which is likely due to the larger distance from the gate. 

\subsection{Source Drain Induced Local Doping (SDILD)}
\label{subsec_SDILD}

As discussed in the main text, the doping distribution $n(x,y)$ is subject to a linear gradient at finite $V_{\rm SD}$. The gate voltage reference at the graphene is ground for the drain electrode and shifted by $V_{\rm SD}$ with respect to ground for the source electrode (Fig.~\ref{fig2:SDind_doping}a, main text). Hence, we define a local gate voltage $V_{\rm gate}^{\rm local}(x,V_{\rm SD}) = V_{\rm gate} - \frac{V_{\rm SD}}{L} (x-x_0)$, where $x=x_0$ is the edge of the drain electrode and $x=0$ the left side of the image (closest to the drain). 
We will use $V_{\rm gate}^{\rm local}(x,V_{\rm SD})$ to calculate the resulting $n(x,y)$ iteratively. Firstly, we regard
the doping $n_0(x,y)$ at $V_{\rm SD}=0$\,V. Its fluctuations change the local charge neutrality point at finite $V_{\rm gate}$ reading $V_{\rm D}^{\rm local}(x,y) = V_{\rm gate} - e \frac{n_0(x,y)}{C_{\rm eff}(x,y,V_{\rm SD}=0\,{\rm V})}$. Here, $C_{\rm eff}(x,y,V_{\rm SD}=0\,{\rm V}) = \left(\frac{1}{C_{\rm gate}} + \frac{1}{C_{\rm Q} (x,y,V_{\rm SD}=0\,{\rm V})}\right)^{-1}$ is the effective capacitance consisting of geometric capacitance $C_{\rm gate}$ and quantum capacitance $C_{\rm Q}(x,y,V_{\rm SD}=0\,{\rm V}) = \frac{2e^2}{\pi}\frac{\sqrt{\pi |n_{\rm 0}(x,y)|}}{\hbar v_{\rm F}}$.

In a second step, we set up the iterative loop for $n(x,y)$ at finite $V_{\rm SD}$ reading

\begin{align}
n(x,y,V_{\rm SD}) & = \frac{C_{\rm eff}(x,y,V_{\rm SD})}{e} \left( V_{\rm gate}^{\rm local}(x,V_{\rm SD}) - V_{\rm D}^{\rm local}(x,y) \right) \nonumber \\
& = \frac{C_{\rm eff}(x,y,V_{\rm SD})}{C_{\rm eff}(x,y,V_{\rm SD}=0\,{\rm V})}n_{0} (x,y) - \left(\frac{V_{\rm SD}}{eL}\right) C_{\rm eff}(x,y,V_\mathrm{SD}) (x-x_0),
\label{eq:SDILD}
\end{align}

\noindent
where $C_{\rm eff}(x,y,V_{\rm SD})$ is the effective capacitance for the changed doping $n(x,y,V_{\rm SD})$ via $V_{\rm SD}$. Thus, both sides of the equation depend on $n(x,y,V_{\rm SD})$ suggesting a self-consistent loop that we applied for solving. Far away from charge neutrality, the second term is negligible, since $|V_{\rm SD}|\ll V_{\rm gate}$ and $C_{\rm eff}(x,y,V_{\rm SD})\approx C_{\rm gate}$ implying $n(x,y)\approx n_0(x,y)$, i.e., SDILD is negligible. However, when $V_{\rm gate} \rightarrow V_{\rm D}$, the first term is almost zero raising the importance of the second term. Consequently, $V_{\rm SD}$ induced gating (SDILD) has to be  considered carefully, in particular, close to charge neutrality.

\section{Electric Field from SDILD}
\label{Sec:EField_SDILD}

In section \ref{sec4_EField}, we assigned the difference between biased and unbiased $V_{\rm CPD}(x,y)$ maps to the current induced voltage drop $V_{\rm transport}(x,y)$ (eq.~(\ref{eq10:Vtransport})) using the assumption that the doping distribution in the sample remains unchanged. However,  section \ref{subsec_SDILD} reveals that $V_{\rm SD}$ changes the doping via SDILD (eq.~(\ref{eq:SDILD})), most strongly  close to charge neutrality. Even for a linear potential drop along the sample by $V_{\rm SD}$, we get non-linearities in $V_{\rm transport}(x,y)$ (eq.~(\ref{eq10:Vtransport})) via the non-linear local doping due to quantum capacitance (eq.~(\ref{eq:SDILD})). We dub the resulting electric field via SDILD   $\mathbf{E}^{\rm SDILD}(x,y)$ (eq.~(\ref{eq11:EField_BG_direction})). Here, we present a formalism to calculate it for known $V_{\rm SD}$ and  $n_0(x,y)$ (eq.~(\ref{DopingCPDrelation})). A comparison of $\mathbf{E}^{\rm SDILD}(x,y)$ with $\mathbf{E}^{\rm meas}(x,y)$ then allows to distinguish between non trivial transport features and those due to SDILD.

The Fermi energy of graphene reads $E_{\rm F} -E_{\rm D} = {\hbar v_{\rm F}\;\rm sign}(n) \sqrt{\pi |n|}$ with $n(x,y,V_{\rm SD})$ being the doping distribution including SDILD (eq.~(\ref{eq:SDILD})). If we assume that the applied $V_{\rm SD}$ drops linearly along the graphene, implying a potential $\frac{V_{\rm SD}}{L}(x-x_0)$ ($x_0$: edge of the drain electrode), we obtain  

\begin{align}
V_{\rm CPD}(x,y,V_{\rm SD}) & = V_{\rm CPD}^0 + \frac{\left(E_\mathrm{F}(x,V_{\rm SD}) - E_{\rm D}(x,y,V_{\rm SD}) \right)}{|e|} + \frac{V_{\rm SD}}{L}(x-x_0) \nonumber \\
& = V_{\rm CPD}^{0} + \frac{\hbar v_{\rm F}}{|e|} \, {\rm sign}(n(x,y,V_{\rm SD}))\sqrt{\pi |n(x,y,V_{\rm SD})| } + \frac{V_{\rm SD}}{L}(x-x_0).
\label{eq:tilted_VCPD}
\end{align}

\noindent

The contact potential difference in the unbiased case is (section \ref{Sec5_DopingDistr_CPD})
\begin{equation}
V_{\rm CPD}(x,y,0) = V_{\rm CPD}^0 + {\frac{\hbar v_{\rm F}}{|e|}\rm sign}(n_0(x,y))  \sqrt{\pi |n_0(x,y)|}
\nonumber.
\end{equation}

The resulting $x$ component of the electric field (eqs.~(\ref{eq10:Vtransport}), (\ref{eq11:EField_BG_direction})) becomes 

\begin{align}
E_{x}^{\rm SDILD}(x,y,V_{\rm SD}) & = -\frac{d}{dx} [V_{\rm CPD} (x,y,V_{\rm SD}) - V_{\rm CPD} (x,y,0)] \nonumber \\
& = -\frac{V_{\rm SD}}{L} - \frac{\hbar v_{\rm F}}{|e|} \frac{d}{dx} [{\rm sign} (n(x,y,V_{\rm SD}))\sqrt{\pi |n(x,y,V_{\rm SD})|} - {\rm sign}(n_0 (x,y))\sqrt{\pi |n_0 (x,y)|}] \nonumber \\
&  = E_{x}^0 - \frac{\hbar v_{\rm F} \sqrt{\pi}}{2|e|} \left( \frac{1}{\sqrt{|n(x,y,V_{\rm SD})|}}  \frac{dn(x,y,V_{\rm SD})}{dx} - \frac{1}{\sqrt{|n_0(x,y)|}} \frac{dn_0(x,y)}{dx} \right),
\label{eq:E_SDLD}
\end{align}

\noindent
using $\frac{d}{dx}\left({\rm sign}(n)\;\sqrt{\pi |n|}\right)= \frac{\sqrt{\pi}}{2} \frac{1}{\sqrt{|n|}} \frac{dn}{dx}$. The term $E_{x}^0 = -V_{\rm SD}/L$ is the trivial electric field by the linear voltage drop due to $V_{\rm SD}$. Normalising the electric field to $V_{\rm SD}$, we get

\begin{align}
\widehat{E}_{x}^{\rm SDILD}(x,y,V_{\rm SD}) & = -\frac{1}{L} - \frac{\beta}{V_{\rm SD}} \left( \frac{1}{\sqrt{|n(x,y,V_{\rm SD})|}}  \frac{dn(x,y,V_{\rm SD})}{dx} - \frac{1}{\sqrt{|n_0(x,y)|}} \frac{dn_0(x,y)}{dx} \right) \nonumber \\
& := \widehat{E}_{x}^0 - \frac{\beta}{V_{\rm SD}} \chi(x,y,V_{\rm SD})
\label{Eq_NegField_TrivialEff}
\end{align}

\noindent
with 
$\beta = \hbar v_\mathrm{F} \sqrt{\pi}/2|e| = 6.425 \cdot 10^{-10}$\,Vm and $\widehat{E}_{x}^0 = -1/L \approx -0.055\,\mu {\rm m^{-1}}$. 

Obviously, $\widehat{E}_{x}^{\rm SDILD}$ diverges, if $n_0(x,y)$ or $n(x,y,V_{\rm SD})$ crosses zero as naturally appearing close to charge neutrality. This implies pronounced local maxima and minima in the measured electric field maps that are not caused by the current flow.

\begin{figure}
	\centering
	\includegraphics[width=165 mm]{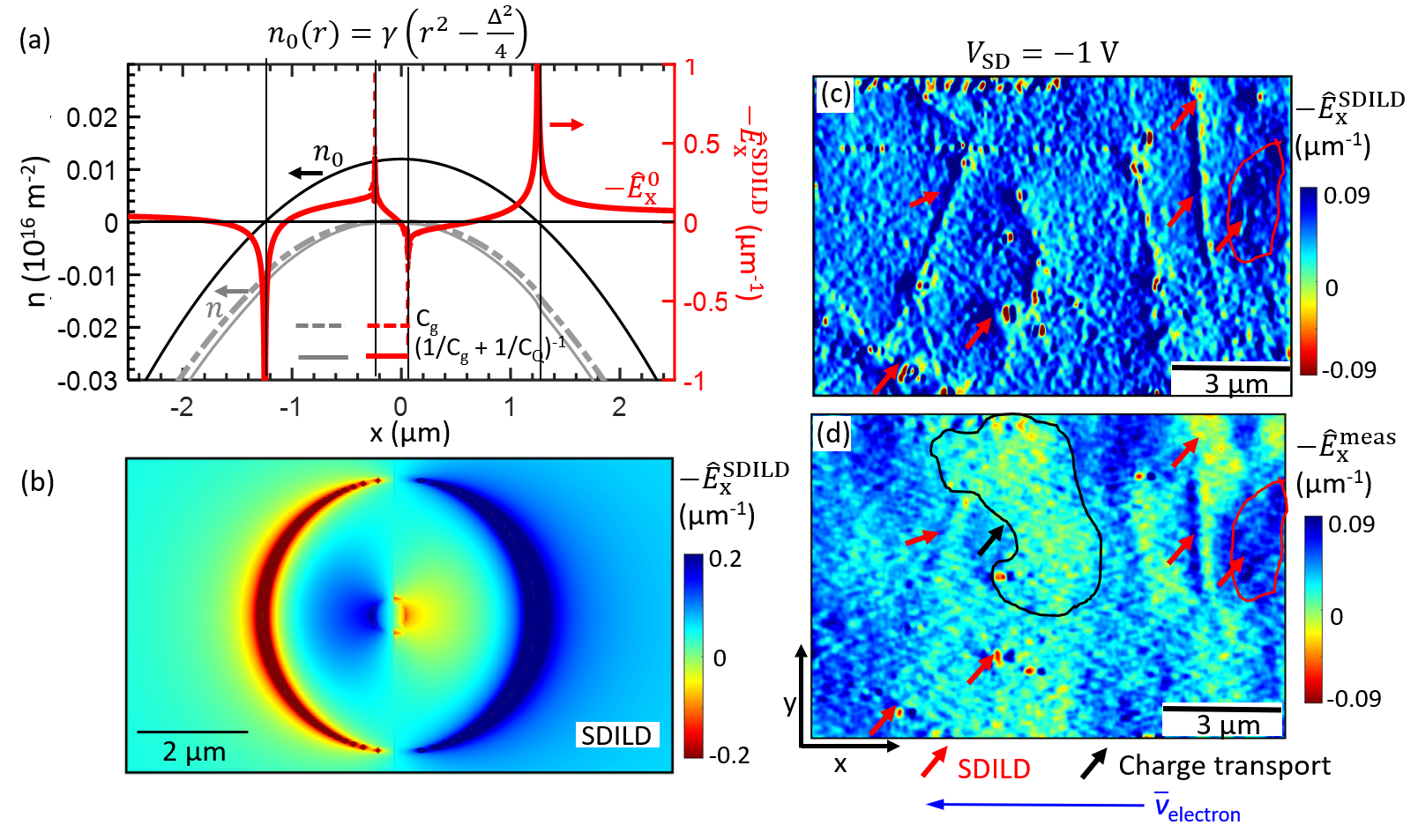}
	\caption{{\bf Electric fields from SDILD.} (a) Doping profiles (left axis): parabolic $n_0(x)$ at $V_{\rm SD} = 0\,$V (black line) using the equation above with $\Delta=2.5$\,$\mu$m, $\gamma= -8.3 \cdot 10^{25}\,{\rm m^{-4}}$ and resulting $n(x,V_{\rm SD}= 0.1\,$V) (grey lines) according to eq.~(\ref{eq:SDILD}) either neglecting quantum capacitance (dashed line) or including it (full line). In addition, the deduced $\widehat{E}_{x}^{\rm SDILD}(x)$ (red line, right axis) is shown using eq.~(\ref{Eq_NegField_TrivialEff}) without quantum capacitance (dashed) or with it (full). 
		(b) 2D color map of $\widehat{E}_{ x}^{\rm SDILD}(x,y,V_{\rm SD}= 0.1\,$V) for a rotational symmetric parabolic doping profile  (formula and parameters as in a) located in the center between source and drain electrode. 
		The trivial field $\widehat{E}_{x}^{0}$ has been subtracted. (c) $\widehat{E}_{x}^{\rm SDILD}(x,y,V_{\rm SD})$ map calculated from a measured $V_{\rm CPD}(x,y,V_{\rm SD}=0\,{\rm V})$ map (KPFM) at charge neutrality ($V_{\rm gate} = 72\,$V) for $V_{\rm SD}=-1\,$V. (d) Measured $\widehat{E}_{x}^{\rm meas}(x,y)$ of the same area as c. The red arrows in c, d  mark features present in both images and, hence, attributed to SDILD.  The black arrow in d points at an encircled feature not present in c, and, hence, assigned to the current flow.}
	\label{fig:NegEfields_SDILD}
\end{figure}

As an example, we consider a one-dimensional parabolic profile  $n_0(x)$ along $x$ with negative curvature crossing zero twice and being located in the center between source and drain electrode (black line, Fig.~\ref{fig:NegEfields_SDILD}a). Applying $V_{\rm SD}$ results in a vertical and horizontal shift (grey line, Fig.~\ref{fig:NegEfields_SDILD}a). The vertical shift is more pronounced than the horizontal one, since the total length within Fig.~\ref{fig:NegEfields_SDILD}a is much smaller than the distance to the drain electrode. Since the drain is grounded, half of the $V_{\rm SD}$ induced potential drops towards the displayed center region. Naturally, the two zero crossings of the $n(x)$ parabola are shifted inwards with respect to the zero crossings of $n_0(x)$ such that eq.~(\ref{Eq_NegField_TrivialEff}) implies four distinct divergences of $\widehat{E}_{x}^{\rm SDILD}(x)$ along $x$. The direction of divergence, peak or dip, changes between the zeroes of $n(x)$ and the zeroes of $n_0(x)$ at the same slope of the parabola (eq.~(\ref{Eq_NegField_TrivialEff})). It also changes with direction of slope for the same density, either $n(x)$ or $n_0(x)$. Consequently, a quartet of $\widehat{E}_{x}^{\rm SDILD}(x)$ divergences appears for the inverted $n_0(x)$ parabola with order dip-peak-dip-peak from left to right (red line, Fig.~\ref{fig:NegEfields_SDILD}a). Quantum capacitance barely changes this scenario (full and dashed grey line, Fig.~\ref{fig:NegEfields_SDILD}a).

Figure~\ref{fig:NegEfields_SDILD}b shows a 2D plot of $\widehat{E}_{x}^{\rm SDILD}(x,y)$ for a rotationally symmetric parabolic doping profile $n_0(x,y)$ with the same apex and curvature as in Fig.~\ref{fig:NegEfields_SDILD}a. It features four lobes, one for each of the four divergences that we have already discussed in Fig.~\ref{fig:NegEfields_SDILD}a.
The inner lobes appear more extended along $x$ than the outer lobes as consequence of the weaker slopes $dn(x,y)/dx$ at the zeroes of $n(x,y)$ compared to the stronger slopes $dn_0(x,y)/dx$ at the zeroes of $n_0(x,y)$. 


Figure~\ref{fig:NegEfields_SDILD}c displays an $\widehat{E}_{x}^{\rm SDILD}(x,y)$ map deduced from a measured $V_{\rm CPD}(x,y, V_{\rm SD}=0\,{\rm V})$ map by KPFM using eqs.~(\ref{DopingCPDrelation}), (\ref{eq:SDILD}) and (\ref{eq:E_SDLD}). For comparison,
the measured $\widehat{E}_{x}^{\rm meas}(x,y)$ map of the same area is displayed in Fig.~\ref{fig:NegEfields_SDILD}d. The features that are similar in both images (red arrows) are attributed to artifacts from SDILD, while additional features in the $\widehat{E}_{x}^{\rm meas}(x,y)$ map, such as the encircled yellowish area of inverted electric field, can be assigned to current induced features. 

\subsection{Origin of discrepancies between  $\widehat{E}_{x}^{\rm SDILD}(x,y)$ and $\widehat{E}_{x}^{\rm meas}(x,y)$}  
\label{sec:fig2_discrepancies}

\begin{figure}
	\centering
	\includegraphics[width=\textwidth]{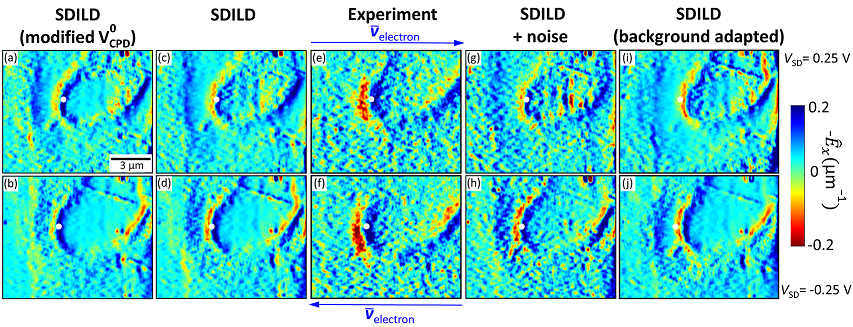}
	\caption{{\bf Sources of Error in SDILD images.} (a), (b) $\widehat{E}_{x}^{\rm SDILD}(x,y)$ deduced from Fig.~\ref{fig2:SDind_doping}e, main text, according to eqs.~(\ref{eq:tilted_VCPD})--(\ref{Eq_NegField_TrivialEff}), but with a manually optimized $V_{\rm CPD}^0=-14$\,meV. (c), (d) Same as a, b for  $V_{\rm CPD}^0=-4$\,meV as deduced from the average $V_{\rm CPD}$ of the $V_{\rm CPD}(x,y)$ map recorded via $V_{\rm gate} = V_{\rm D}$ (same as Fig.~\ref{fig2:SDind_doping}f, h, main text). (e), (f) Experimentally determined  $\widehat{E}_{x}^{\rm meas}(x,y)$ using eqs.~(\ref{eq:1}) and (\ref{eq:3}), main text (same as  Fig.~\ref{fig2:SDind_doping}g, i, main text). (g), (h) Same as c, d, but adding the noise of the experimental $V_{\rm CPD} (x,y,V_{\rm SD})$ to the artificially tilted $V_{\rm CPD} (x,y,V_{\rm SD}=0)$ (eq.~(\ref{eq:2}), main text) prior to calculating $\widehat{E}_{x}^{\rm SDILD}(x,y)$. (i), (j) Same as c, d with modified background subtraction of the KPFM images (see text). Upper row: $V_{\rm SD}=0.25 V$, lower row: $V_{\rm SD}=-0.25 V$. Average direction of electron flow $\overline{v}_{\rm electron}$ is indicated. White dots mark the same position in all images.}
	\label{fig:SDILD_discrepancy}
\end{figure}

As discussed in the main text, there are some remaining discrepancies between the calculated SDILD images $\widehat{E}_{x}^{\rm SDILD}(x,y)$ (Fig.~\ref{fig2:SDind_doping}f,h, main text, and Fig.~\ref{fig:SDILD_discrepancy}e, f) and the measured $\widehat{E}_{x}^{\rm meas}(x,y)$ (Fig.~\ref{fig2:SDind_doping}g,i, main text, and Fig.~\ref{fig:SDILD_discrepancy}c, d). The rms values of  difference images $\widehat{E}_{x}^{\rm meas}(x,y)-\widehat{E}_{x}^{\rm SDILD}(x,y)$ are 0.04\,$\mu {\rm m}^{-1}$, i.e. about 10 \% of the difference between maximum and minimum of the dominating structures (e.g. around the white dot in Fig.~\ref{fig:SDILD_discrepancy}) that we regard as SDILD features. The possible origin of these discrepancies is discussed in the following.

Firstly, the noise within the two KPFM images (section~\ref{subsec_Noise_Sensitivity}), that are used to determine $\widehat{E}_{x}^{\rm meas}(x,y)$ (Fig.~\ref{fig:SDILD_discrepancy}e, f), is uncorrelated. In contrast, the determination of $\widehat{E}_{x}^{\rm SDILD}(x,y)$ employs the same noise twice by tilting the recorded $V_{\rm CPD}(x,y,V_{\rm SD}=0)$ to mimic $V_{\rm CPD}(x,y,V_{\rm SD})$. This leads to cancellation, i.e. to a reduced noise in $\widehat{E}_{x}^{\rm SDILD}(x,y)$ (Fig.~\ref{fig:SDILD_discrepancy}c, d) with respect to the experiment (Fig.~\ref{fig:SDILD_discrepancy}e, f). In order to compensate this error, we deduce the noise map of the experimental $V_{\rm CPD}(x,y,V_{\rm SD})$ via Gaussian smoothing ($\sigma = 200$\,nm) of the recorded $V_{\rm CPD}(x,y,V_{\rm SD})$ and subsequently subtracting the smoothed image from the recorded one. The resulting noise map is then added to the tilted $V_{\rm CPD}(x,y,V_{\rm SD}=0)$ prior to calculating $\widehat{E}_{x}^{\rm SDILD}(x,y)$. The result is shown in Fig.~\ref{fig:SDILD_discrepancy}g, h. It adequately accounts for the noise strength in the measured $\widehat{E}_{x}^{\rm meas}(x,y)$ (Fig.~\ref{fig:SDILD_discrepancy}e, f). In line, the rms discrepancy of these $\widehat{E}_{x}^{\rm SDILD}(x,y)$ maps to $\widehat{E}_{x}^{\rm meas}(x,y)$ drops to 0.02\,$\mu {\rm m}^{-1}$, hence, improving by a factor of two.

Secondly, there is a slight lateral offset ($\sim 0.5$\,$\mu$m) between the observed SDILD features in $\widehat{E}_{x}^{\rm meas}(x,y)$ and the calculated ones (white dots in Fig.~~\ref{fig:SDILD_discrepancy}c--h). Using the simultaneously recorded topography images, we checked that this offset is not caused by lateral drift or creep that amounts to below 250 nm between subsequent images. The lateral offset can be avoided by changing $V_{\rm CPD}^0$ by $\sim 10$\,meV (Fig.~\ref{fig:SDILD_discrepancy}a,b). However, such change is larger than the precision in determining $V_{\rm CPD}^0$ of 2\, meV 
(Fig.~\ref{DopingFromCPD_fig}). Hence, we assume that the work function of the tip slightly changes during the recording of subsequent images\cite{Dombrowski1999}, which is reasonable since we are operating at ambient conditions on a sample partially covered by polymers. 

Thirdly, the width of the experimentally observed features (Fig.~\ref{fig:SDILD_discrepancy}e, f) is slightly larger than in the simulation (Fig.~\ref{fig:SDILD_discrepancy}g, h) and also the intensity is partially larger  in the experiment (right feature in Fig.~\ref{fig:SDILD_discrepancy}e, g). These deviations can be reduced by manually adapting the background subtraction of the KPFM images   with respect to
the numerically determined background $V_{\rm CPD}^{\rm BG}(x,y)=\sum_{i=1}^3\sum_{j=1}^3 a_{ij}x^{i-1}y^{j-1}$ (section~\ref{subsec_BG}, Fig.~\ref{fig2_KPFM}c--d). As an example, Fig.~~\ref{fig:SDILD_discrepancy}i, j show $\widehat{E}_{x}^{\rm SDILD}(x,y)$ calculated after reducing $a_{13}$ by 8\,\% with respect to the numerically determined value as employed in Fig.~~\ref{fig:SDILD_discrepancy}c, d. Obviously, the right feature in Fig.~\ref{fig:SDILD_discrepancy}i becomes more prominent and the right feature in Fig.~\ref{fig:SDILD_discrepancy}j becomes slightly wider, both, improving the agreement with the experimental data, but at the expense of a lateral shift of the features not matching the experiment. Since we assume that temporal fluctuations of the doping profile (Fig.~\ref{fig8:ElecField_ImagingConditions}) are, at least, similarly important for the detailed shape of the SDILD features, we refrain from a manual adaption of $a_{ij}$ to optimze $\widehat{E}_{x}^{\rm SDILD}(x,y)$.    

Most importantly, the discrepancies between $\widehat{E}_{x}^{\rm meas}(x,y)$ and the various $\widehat{E}_{x}^{\rm SDILD}(x,y)$ in Fig.~\ref{fig:SDILD_discrepancy} are of order 10\,\% of the $\widehat{E}_{x}^{\rm meas}(x,y)$ variations featured by the prominent SDILD features. On the other hand, the features expected from SDILD in Fig.~\ref{fig3:NegFields_ViscousFlow} are nearly one order of magnitude weaker in strength than the observed inverted fields that we attribute to hydrodynamic electron flow. Thus, the minor discrepancies between
$\widehat{E}_{x}^{\rm meas}(x,y)$ and  $\widehat{E}_{x}^{\rm SDILD}(x,y)$
cannot account for any of the features attributed to electron viscosity.   

\subsection{Simulation of the Measured Electric Fields in Fig.~\ref{fig3:NegFields_ViscousFlow} by SDILD}

Figure~\ref{fig3:NegFields_ViscousFlow}e, f, main text, showcase a doublet $\widehat{E}_{x}^{\rm meas}(x,y)$ structure as determined by EFM that exhibits a spatial extent and an intensity of its two lobes much larger than the apparent electric field generated by SDILD (Fig.~\ref{fig3:NegFields_ViscousFlow}c, d, main text). 
Hence, we attribute this feature to viscous electron flow.
This structure is shown again in Fig.~\ref{fig:SDILD_PatchedDoping}e.
Since we can not avoid small temporal fluctuations of the doping profile $n_0(x,y)$  (section~\ref{subsec_DopingStability}),  
we attempted to reproduce the recorded doublet $E_x^{\rm meas}(x,y)$  structure by an arbitrary, artificial doping profile $n_0(x,y)$. Eventually, we found such a doping profile (black line, Fig.~\ref{fig:SDILD_PatchedDoping}d) that, however, appears to be impossible in reality. We describe the reasoning in the following
after recalling that the doping profiles recorded prior and after the $V_{\rm CPD}(x,y,V_{\rm SD}=0.1\,{\rm V})$ map necessary for $E_x^{\rm meas}(x,y)$ are, both, not capable to reproduce  $E_x^{\rm meas}(x,y)$ via SDILD (Fig.~\ref{fig3:NegFields_ViscousFlow}, main text).

\begin{figure}
	\centering
	\includegraphics[width = 160 mm]{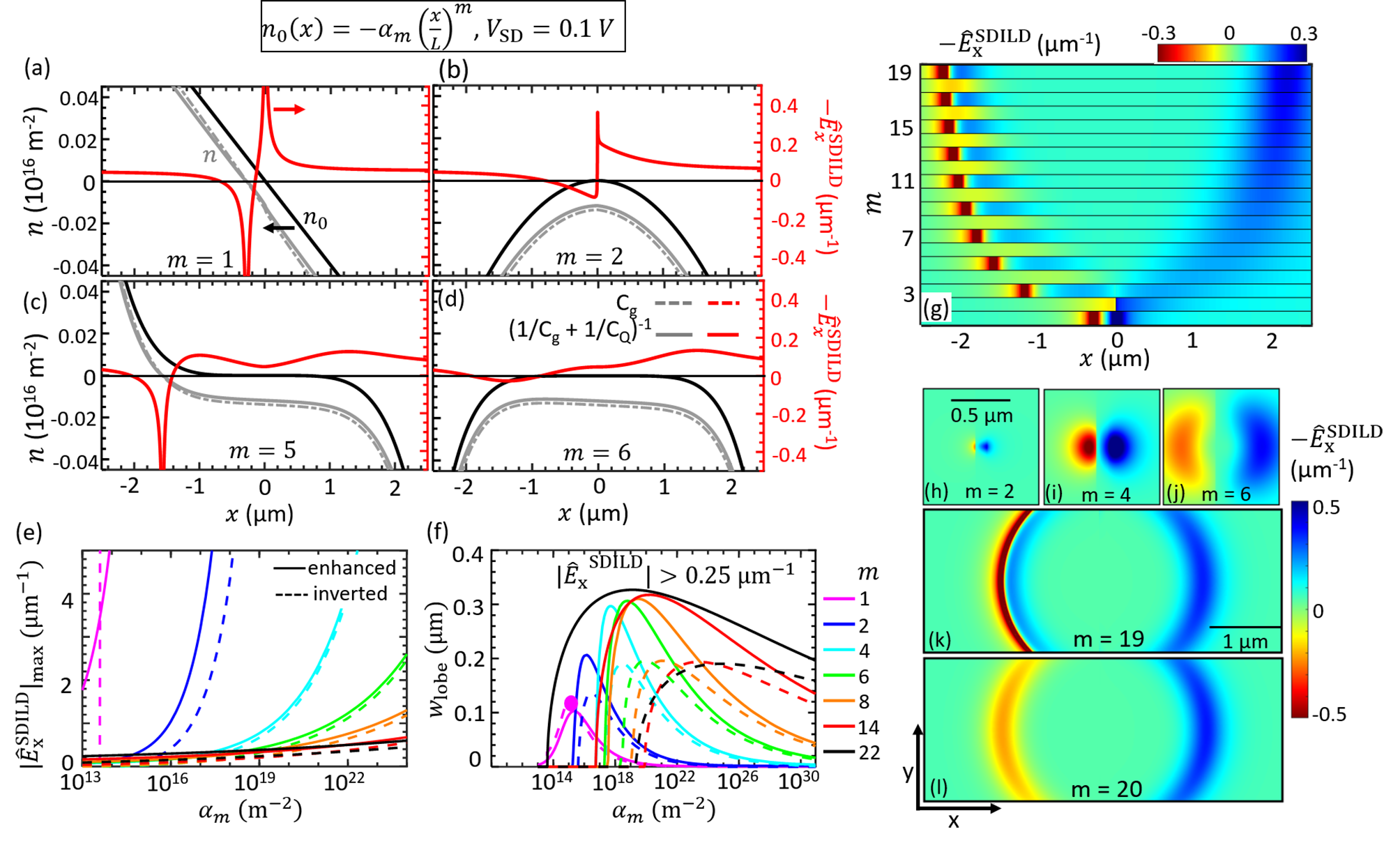}
	\caption{{\bf $\widehat{E}_{x}^{\rm SDILD}(x,y)$ for different exponents of $n_0(x,y)$.} (a)--(d) Simulated doping profiles  $n_0(x)$ (black lines) according to the formula on top with $\alpha_m =1\cdot 10^{15}$/m$^2$ and $n(x)$ at $V_{\rm SD}=0.1$\,V (grey lines) according to eq.~(\ref{eq:SDILD}) for different exponents $m$ as marked (left axis).  The resulting electric field from SDILD $\widehat{E}_{x}^{\rm SDILD}(x)$ according to eq.~(\ref{Eq_NegField_TrivialEff}) is drawn in red  (right axis). Full (dashed) lines include (do not include) quantum capacitance. 
		(e) Extremal electric field within the two lobes of the simulated dipolar structures (as shown in a, b, d) for different $m$. Solid and dashed lines refer to the lobe with enhanced and inverted electric field, respectively. (f) Width along $x$ of the simulated two lobes with $|\widehat{E}_{x}^{\rm SDILD}(x,y)|\ge 0.25\,{\mathrm \mu}$m. Large pink dot marks the width deduced from a.
		(g) $\widehat{E}_{x}^{\rm SDILD}(x)$ for different $m$ using $n_0(x)$ as given above a, b, $\alpha_m =1\cdot 10^{15}$/m$^2$. (h)--(l)  $\widehat{E}_{x}^{\rm SDILD}(x,y)-\widehat{E}_x^0$ (eq.~(\ref{Eq_NegField_TrivialEff})) for different exponents $m$ as marked using a rotational symmetric $n_0(|\vec{x}|)$ with radial dependence as given above a, b, $\alpha_m=5\cdot 10^{18}$/m$^2$ (b), h--j have the same size, k--l have the same size. All images use $V_{\rm SD}=0.1$\,V.}
	\label{fig:SDILD_exponents}
\end{figure}

\begin{figure}
	\centering
	\includegraphics[width = 160 mm]{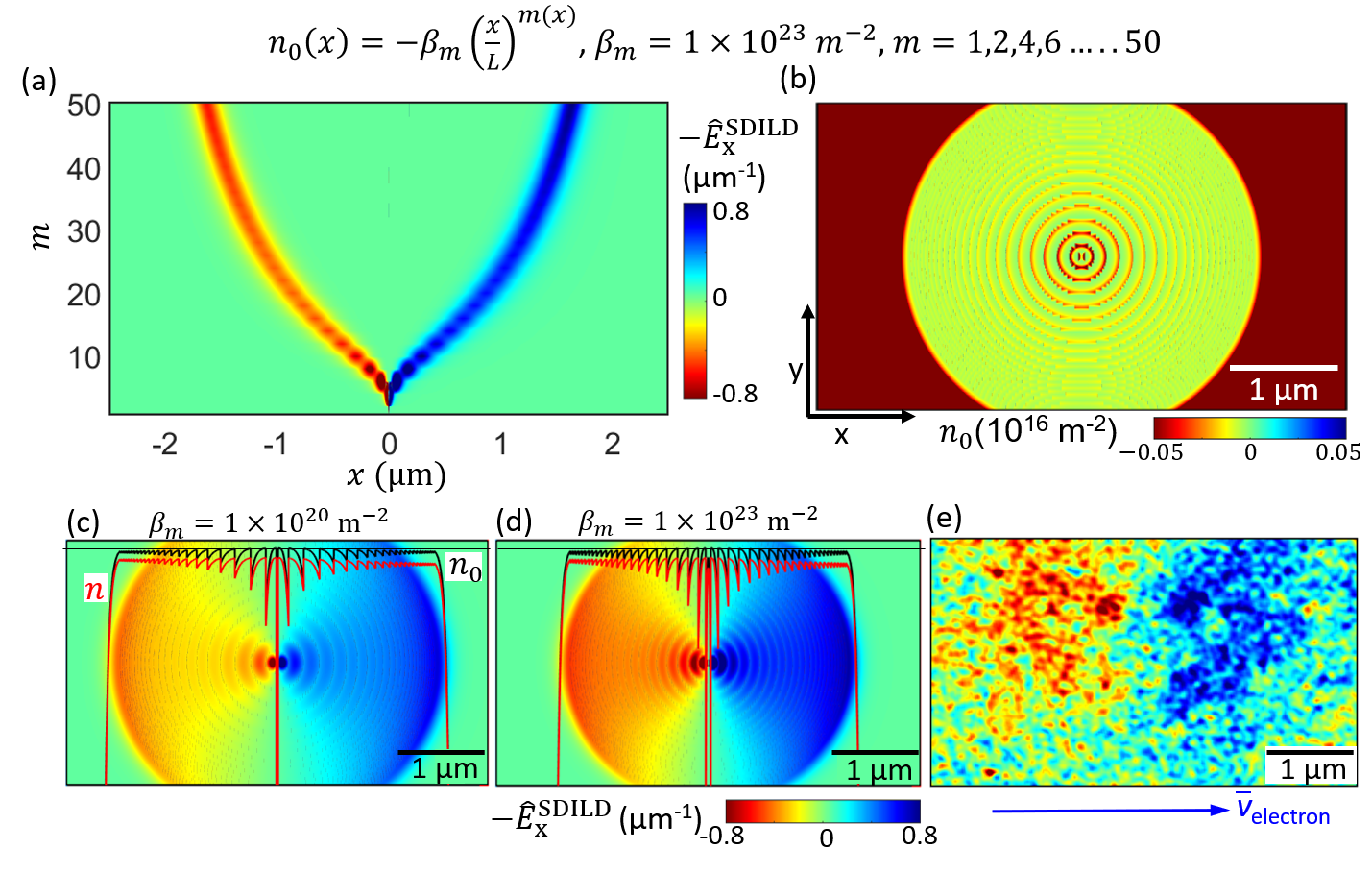}
	\caption{{\bf Patching doping profiles to reproduce  $\widehat{E}_{x}^{\rm meas}(x,y)$ by SDILD.} (a) $\widehat{E}_{ x}^{\rm SDILD}(x)$ at $V_{\rm SD}=0.1$\,V for different exponents $m$ of the employed one-dimensional $n_0(x)$ function given on top using eqs.~(\ref{eq:SDILD}). (\ref{Eq_NegField_TrivialEff}). 
		(b) Two-dimensional doping distribution $n_0(x,y)$ using eq.~(\ref{eq:Doping_Patched}) and the patched $m$ sequence as in a, but rotationally symmetric. (c)--(d) $\widehat{E}_{ x}^{\rm SDILD}(x,y)$ using $n_0(x,y)$ from b, but with different constants $\beta_m$ as marked. Black and red curves are cross sections through the doping profiles $n_0(x,y=0)$ and $n(x,y=0,V_{\rm SD})$ along $x$ at the horizontal center of the image.
		(e) $\widehat{E}_{x}^{\rm meas}(x,y)$ at $V_{\rm SD} = 0.1$\,V via eq.~(\ref{eq:Ex_VtrDer_filter}), (\ref{eq10:Vtransport}) (same as Fig.\ref{fig3:NegFields_ViscousFlow}e, main text), same colorbar for c--e.}
	\label{fig:SDILD_PatchedDoping}
\end{figure}

We firstly consider a one dimensional doping profile $n_0(x)$ and later extend the analysis to two dimensions. To observe a divergence of the electric field via SDILD, either $n_0(x)$ or $n(x)$ must cross zero (eq.~(\ref{Eq_NegField_TrivialEff})). As shown in Fig.~\ref{fig:NegEfields_SDILD}a--b, $\widehat{E}_x^{\rm SDILD}$ rapidly decreases away from such crossings. Hence, we firstly focus on the area close to such crossings.
If, both, $n_0(x)$ and the shifted $n(x)$ cross zero with the same direction of slope, one gets a dipolar divergence of $\widehat{E}_x^{\rm SDILD}(x)$ (Fig.~\ref{fig:NegEfields_SDILD}a). Around the crossing point, we apply a Taylor expansion of $n_0(x)$ and $n(x)$, here given for $n_0(x)$ around a crossing at $x=0$, to deduce leading terms reading %
\begin{equation}
n_0 (x) = \sum_m \alpha_m \left(\frac{x}{L}\right)^m,\,\,\, m = 1,2,3.....
\end{equation}
Here, $m$ is an integer exponent, $\alpha_m$ the corresponding prefactor and $L$ a constant length.
Considering eq.~(\ref{Eq_NegField_TrivialEff}), we find for the individual terms of the expansion
\begin{equation}
\frac{dn_0(x)/dx}{\sqrt{n_0(x)}} =\frac{m\sqrt{\alpha_m}}{L}\cdot \left(\frac{x}{L}\right)^{m/2-1}.
\label{eq:S19}
\end{equation}
Consequently, the linear term of the expansion ($m=1$) produces the divergence, while the other terms contribute (in first order) by 
zero ($m>2$) or $m\sqrt{\alpha_m}/L$ ($m=2$) to $\widehat{E}_{x}^{\rm SDILD} (x)$ at the crossing point. Thus, the width of the prominent SDILD feature around $n_0(x)=0$ is largely dominated by the linear term of the Taylor expansion (see discussion below for the influence of higher order terms). 
A resulting dipolar $\widehat{E}_{x}^{\rm SDILD} (x)$ for a linear $n_0(x)$ term, as calculated numerically via eq.~(\ref{Eq_NegField_TrivialEff}), is shown in Fig.~\ref{fig:SDILD_exponents}a. It exhibits a very small width $w_{\rm lobe}$ of its two lobes at the required strength $|\widehat{E}_{x}^{\rm SDILD} (x)|>0.25/{\mathrm \mu}$m that has been found in the experiment (Fig.~\ref{fig:SDILD_PatchedDoping}e). This width $w_{\rm lobe}$ in $x$ direction, where $|\widehat{E}_{x}^{\rm SDILD} (x)|>0.25/{\mathrm \mu}$m, is plotted as a function of the linear prefactor $\alpha_1$ in Fig.~\ref{fig:SDILD_exponents}f (pink lines) revealing a maximum of $w_{\rm lobe}$ for both lobes (dashed, full line). Thus, albeit the extremal $|\widehat{E}_{x}^{\rm SDILD} (x)|$ increases monotonously with $\alpha_1$
(Fig.~\ref{fig:SDILD_exponents}e), $w_{\rm lobe}$ is non-monotonous with a
maximum at $w_{\rm lobe}\approx 0.1\,\mathrm{\mu m}$, exactly for the $n_0(x)$ slope as used in Fig.~\ref{fig:SDILD_exponents}a. Thus, a linear zero crossing of $n_0(x)$ can not reproduce the width of the experimentally observed  $\widehat{E}_{x}^{\rm meas} (x)$ dipolar feature with $w_{\rm lobe} \simeq 1.7\,{\mathrm \mu}$m. 
Numerical tests of various shapes around the $n_0(x)=0$ crossing  corroborated this result for more general crossings within reasonable limits.  


As alternative, we consider $n_0(x)$ functions that are touching $n_0(x)=0$ with zero slope $dn_0(x)/dx=0$ (argument applies analogously for $n(x)$).  Again, we apply a Taylor expansion of $n_0(x)$ around $n_0(x)=0$. This also adresses the higher order terms of the zero crossings on the same footing. We realize that the odd powers of the Taylor expansion with $m>1$ cannot produce a dipolar structure at all (Fig.~\ref{fig:SDILD_exponents}c), while for $m=1$ the same argument as above applies. As illustrated for $m=5$ in Fig.~\ref{fig:SDILD_exponents}c, the functions with odd exponent $m>1$ produce a strong divergence of $\widehat{E}_{x}^{\rm SDILD}(x)$ at the shifted $n(x)=0$ due to its large slope $|dn/dx|$. In addition, two weaker features appear that are symmetric around the touching point $n_0(x)=0$. These two weaker features originate from $n_0(x)$ and naturally exhibit the same strength on both sides of $n_0(x)=0$ and the same distance from $x=0$ (eq.~(\ref{Eq_NegField_TrivialEff})). Thus, each odd power produces either a monopolar or a tripolar structure, but not a dipolar one as crosschecked numerically. In contrast, the even powers of $n_0(x)$ result in an antisymmetric $\widehat{E}_{x}^{\rm SDILD}(x)$ feature around the touching point  provided that $n(x)$ is moved completely away from $n(x)=0$ via $V_{\rm SD}$
(Fig.~\ref{fig:SDILD_exponents}b,d).
Hence, a dipolar structure naturally appears for an even power function of $n_0(x)$ ($n(x)$)  with negative (positive) curvature in case of the applied positive $V_{\rm SD}$. However, numerically, it turns out that also the lobes of these dipolar $\widehat{E}_{x}^{\rm SDILD}(x)$ structures exhibit a maximum width of 0.3\,$\mu$m at the required $\widehat{E}_{x}^{\rm SDILD}>0.25/\mu$m (Fig.~\ref{fig:SDILD_exponents}f). This is again much too small to reproduce the experiment (Fig.~\ref{fig:SDILD_PatchedDoping}e). 

However, the numerical analysis of the different exponents reveals that the extrema of the dipolar (even $m$) and tripolar (odd $m>1$) $\widehat{E}_{x}^{\rm SDILD}(x)$ structures move continuously outwards with increasing $m$. This is displayed in Fig.~\ref{fig:SDILD_exponents}g showing $\widehat{E}_{x}^{\rm SDILD}(x)$ profiles for each exponent $m$, in this case using $\alpha_m=1\cdot10^{15}$/m$^2$ independent of $m$. 
The systematic shift to larger distances  with $m$ is due to the increasingly flat $n_0(x)$ part in the center (Fig.~\ref{fig:SDILD_exponents}a--d). While eq.~(\ref{eq:S19}) indicates a continuously increasing $\widehat{E}_{x}^{\rm SDILD}(x)$ with increasing $|x|$ and increasing $m$, this increasing strength due to $n_0(x)$ gets increasingly compensated by the additional $n(x)$ term in eq.~(\ref{Eq_NegField_TrivialEff}). 
Basically, the offset between $n_0(x)$ and $n(x)$ within the square root denominators looses its importance. Very roughly, the extrema appear at the $x$ positions, where $n_0(x)$ has a similar value than $n(x=0)$. This naturally explains that the extrema move outwards with increasing $m$.  Nevertheless, its width $w_{\rm lobe}$ always remains far below the experimental $w_{\rm lobe}\simeq 1.7\mathrm{\mu}$m (Fig.~\ref{fig:SDILD_exponents}f). 
In addition, Fig.~\ref{fig:SDILD_exponents}g nicely shows the alternating tripolar and dipolar structures for increasing $m$.

The extension to 2D isotropic profiles does not change the above arguments as shown exemplarily in Fig.~\ref{fig:SDILD_exponents}h--l. 


Since the single components of the Taylor series are not able to reproduce inverted field lobes with intensity above $ 0.25\,\mu {\rm m}^{-1}$ and width $w_{\rm lobe}>1.7\,\mu$m, we have to patch $n_0(x)$ piecewise in order to get a large enough width of the dipolar structure.
Since a patched sequence of crossings would produce a sequence of dipolar structures via their dominating linear term instead of a single extended dipolar structure,  the option of patching crossing points does not exist.

The fact that the extrema of the dipolar structure for each even $m$ shift outwards with increasing $m$ (Fig.~\ref{fig:SDILD_PatchedDoping}a) suggests a natural way of patching by choosing $x$ ranges of extremal $\widehat{E}_{x}^{\rm SDILD}$ for each even $m$ until the extrema position of the largest $m$ matches the $x$ extension of the experimental dipolar $\widehat{E}_{x}^{\rm meas} (x,y)$ structure. More formally, we use
\begin{equation}
n_0(x) = - \beta_m \cdot \left( \frac{x}{L} \right)^{m(x)}
\label{eq:Doping_Patched}
\end{equation}
with $m(x)$ being a step function that is a piecewise constant integer along $x$, featuring subsequently $m=1,2,4,6,8,...$ and using $L=2.5\,\mu$m (Fig.~\ref{fig:SDILD_PatchedDoping}a,b).
Increasing the parameter $\beta_m$ tunes the intensity of the dipolar structure (Fig.~\ref{fig:SDILD_exponents}e), but increases the number of required patches simultaneously by decreasing $w_{\rm lobe}$ (Fig.~\ref{fig:SDILD_exponents}f).
Using this 
construction, we found that $\sim 50$ patches are required to reproduce the width and strength of the experimentally observed dipolar structure (Fig.~\ref{fig:SDILD_PatchedDoping}c--e). 
However, this necessarily requires that each area of different $m$ is adjusted to $n_0(x)=0$/m$^2$ in the center (at $x=0$) implying jumps of $n_0(x)$ as displayed by the black line in Fig.~\ref{fig:SDILD_PatchedDoping}c, d. 
It implies $\sim 50$ jumps in $n_0 (x)$ on a width of $~3\,\mu$m (lines in Fig.~\ref{fig:SDILD_PatchedDoping}c, d). Each jump returns $n_0(x)$ basically back to zero. Such doping profiles are experimentally very unlikely, in particular, to be present in a certain region but not in its surrounding. 

Importantly, albeit distinct patching profiles might reproduce the experimental dipolar structure as well, the patching method with its multiple jumps back to $n_0(x)\approx 0$ cannot by avoided. Any type of a smooth return to $n_0(x)\approx 0$  would lead to the opposite electric field such that instead of an extended lobe, one would get multiple dipolar structures within the lobe area.   
The resulting requirement of the jumps strongly excludes a physical possibility that the observed dipolar structure is caused by SDILD, even in the unlikely case that the doping profile during recording $V_{\rm CPD}(x,y,V_{\rm SD})$ is substantially different from the one prior and after the recording. 

\section{Influence of Dirt on Potential Maps and Current Induced Electric Field Maps}
\label{sec:dirt}

\begin{figure}
	\centering
	\includegraphics[width = 160 mm]{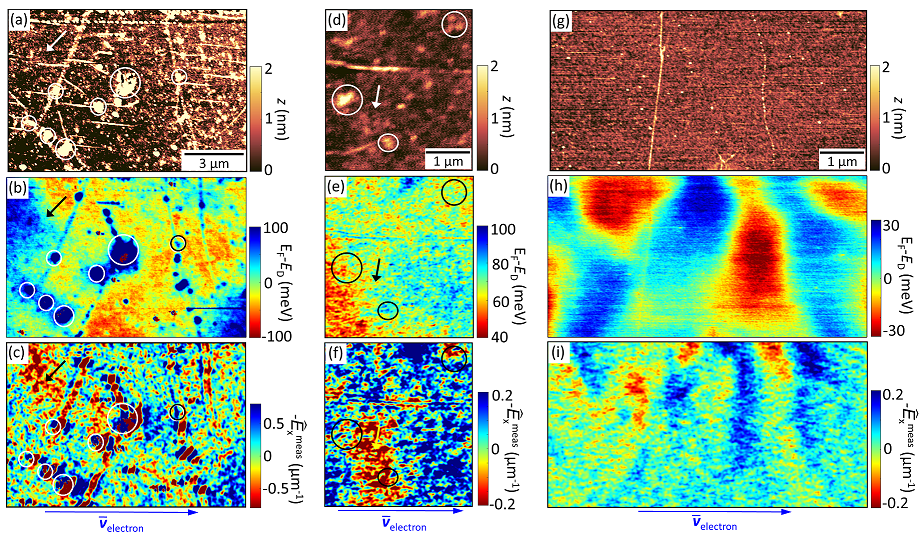}
	\caption{{\bf Influence of Polymer Residues.} (a) Topography of a graphene area with multiple larger clusters of dirt (height $> 10$\,nm) partly marked by circles, tapping mode AFM. The arrow points to an inverted field area visible in c. (b) $V_{\rm CPD}(x,y)$ map displayed as $(E_{\rm F}-E_{\rm D})(x,y)$ of the same area as in a with the same circles and arrow, $V_{\rm Gate}=V_{\rm D}=82$\,V. Large clusters exhibit an increased potential. (c) $\widehat{E}_{x}^{\rm meas}(x,y)$ map of the same area as a--b revealing inverted electric field areas (red) that are not related to dirt clusters (arrow), same circles and arrow as in a--b, $V_{\rm SD}= 0.1\,$V, $V_{\rm Gate}=82$\,V. (d)--(f) Same as a--c recorded on an area without larger clusters of dirt (height corrugations $<3$\,nm). Smaller clusters are marked by circles. No clear fingerprints of polymer residues appear in the potential map, while inverted fields are still observed in the $\widehat{E}_{x}^{\rm meas}(x,y)$ map, $V_{\rm SD}= 0.1\,$V, $V_{\rm Gate}=V_{\rm D}=86$\,V. (g)--(i) Same as a--c recorded after sweeping the area by contact mode AFM. Polymer residues largely disappeared and the roughness drops to 0.36\,nm leading to a similar potential map as in e and still to inverted current induced electric fields, $V_{\rm SD}= 0.2\,$V, $V_{\rm Gate}=57$\,V, $V_{\rm D}=52$\,V.}
	\label{fig:S12c}
\end{figure}

The transfer process of graphene from Cu to SiN as well as the subsequent lithography (section~\ref{sec:prep}) leaves polymer residues on the surface that might influence the surface potential $V_{\rm CPD}(x,y)$. These polymers are visible in  topography (tapping mode AFM) as corrugations with rms values $ 0.3-4$\,nm that are varying between images. The lowest corrugation (rms value 0.3\,nm) remains identical after sweeping the graphene carefully in AFM contact mode (section~\ref{sec:sweep}) indicating that it is induced by the substrate. Individual larger clusters of residues with heights up to 60\,nm appear in some areas prior to sweeping, mostly close to folds and bubbles (Fig.~\ref{fig:S12c}a).
Since all residues result from resists and, hence, are insulating, they affect the work-function of graphene by the formation of interface dipoles \cite{Loppacher2004, Pivetta2005, Ploigt2007,Prada2008} or by trapped charges \cite{Teyssedre2021}. 


Dipoles from residues on top or below graphene are known to locally dope graphene.\cite{Melios2016} This has been evidenced, e.g., via intentionally preparing self-assembled polymer films on the substrate prior to graphene deposition \cite{Wang2011}. A self-assembled PMMA film below graphene changes the work function by $\sim 10$\,meV only. It is, moreover, well known that such  small $V_{\rm D}$ shifts barely affect the square-root dependence of $V_{\rm CPD}(V_{\rm gate}-V_{\rm D})$ arising from the Dirac cone dispersion  \cite{Yu2009, Samaddar2016,Behm2021}. Hence, the interface dipoles only lead to a local shift of $V_{\rm CPD}$ (see also section~\ref{sec:compressibility}). This can not lead to inverted electric fields induced by $V_{\rm SD}$ except by SDILD (section~\ref{Sec:EField_SDILD}).     

However, the residues on the surface could also trap charges changing the doping in graphene abruptly. Residues that can be charged imply a stronger change in doping and, thus, in $V_{\rm CPD}(x,y)$ and, hence, can be identified. Generally, we find that residues with height appearing lower than 2\,nm in tapping mode AFM can not be identified in $V_{\rm CPD}(x,y)$ (Fig.~\ref{fig:S12c}d-e) independent of $V_{\rm gate}$ such that their doping effect is negligible.
Clusters larger than 2\,nm in height
cause a local increase of the Fermi level by $20-100$\,meV corresponding to electron doping
(Fig.~\ref{fig:S12c}a-b) with a tendency of larger doping for larger height. Figure~\ref{fig:S12c}a shows a rather dirty area of graphene. Such areas have been discarded for analysis in the main text. Large clusters of polymers with height up to $60$\,nm appear in the tapping mode AFM image as partly encircled. The rms roughness in this area amounts to $4$\,nm. The measured surface potential map of the same area  is displayed as $(E_{\rm F}-E_{\rm D})(x,y)$ in Fig.~\ref{fig:S12c}b. 
Dark spots appear at the positions of the clusters as well as at the wrinkles that appear to be decorated by polymers. In line with the above reasoning, we deduce that the polymers attract electrons in the graphene indicating that they are positively charged. The observed change in $E_{\rm F}$ by $20-100$\,meV around the clusters corresponds to a local electron doping $n_0(x,y)= 0.3-6\cdot 10^{15}$\,/m$^2$, that is surrounded by areas of hole doping. 
Indeed, we observe the typical doublet structures caused by SDILD (Fig.~\ref{fig3:NegFields_ViscousFlow}, section~\ref{Sec:EField_SDILD}) in $\widehat{E}_{x}^{\rm meas}(x,y)$ (Fig.~\ref{fig:S12c}c) at the positions of the large polymer structures, respectively, around the positions of $E_{\rm F}\approx E_{\rm D}$. However, in the upper left of Fig.~\ref{fig:S12c}c, a larger patch of inverted field (red area marked by black arrow) appears that is not correlated to features in the topography with heights larger than 2\,nm or to features in the surface potential map. Thus, the presence of the dirt clusters does not prohibit the appearance of current-induced inverted fields close to charge neutrality within its surrounding.

Figure~\ref{fig:S12c}d shows an area where the largest polymer cluster is smaller than $\sim 3$\,nm and the surface roughness is reduced to $0.35$\,nm. Here, the fingerprints of the remaining dirt in the surface potential map (Fig.~\ref{fig:S12c}e) are negligible and again patches of inverted field not correlated with the positions of dirt are apparent in the $\widehat{E}_{x}^{\rm meas}(x,y)$ map (yellow arrow in Fig.~\ref{fig:S12c}f). To exclude the influence of the polymer residues with smaller height completely, we swept a larger area ($10\times 10$\,$\mu$m$^2$) in contact mode AFM (section~\ref{sec:sweep}) removing the polymers that afterwards appear as ridges surrounding the swept area \cite{Goossens2012,Schweizer2020,Lindvall2012}. An inner part of the swept area is shown in Fig.~\ref{fig:S12c}g-i. It barely exhibits residues on the surface and a rms roughness of $0.35$\,nm, probably dominated by roughness of the substrate. None of the residues has a height large than 2\,nm. Nevertheless, the area exhibits a very similar potential fluctuation (Fig.~\ref{fig:S12c}h) as with remaining minor residues on the surface (Fig.~\ref{fig:S12c}e) and areas of inverted field in the $\widehat{E}_{x}^{\rm meas}(x,y)$ map (Fig.~\ref{fig:S12c}i). This swept area of graphene has also been used to demonstrate the removal of current-induced inverted field areas by ion bombardment (Fig.~\ref{fig5:ion_bombardment}, main text). 

Thus, we conclude that the presence of polymers on the surface does not prohibit the observation of current induced potential maps that show fingerprints of hydrodynamic electron flow. Larger clusters of polymers imply a relatively strong local doping that causes pronounced SDILD effects overlapping with current induced features, but are avoided by preselecting adequate areas as for the images discussed in the main text.    

Table~\ref{table_Residues_Doping} summarizes the roughnesses and the relative areas covered with clusters larger than 2\,nm  for all figures evaluated in the main text.
The rms roughness of the whole area is dubbed $\delta z$ and the percentage covered with clusters larger than 2\,nm is dubbed $\Delta A$. Additionally, we show the same numbers for the areas exhibiting inverted electric field. Therefore, we firstly applied a stronger smoothing, for obtaining $\widehat{E}_{x}^{\rm meas}(x,y)$, than described in section~\ref{Efield_noise_filter} by using $\Delta x =1.5\,\mu$m (eq.~\ref{eq:Ex_VtrDer_filter})
in order to remove all inverted fields caused by noise, cut out the remaining areas of inverted $\widehat{E}_{x}^{\rm meas}(x,y)$
and evaluated its rms roughness and its areas with clusters larger than 2\,nm. It appears that the areas of inverted field barely show larger clusters and, moreover, less clusters and roughness than observed in general on the surface. 

\begin{table}[h!]
	\centering
	\begin{tabular}{|c|c|c|c|c|} 
		\hline
		Figure & $\delta z$ (nm) & $\Delta A$ (\%) & $\delta z_{\rm inv}$ (nm)  & $\Delta A_{\rm inv}$ (\%) \\
		\hline
		\ref{fig1:ExptProcedure}f,g & 1.6 & 10.4 & 0.97 & 5.6\\
		\ref{fig2:SDind_doping}d-i & 1.9 & 0.6 & 1.7 & 0.25\\
		\ref{fig3:NegFields_ViscousFlow}a-f & 0.3 & 0.14 & 0.14 & 0\\
		\ref{fig4:Relevance_Viscous}a-f & 1.2 & 14 & 0.9 & 4.2 \\
		\ref{fig5:ion_bombardment}a-d & 0.35 & 0.5 & 0.35 & 0.26 \\
		\hline
	\end{tabular}
	\caption{Contamination by polymers for the figures in the main text: $\delta z$ describes the rms roughness of the total area of the image. $\Delta A$ describes the percentage of area that is covered with clusters larger than 2\,nm in height. The same numbers are given for the preselected areas with inverted current-induced electric fields dubbed $\delta z_{\rm inv}$ and $\Delta A_{\rm inv}$.}.
	\label{table_Residues_Doping}
\end{table}

\section{Influence of Penetration Fields and Compressibility}
\label{sec:compressibility}
Generally, graphene is not a perfect metal such that electric fields penetrate the material as captured in first order by the quantum capacitance model.\cite{Xia2009} Moreover, the compressibility can be influenced by electron-electron interactions, in particular, at low charge carrier densities.\cite{Elias2011,Li2009b,Siegel2011} In other materials such as GaAs, this leads to negative compressibility,\cite{Eisenstein1992} as not expected for graphene due to its linear dispersion.\cite{Li2011,Sheehy2007} Indeed, so far negative compressibilities have not been reported for monolayer graphene \cite{Martin2007} except if  combined with other conducting 2D materials\cite{Larentis2014}.

\begin{figure}
	\centering
	\includegraphics[width = 120 mm]{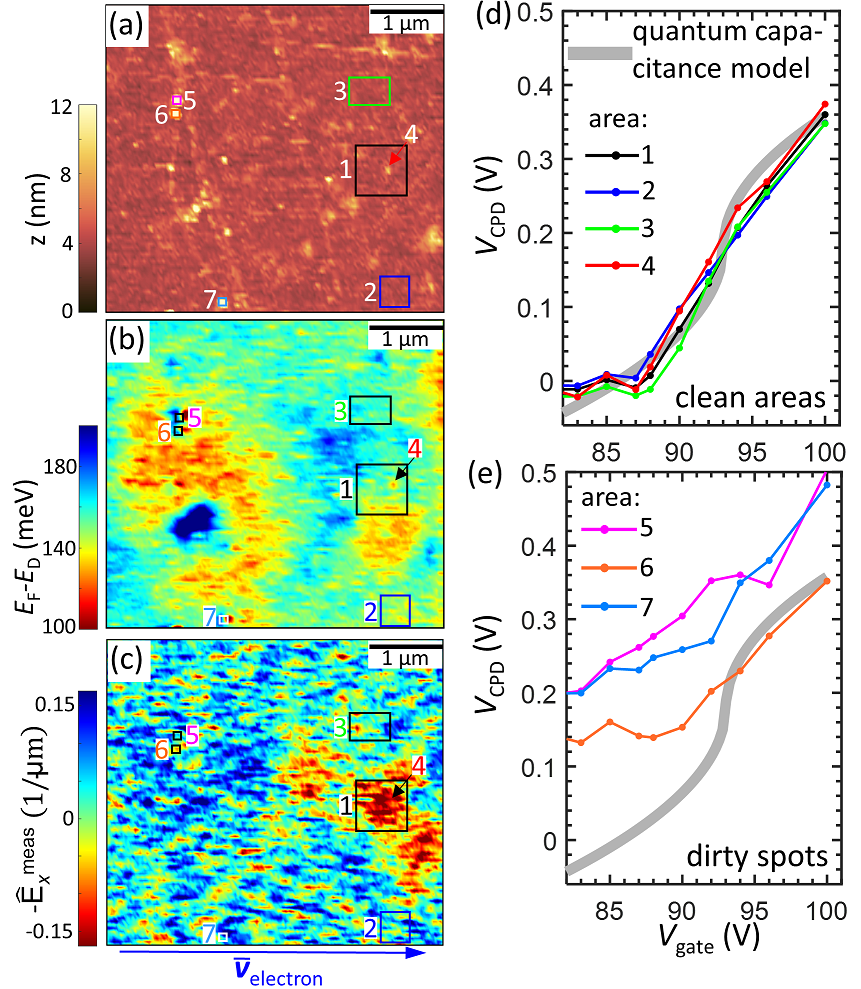}
	\caption{{\bf Applicability of the Quantum Capacitance Model.} (a) Topgraphy of graphene recorded by tapping mode AFM. (b) $V_{\rm CPD}(x,y)$ map of the same area displayed as $(E_{\rm F}-E_{\rm D})(x,y)$, $V_{\rm Gate}=86$\,V. (c) $\widehat{E}_{x}^{\rm meas}(x,y,V_{\rm SD}= 0.4\,$V) map of the same area as a--b revealing inverted electric fields (red). Marked rectangles in a--c refer to the positions where curves in d--e are recorded. (d) Colored dots connected by lines: $V_{\rm CPD}(V_{\rm gate})$ probed by EFM and averaged across the areas marked in a--c in comparison with the result from the quantum capacitance model (eq.~\ref{eq:QCM}) using $V_{\rm CPD}^0=0.18$\,V and $V_{\rm D}=93$\,V as fit parameters (grey line). Only areas of graphene, where contaminations have less height than 5\,nm are considered showing only minor deviations from the quantum capacitance model. Area 4 is recorded on a small topographic bump in the region of inverted electric field. It exhibits slightly reduced $E_{\rm F}-E_{\rm D}$ with respect to its surroundings (b), but no feature in the $\widehat{E}_{x}^{\rm meas}(x,y)$ map (c). (e) Same as d, but measured at spots of larger clusters (height $> 8$\,nm). A strong deviation from the quantum capacitance model appears including negative $V_{\rm CPD}(V_{\rm gate})$ slopes.}
	\label{fig:S12b}
\end{figure}

However, since deviations of compressibility from the quantum capacitance model could vary locally, it is important to rule out that the observed inverted electric fields are caused by such artifacts.
Therefore, we recorded maps of $V_{\rm CPD}(V_{\rm gate})$ curves by EFM in an area covering inverted electric fields as displayed in Fig.~\ref{fig:S12b}. We display the $V_{\rm CPD}(V_{\rm gate})$ data separately for more clean graphene areas with small clusters of residues only (Fig.~\ref{fig:S12b}d) and areas with polymer clusters of large heights $> 8$\,nm (Fig.~\ref{fig:S12b}e) (Section~\ref{sec:prep}).
It is obvious that the more clean areas do not exhibit negative slopes beyond error bars that would imply negative compressibility. This holds true for all graphene areas with corrugation below 5\,nm, but is often different on top of larger contaminations that must, hence, be excluded from the analysis of negative electric fields.

On the more clean areas, we compare the experimental data with the quantum capacitance model reading:
\begin{equation}
\label{eq:QCM}
V_{\rm CPD} (V_{\rm gate}) = \frac{\hbar v_{\rm F}}{e}\cdot {\rm sign}(V_{\rm gate}-V_{\rm D})\cdot \sqrt{\frac{\pi C_{\rm eff}|V_{\rm gate}-V_{\rm D}|}{e}}+V_{\rm CPD}^0   
\end{equation}
with effective capacitance per area $C_{\rm eff}$ calculated by
\begin{equation}
\frac{1}{C_{\rm eff}}=\frac{1}{C_{\rm gate}}+\frac{1}{C_{\rm Q}}    
\end{equation}
using geometric capacitance $C_{\rm gate}=\frac{\epsilon \epsilon_0}{d_{\rm SiN}}$ ($d_{\rm SiN}=151$\,nm,  $\epsilon = 7.6$, section~\ref{sec:prep}) and quantum capacitance $C_{\rm Q}$ as determined iteratively (section~\ref{subsec_SDILD}).
The result is displayed as a grey line in Fig.~\ref{fig:S12b}d matching the experimental curves on the more clean graphene areas satisfactorily.
The experimental curves are more straight than the model calculation. This is expected due to the influence of disorder, respectively puddles, that implies a cut-off density  prohibiting steeper slopes of $V_{\rm CPD}(V_{\rm gate})$.\cite{Li2013b,DasSarma2011,Hu2008} The offsets between different experimental $V_{\rm CPD}(V_{\rm gate})$ curves largely match the differences in $E_{\rm F}-E_{\rm D}$
(Fig.~\ref{fig:S12b}b). For two selected curves, the comparative variation of $V_{\rm CPD}$ distance along $V_{\rm gate}$ is about $20$\,mV, such that penetration voltages not covered by the quantum capacitance model are on the order of only  10\,\% of $V_{\rm CPD}$ within the $V_{\rm gate}$ regime of inverted electric fields (e.g. Fig.~\ref{fig:S12b}b).
The slope variations between different $V_{\rm CPD}(V_{\rm gate})$ curves at the same $V_{\rm gate}$ are about 50 \%. This is the relevant number for describing possible spatial variations of the screening properties that are relevant during application of $V_{\rm SD}$. Only these deviations could imply electric fields due to inhomogeneities in the compressibility at finite $V_{\rm SD}$ that are not captured by the SDILD model and are not caused by the current flow (Section~\ref{Sec:EField_SDILD}). They are compared with the observed inverted fields in the following.  

In the investigated area of Fig.~\ref{fig:S12b}, located at the center of the graphene device, the applied $V_{\rm SD}= 400$\,mV  changes the effective gate voltage by $\Delta V_{\rm gate}^{\rm SD} \approx V_{\rm SD}/2 =200$\,mV such that the average shift of $V_{\rm CPD}$ by $V_{\rm SD}$ amounts to  
$\Delta V_{\rm CPD}^{\rm SD} \approx \Delta V_{\rm gate}^{\rm SD}\cdot \overline{\frac{dV_{\rm CPD}}{dV_{\rm gate}}}\approx 8$\,mV only, extracting $\overline{\frac{dV_{\rm CPD}}{dV_{\rm gate}}}$ from Fig.~\ref{fig:S12b}d straightforwardly. The spatial variations of $\Delta V_{\rm CPD}^{\rm SD}$ via the local slope differences are about 50 \% of this value, i.e 4 mV.  
This must be compared with the observed inverted electric fields in Fig.~\ref{fig:S12b}c that amount up to $|E_{x}^{\rm meas}(x,y)|\approx 60$\,mV/$\mu$m (Fig.~\ref{fig:S12b}c). Using the selected point distance $\Delta x=0.39\,{\rm \mu}$m for determination of $E_{x}^{\rm meas}(x,y)$ (table~\ref{table1_Filter_Pmts}), the respective $V_{\rm CPD} (x,y)$ differences are $-24$\,mV
as crosschecked directly from the recorded $V_{\rm CPD} (x,y)$ maps. This is a factor of six larger than the observed slope variations. Moreover, the variations have to oppose the applied $V_{\rm SD}$ making the discrepancy even larger. Correspondingly strong spatial variations in slope of $V_{\rm CPD}(V_{\rm gate})$  by 300-350\,\% are not found experimentally for the equilibrium $V_{\rm CPD}(V_{\rm gate})$ curves, at $V_{\rm SD}=0$\,V in the regimes that are analyzed in the main text. Hence, we can safely exclude that the observed inverted fields are caused by spatial variatons of the screening properties of graphene or by deviations from the quantum capacitance model.

\section{Estimates of Local Scattering Lengths}
\label{sec:ScatteringLength}

\subsection{Electron Electron Scattering Length $l_{\rm ee}$}

The electron-electron scattering length $l_{\rm ee}$ is associated with the quasiparticle decay rate $\tau_{\rm ee}^{-1}$ due to inelastic electron electron scattering via $l_{\rm ee} = v_{\rm F} \tau_{\rm ee}$. This is directly related to the imaginary part of the retarded self energy $\Sigma_{s}^{\rm R}(\mathbf{k},\xi_{\rm \mathbf{k},s})$ of monolayer graphene as \cite{Hwang2007}

\begin{equation}
\frac{1}{\tau_{\rm ee}} = 2\,{\rm Im} \left[\Sigma_{s}^{\rm R}(\mathbf{k},\xi_{\rm \mathbf{k},s}), \right]
\label{eq:eeScatter_time_selfenergy}
\end{equation}
\noindent

where $s = +/-$ is the conduction/valence band index and $\xi_{\rm \mathbf{k},s} = s \hbar v_{\rm F} |\mathbf{k}| - \mu$ is the single particle band energy relative to the non-interacting chemical potential $\mu$ at finite temperature. Since graphene on SiN is a weakly correlated material (interaction parameter $r_{\rm s} = 0.51$)\cite{DasSarma2011}, the $GW$ approximation is adequate \cite{Giuliani_Vignale_2005} leading to \cite{DasSarma2007}

\begin{align}
{\rm Im}[\Sigma_{s}^{\rm R}(k,\omega)] = \sum_{\mathbf{q}, s'=\pm}  V_q \left[n_{\rm B} \left( \xi_{\mathbf{k+q}, \,s'} - \hbar \omega \right)+ n_{\rm F}\left( \xi_{\mathbf{k+q}, \,s'} - \hbar \omega \right) \right] \nonumber \\ 
\times \left(1+ss' \cos{\theta} \right) {\rm Im} \left[\frac{1}{\epsilon \left(q, \xi_{\mathbf{k+q}, \,s'}/\hbar  - \omega \right)} \right],
\label{eq:SelfEnergy_GW}
\end{align}

\noindent where $V_q = \frac{2\pi e^2}{\kappa q}$ is the Coulomb interaction for momentum transfer $q$, $\kappa=4.25$ is the background dielectric constant combining SiN and vacuum, $n_{\rm B}$ and $n_{\rm F}$ are the Bose and Fermi distribution functions, respectively, and $\theta$ is the angle between electron wave vectors $\mathbf{k}$ and $\mathbf{k+q}$. The finite temperature dynamic dielectric function  $\epsilon(q,\omega)$ reads within random phase approximation (RPA) $\epsilon(q,\omega) = 1 + V_q \Pi (q,\omega)$, where $\Pi(q,\omega)$ is the irreducible polarizability.

In the limit of small $\xi_{\rm k}$ and low temperature, $\xi_{\mathbf{k}} \ll k_{\rm B}T \ll E_{\rm F}$, where $E_{\rm F}=\hbar v_{\rm F}\cdot {\rm sign}(n) \sqrt{\pi |n|}$ is the Fermi energy for charge carrier density $n$, an asymptotic form of eq.~(\ref{eq:SelfEnergy_GW}) has been derived by Li et al. \cite{Li2013} (eq.~(10), there), that by substituting into eq.~(\ref{eq:eeScatter_time_selfenergy}) results in an electron-electron scattering length $l_{\rm ee}$ \cite{Polini2016_NoNonsensePhysicist, Kim2020, Kumar2017}:

\begin{equation}
l_{\rm ee} = \frac{4}{\pi} \left( \frac{\hbar v_{\rm F}}{k_{\rm B} T} \right)^2 \sqrt{\pi |n|} \frac{1}{\ln{\left( \frac{2E_{\rm F}}{k_{\rm B} T} \right)}}
\label{eq:lee}
\end{equation}
as also given as eq.~(\ref{eq:4}), main text.
For the gate voltages of Fig.~\ref{fig4:Relevance_Viscous}g, main text, where $l_{\rm ee} (V_{\rm gate})$ is plotted for $T=298\,$K, we have $k_{\rm B} T/E_{\rm F} \in [0.08 - 0.24]$, except at charge neutrality,   such that the required limit of eq.~(\ref{eq:lee}) applies.\cite{Li2013}.

The formalism is, moreover,  not applicable in the quantum critical regime that appears at room temperature for carrier concentrations $n<2 \times 10^{14}\,{\rm m^{-2}}$ \cite{Sheehy2007}, far lower than $n$ for all data points in Fig.~\ref{fig4:Relevance_Viscous}g, main text, and Fig.~\ref{fig:Relevance_Hydrodynamics_II}e, except at charge neutrality. But even at charge neutrality, the lateral charge fluctuations $\Delta n_{\rm 0} \approx 1 \times 10^{15}\,{\rm m^{-2}}$ (puddles) are one order of magnitude larger than the threshold for quantum criticality. Finally, the threshold carrier concentration $n_{\rm Th}$ above which eq.~(\ref{eq:lee}) is valid ( $E_{\rm F}(n_{\rm Th})= k_{\rm B} T$) is  $n_{\rm Th}\approx 4 \times 10^{14}\,{\rm m^{-2}}$, i.e., larger than the quantum critical threshold, but lower than $\Delta n_{\rm 0}$.

\subsection{Electron Disorder Scattering Length $l_{\rm dis}$}
\label{subsec_disorderLength}

\begin{figure}
	\centering
	\includegraphics[width = 160 mm]{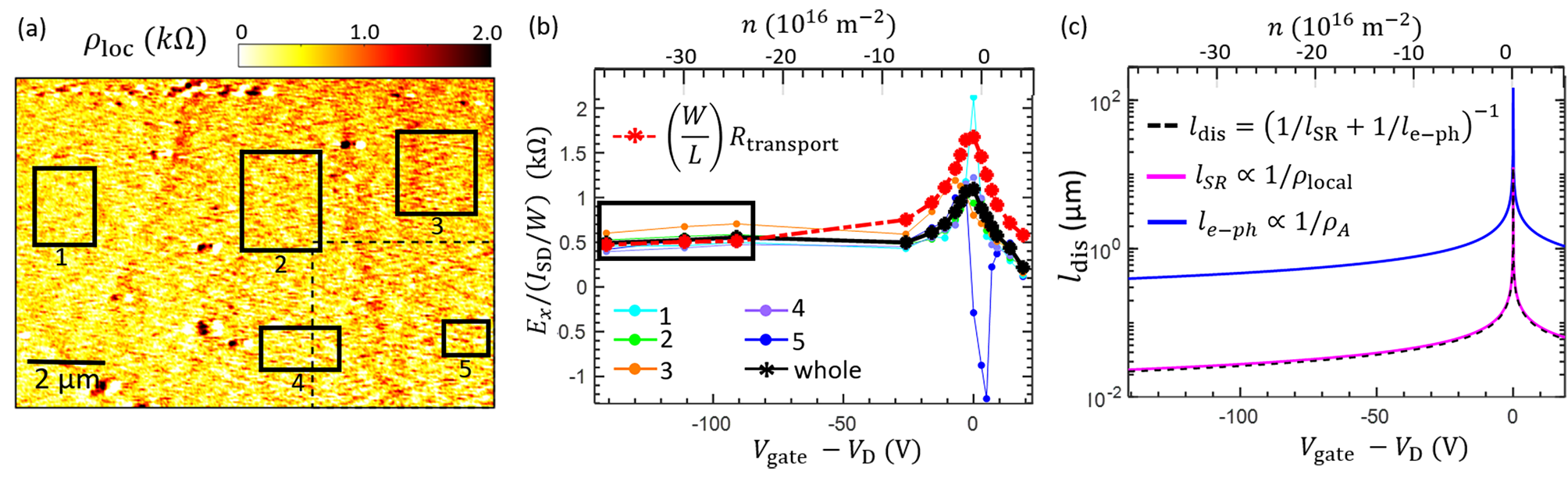}
	\caption{{\bf Disorder Scattering Length.} (a) Map of local resistivity $\rho_{\rm loc}(x,y) = \left<E_{ x}^{\rm meas}(x,y,V_{\rm SD} = 0.4\,{\rm V}, V_{\rm gate})\right>_{V_{\rm gate}}/(I_{\rm SD}/W)$  with $E_{x}^{\rm meas}(x,y)$ averaged across $V_{\rm gate} = -141 - -91\,$V (rectangle in b), $I_{\rm SD}$: source-drain current during imaging, $W$: sample width. Corresponding topography: Fig.~\ref{fig5:Gallery_FieldInversion}h. Black rectangles mark the spatial averaging areas for the data in b. (b) Spatially averaged $\overline{\rho}_{\rm loc}(V_{\rm gate}) = \left<E_{x}^{\rm meas}(x,y,V_{\rm SD} = 0.4 \,{\rm V}, V_{\rm gate})\right>_{x,y}/(I_{\rm SD}/W)$. Full colored lines: areas marked in a, black line: whole region of a, red dashed line: two-terminal device resistivity, $L$: sample length. (c) Simulated $V_{\rm gate}$ dependence (eq.~\ref{eq:Graphene_Conductivity}) of the electron-phonon scattering length $l_{\rm e-ph}$ due to gate independent resistivity $\rho_{\rm A} = 30\,{\Omega}$ from longitudinal accoustic phonon scattering\cite{Chen2008, Li2013} (blue), the complete electron disorder scattering length $l_{\rm dis}$  deduced from the measured $\rho_{\rm local} = 500\,{\Omega}$ at large $|V_{\rm gate}|$ (dashed black), and the deduced
		short range disorder scattering length $l_{\rm SR}$ (magenta) employing the Matthiesen rule shown as inset. The upper electron density axis ($n$) in b and c is deduced from a capacitive model.} 
	\label{fig:rho_loc}
\end{figure}

All scattering processes  of electrons with external perturbations as point defects, Coulomb type impurities, sample edges, or phonons  change the momentum of the electron system. We dub the respective scattering length $l_{\rm dis}$ (disorder scattering length). A straightforward approach to determine $l_{\rm dis}$ uses the resistivity of graphene at high carrier densities where electron-electron scattering is negligible (eq.~(\ref{eq:lee})), such that the effective mean free path reading $l_{\rm MFP}=\left(l_{\rm ee}^{-1} + l_{\rm dis}^{-1}\right)^{-1}$ is dominated by $l_{\rm dis}$. 

Although we are not able to directly map the resistivity with KPFM or EFM, the recorded electric field maps at high doping give a reasonable estimate assuming homogeneous current density
$I_{\rm SD}/W$ ($I_{\rm SD}$: source-drain current, $W$: sample width) in first order, since non-trivial viscous properties disappear at large $|V_{\rm gate}|$ (Fig.~\ref{fig5:Gallery_FieldInversion}a, section~\ref{sec:gallery}, Fig.~\ref{fig4:Relevance_Viscous}, main text). This allows to approximately map local variations in resistivity $\rho_{\rm local}(x,y) = \frac{\left<E_{ x}^{\rm meas}(x,y, V_{\rm gate})\right>_{V_{\rm gate}}}{I_{\rm SD}/W}$ by averaging $E_{x}^{\rm meas}(x,y)$ across $V_{\rm gate}=-141--91$\,V for each position as displayed in Fig.~\ref{fig:rho_loc}a. Averaging instead across a spatial area results in $\overline{\rho}_{\rm local}(V_{\rm gate}) =\frac{\left<E_{ x}^{\rm meas}(x,y, V_{\rm gate})\right>_{x,y}}{I_{\rm SD}/W}$ as shown in Fig.~\ref{fig:rho_loc}b for a few areas and compared to the measured two probe resistivity of the device. While  differences appear around charge neutrality, all curves display a largely gate independent similar resistivity for $n\le -2\times 10^{17}\,{\rm m^{-2}}$. 

Independence of resistivity from $V_{\rm gate}$, respectively from carrier concentration $n$, is indicative of scattering dominated by short range disorder, while $\rho \propto n^{-1}$ indicates dominant long range disorder scattering \cite{DasSarma2011}. Therefore, we deduce that the former dominates at large hole density. Electron-phonon scattering at longitudinal accoustic phonons results as well in $n$-independent resistivity \cite{Li2013}, but can be estimated as $\rho_{\rm A} \approx 30\,{\rm \Omega}$ \cite{Chen2008, Li2013}, i.e., much lower than the measured resistivity of $\rho=500\,\Omega$. Finally, scattering at remote interfacial phonons (polar optical phonons of the substrate) can play a role for graphene at room temperature \cite{DasSarma2011, Chen2008} exhibiting a resistivity contribution $\rho_{\rm remote} \propto \frac{\hbar\omega_{\rm s}}{e^{\hbar\omega_{\rm s}/k_{\rm B}T} -1}$, \cite{Fratini2008} where $\omega_{\rm s}$ is the frequency of the remote surface polar optical phonon being $110\,{\rm meV}$ for ${\rm Si_3N_4}$ \cite{Zhu2010} ($59\,{\rm meV}$ for ${\rm SiO_2}$) \cite{Chen2008, Fratini2008}. Since the remote phonon contribution on ${\rm SiO_2}$ has been measured as $\rho_{\rm remote} \approx 26\,{\rm \Omega}$,\cite{Chen2008} we get for ${\rm Si_3N_4}$: $\rho_{\rm remote} \approx 26\,{\rm \Omega} \left(\frac{110\,{\rm meV}}{59\,{\rm meV}}\right)  \frac{e^{59\, {\rm meV}/k_{\rm B}T} -1}{e^{110\, {\rm meV}/k_{\rm B}T} -1} = 5 \,{\Omega} \ll \rho_{\rm A } \ll \rho_{\rm loc}$. Thus, the remote phonon contribution can be neglected as well. This implies a dominating scattering at short range defects for large hole doping with scattering length $l_{\rm SR}\approx l_{\rm dis}$.

To calculate the gate dependence of the resulting $l_{\rm dis}(E_{\rm F})$, we use the semiclassical Boltzmann transport equation for the scattering time $\tau(E_{\rm F}) = l_{\rm dis}(E_{\rm F})/v_{\rm F}$ reading\cite{DasSarma2011}

\begin{equation}
\frac{1}{\rho} = \frac{e^2 v_{\rm F}^2}{2} D(E_{\rm F}) \tau (E_{\rm F})= \frac{e^2 v_{\rm F}}{2} D(E_{\rm F}) l_{\rm dis} (E_{\rm F})
\label{eq:Graphene_Conductivity}
\end{equation}

\noindent
with density of states of graphene $D(E_{\rm F}) = 4\cdot\frac{ \sqrt{\pi |n(E_{\rm F})|}}{h v_{\rm F}} $. Substituting $D(E_{\rm F})$ into eq.~(\ref{eq:Graphene_Conductivity}) and solving for $l_{\rm dis}$, we obtain eq.~(\ref{eq:5}), main text, when using $\rho = \rho_{\rm local}(x,y)$ in order to maintain the spatial character of $l_{\rm dis}(x,y)$ and $n(x,y)$.  Figure~\ref{fig:rho_loc}c displays the resulting $l_{\rm dis}(n)$, respectively $l_{\rm dis}(V_{\rm gate})$ using a capacitive model  together with its contributions $l_{\rm SR}$ and the electron-phonon contribution $l_{\rm e-ph}\propto 1/\rho_{\rm A}$ as discussed above. Importantly, $l_{\rm dis}$ increases towards charge neutrality ($V_{\rm D}$), oppositely to the behaviour of $l_{\rm ee}$ (eq.~(\ref{eq:lee})),  such that dominating electron-electron scattering naturally results at low doping. 

\subsection{Comparing Local Scattering Lengths}
\label{subsec_CompLocalLengths}

\begin{figure}
	\centering
	\includegraphics[width = 165 mm]{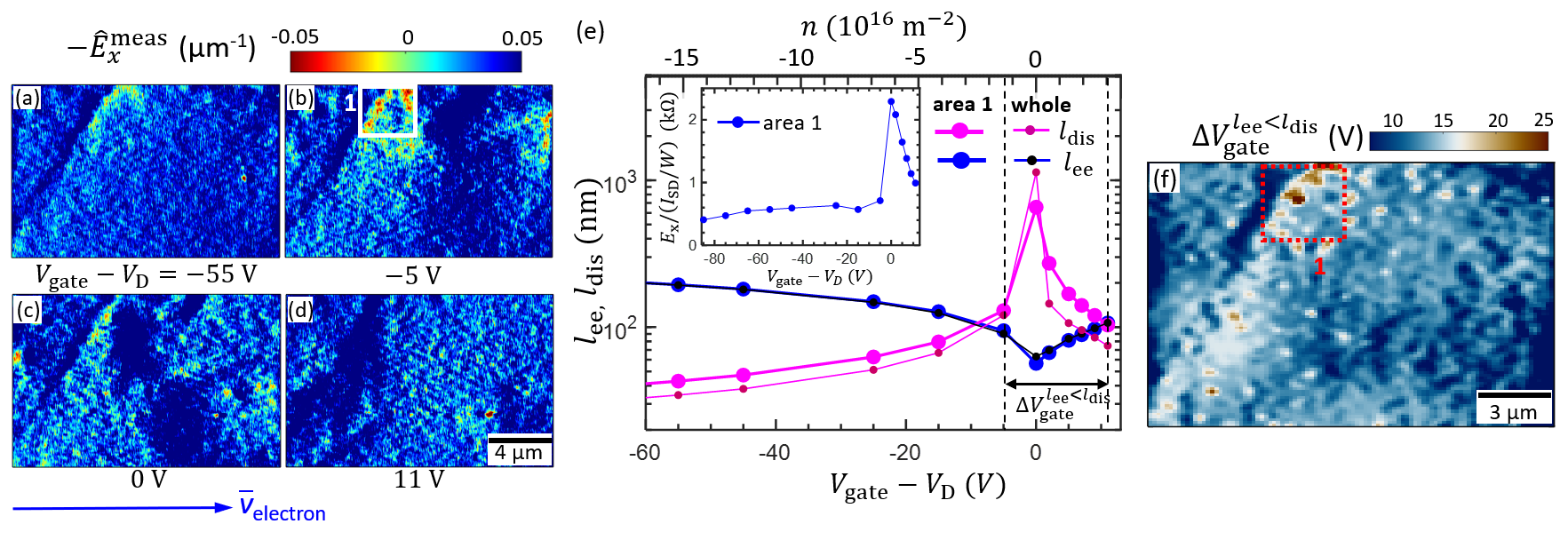}
	\caption{{\bf Comparison of Different Scattering Lengths.} (a--d) $\widehat{E}_{x}^{\rm meas}(x,y,V_{\rm SD}= 0.5\,$V) at the marked gate voltages, $V_{\rm D} = 85\,$V. Corresponding topography: Fig.\ref{fig1:ExptProcedure}f, main text. The white rectangle highlights an area with patches of inverted electric fields. (e) Dependence on $V_{\rm gate}$ (bottom axis)  and carrier concentration (top axis) of electron-electron scattering length $l_{\rm ee}$ (dark lines) and electron-disorder scattering length $l_{\rm dis}$ (pink lines) for the marked area in b (area 1, thick symbols) and the whole area of a--d (thin symbols). Dashed lines with distance $\Delta V_{\rm gate}^{l_{\rm ee}<l_{\rm dis}}$ mark the range where $l_{\rm ee}< l_{\rm dis}$ for the thick symbols. Inset: $V_{\rm gate}$ dependence of $\overline{\rho}_{\rm loc}$ for area 1. (f) Map of $\Delta V_{\rm gate}^{l_{\rm ee}<l_{\rm dis}}(x,y)$ for the same area as displayed in a--d using a spatial average of $0.2\,\mu{\rm m} \times 0.2\,\mu{\rm m}$ for each point $(x,y)$. The rectangle marks the same area as in b.}
	\label{fig:Relevance_Hydrodynamics_II}
\end{figure}

The ability to map, within the same area, the doping distribution $n_0(x,y)$ and the transport induced electric fields $E_x^{\rm meas}(x,y)$, leading to $\rho_{\rm local}(x,y)$ at high hole doping, allows us to derive corresponding maps of electron-electron scattering length $l_{\rm ee}(x,y)$ and electron-disorder scattering length $l_{\rm dis}(x,y)$ via eq.~(\ref{eq:4}) and eq.~(\ref{eq:5}), main text, respectively. Hence, we are able to locally compare these length scales as shown in Fig.~\ref{fig4:Relevance_Viscous}g, main text, revealing that inverted electric fields appear preferentially in areas where the $V_{\rm gate}$ range with $l_{\rm ee}< l_{\rm dis}$ is largest. In the particular case of  Fig.~\ref{fig4:Relevance_Viscous}, the inverted electric fields appeared on the electron side close to charge neutrality. Here, we add an example, where the inverted fields are observed on the hole side close to charge neutrality (Fig.~\ref{fig:Relevance_Hydrodynamics_II}a--d), in particular, at the right of the fold penetrating the whole image (topography:  Fig.~\ref{fig1:ExptProcedure}f, main text). This area also shows lowest $-E_x^{\rm meas}(x,y)$ far away from charge neutrality (Fig.~\ref{fig:Relevance_Hydrodynamics_II}a) implying low $\rho_{\rm loc}(x,y)$ and, thus, large $l_{\rm dis} (x,y)$ (eq.~(\ref{eq:Graphene_Conductivity})), hence, favoring electron viscosity around charge neutrality. 
In line, again the areas with largest $V_{\rm gate}$ range, where $l_{\rm ee}$ dominates  with respect to $l_{\rm dis}$ (Fig.~\ref{fig:Relevance_Hydrodynamics_II}e), showcase inverted electric fields as more directly corroborated by mapping the $V_{\rm gate}$ range with $l_{\rm ee}< l_{\rm dis}$
called $\Delta V_{\rm gate}^{l_{\rm ee}<l_{\rm dis}}(x,y)$ (Fig.~\ref{fig:Relevance_Hydrodynamics_II}f) and comparing it with the $E_x^{\rm meas}(x,y)$ maps of the identical region (Fig.~\ref{fig:Relevance_Hydrodynamics_II}a--d).

The inset in Fig.~\ref{fig:Relevance_Hydrodynamics_II}e displays $\overline{\rho}_{\rm loc}(V_{\rm gate})$ for the highlighted area 1 exhibiting a wide $V_{\rm gate}$ range with constant $\overline{\rho}_{\rm loc}$. This confirms our previous conclusion of dominating short range disorder scattering for $l_{\rm dis}$, particularly for the area with inverted electric fields.

\section{Origin of Inverted Electric Fields}
\label{sec:origin}
After relating the observed inverted electric fields to a locally increased ratio $l_{\rm dis}/l_{\rm ee}$, we compare the $\widehat{E}_{x}^{\rm meas}(x,y)$ maps with other local properties in more detail.
Generally, for two dimensions, electric fields resulting from viscous flow can be linked to the vorticity  $\omega(x,y) =\nabla \times  \mathbf{v}(x,y)$ of the velocity field $\mathbf{v}(x,y)$ of the charge carriers via\cite{Falkovich2017} 
\begin{equation}
\label{eq:S27}
\frac{d\omega(x,y)}{dx}=\frac{en(x,y)}{\eta(x,y)}E_y(x,y), \hspace{2mm} \frac{d\omega(x,y)}{dy}=-\frac{en(x,y)}{\eta(x,y)}E_x(x,y),   
\end{equation}
where $\eta(x,y)$ is the shear viscosity.
Such vorticity is established by a gradient in velocity perpendicular to the flow direction as naturally appearing at obstacles. Consequently, a transversal curvature of the velocity profile implies electric fields along the current flow as familiar from Poiseuille flow profiles.\cite{Falkovich2017}  
In turn, a positive curvature of flow would imply an inverted electric field. Thus, a quiet area in flow ($\mathbf{v}(x,y)=\mathbf{0}$\,m/s ) caused by an obstacle that is surrounded by laminar flow prone to shear viscosity naturally leads to inverted electric fields.
More intuitively,  inhomogeneous disorder scattering can lead to varying charge carrier velocities that develop viscous flow patterns on the scale of the slip length ($\approx l_{\rm ee}$)\cite{Kiselev2019b}. This eventually exposes relatively quiet areas of low velocity to sucking of their charge density by the strong neighboring currents such that the electric field gets locally inverted by charge rearrangements opposite to the global flow.  

To estimate the length scales of the resulting velocity profiles, we use the result for the kinematic viscosity $\nu$ of graphene\cite{Guo2017,Kumar2017,keser2021}
\begin{equation}
\nu \approx \frac{v_{\rm F}l_{\rm ee}}{4}    
\end{equation}
that describes viscous flow via\cite{Guo2017,Kumar2017} 
\begin{equation}
\frac{e\mathbf{E}(\mathbf{x},t)}{m^*} = \nu \nabla^2 \mathbf{v}(\mathbf{x},t). 
\end{equation}
Here, we neglect the influence of thermodynamic pressure and momentum relaxation scattering for the sake of simplicity. Using the strength of the inverted electric fields from the KPFM and EFM measurements, $E_x^{\rm meas}\approx 0.01$\,V/$\mu$m, and $m^* =\hbar \sqrt{\pi \cdot n}/v_{\rm F}$,\cite{Ariel2013} we get, e.g., for $n\approx 2\cdot 10^{16}/$\,m$^2$ and $l_{\rm ee}\approx 100$\,nm at  $|V_{\rm gate}-V_{\rm D}|=10$\,V (Fig.~\ref{fig4:Relevance_Viscous}g, main text) : 
%
\begin{equation}
\nabla^2 v_x (\mathbf{x})=\frac{e}{\hbar}\cdot\frac{4 E_x^{\rm meas}(\mathbf{x})}{l_{\rm ee}(\mathbf{x})\sqrt{\pi \cdot n(\mathbf{x})}} \approx 2\cdot 10^{18}/{\rm ms}.
\end{equation}
The average charge carrier velocity $\overline{v}_x =V_{\rm SD}/(RWn_0e)$ along $x$ direction amounts to $\overline{v}_x=6\cdot 10^3$\,m/s using $V_{\rm SD}=500$\,mV, sample width $W=26$\,$\mu$m (section~\ref{sec:prep}), and resistance $R=1$\,k$\Omega$ for $|V_{\rm gate}-V_{\rm D}|=10$\,V (Fig.~\ref{fig4:Relevance_Viscous}g, main text). Assuming a parabolic profile $v_x(y)$ connecting an area exhibiting $\overline{v}_x$ with an area of $v_x =0$\,m/s,  we find a length scale $\xi$:
\begin{equation}
\xi=\sqrt{\frac{2\cdot\overline{v}_x}{\nabla^2 v_x(\mathbf{x})}}\approx 80\,{\rm nm}
\end{equation}
This reasonably fits with the lengths scales of observed patches of inverted electric fields that are in the few hundred nm regime.
\begin{figure}
	\centering
	\includegraphics[width = 165 mm]{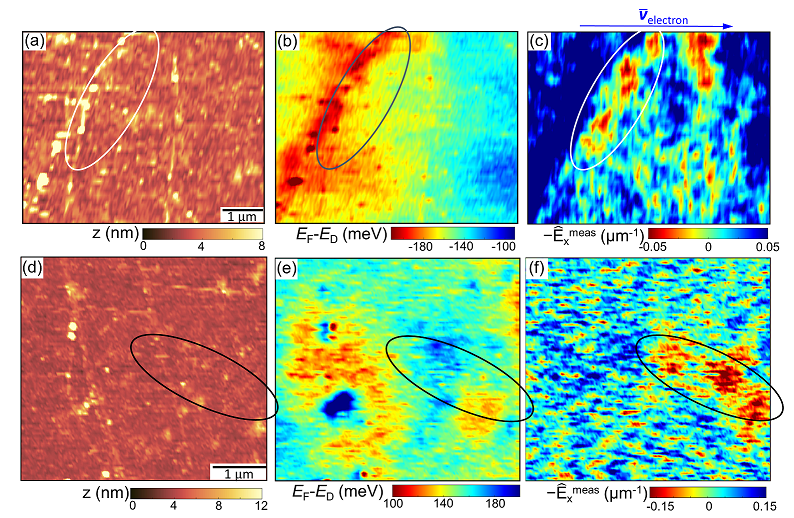}
	\caption{{\bf Comparison of Topography, Electrostatic Potential, and Current Induced Electric Field.} (a) Topgraphy of graphene recorded by tapping mode AFM. (b) $V_{\rm CPD}(x,y)$ map of the same area displayed as $(E_{\rm F}-E_{\rm D})(x,y)$, $V_{\rm gate}=80$\,V, $V_{\rm D}=85$\,V. (c) $\widehat{E}_{x}^{\rm meas}(x,y)$ map of the same area as a--b, $V_{\rm SD}= 0.5$\,V (same as Fig.~\ref{fig1:ExptProcedure}g). (d)--(f) same as a--c, but for electron doping in another area, 
		$V_{\rm gate}=86$\,V, $V_{\rm D}=81$\,V, $V_{\rm SD}=0.4$\,V. The main direction of electron flow $\overline{\mathbf{v}}_{\rm electron}$ is marked on top of c. The identical area exhibiting inverted electric fields (red in c, f) is encircled in a-c and d-f, respectively. 
	}
	\label{fig:S15}
\end{figure}

Figure~\ref{fig:S15} displays a direct comparison
of the measured topography of graphene, the equilibrium potential map and the resulting
$\widehat{E}_{x}^{\rm meas}(x,y)$ map of the same area (see also Fig.~\ref{fig:S12c}). The gate voltages for the two cases are $\pm 5$\,V away from charge neutrality, i.e. in the regime where $l_{\rm ee}>l_{\rm dis}$. Moreover, the images feature a unique type of charge carriers within the whole image being either holes (a-c) or electrons (d-f). Hence, these images do not probe the more complex Dirac fluid regime.\cite{Sheehy2007,Li2020} 

Firstly, we discuss the possibility that the negative electric fields are caused by the Landauer resistivity dipole, i.e. by the diffusive properties of the charge carriers in presence of an obstacle.\cite{Landauer1957} 
This implies a current induced electric field along the flow direction with sequence enhanced field-inverted field-enhanced field resulting from a dipolar distribution of the potential, respectively, the current-induced charge around the obstacle. The inverted field is expected rather precisely at the position of the obstacle.\cite{Landauer1957} 
Resistivity dipoles have been mapped previously by scanning tunneling potentiometry at a step edge of graphene \cite{Wilke2016} and around a hole at a topological insulator surface \cite{Luepke2017}. A region of inverted electric field  has been resolved in the latter case being as large as the diameter of the defect ($\simeq 5$\,nm), as expected, but much shorter than $l_{\rm dis}\simeq 40$\,nm of this sample. 

Thus, one expects a small scale triple structure of field enhancement-inverted field- field enhancement, that is centered at an obstacle, i.e. either at a topographic feature and/or at a feature in the potential map.
Both is not observed, neither in Fig.~\ref{fig:S15} nor in Fig.~\ref{fig:S12c}d--i. Moreover, the inverted fields appear on a scale significantly larger than $l_{\rm dis}$ and not smaller than $l_{\rm dis}$ ruling out that the inverted fields are caused by Landauer dipoles.  We are not aware of other diffusive effects leading to inverted electric fields, hence, discarding ohmic current flow as the origin of field inversion.  

Inverted electric fields also appear in the ballistic regime, if current is injected via a constriction.\cite{Shytov2018,Lent1990,Hui2020} The dilution of current density in the diffraction pattern behind a constriction   
naturally leads to current vorticity that eventually causes inverted electric fields via boundary scattering.\cite{Shytov2018,Lent1990,Hui2020} 
We cannot exclude that such effects contribute to the observed field inversion in our experiments, if one substitutes the boundaries by local obstacles. However, since the patches of inverted fields are significantly larger than the total scattering length $l_{\rm MFP}=(l_{\rm ee}^{-1}+l_{\rm dis}^{-1})^{-1}$, ballistic flow cannot be the main effect. Moreover, since inverted fields are observed preferentially in close vicinity of $V_{\rm D}$ (Fig.~\ref{fig4:Relevance_Viscous}, main text, Fig.~\ref{fig5:Gallery_FieldInversion}a--f), where $l_{\rm MFP}$ is relatively small, we conclude, that a large ratio $l_{\rm dis}/l_{\rm ee}$ is more important for inverted fields than a large $l_{\rm MFP}$ such that viscosity obviously dominates with respect to the consequences of locally ballistic flow.

Thus, we conclude that viscous friction is the main driving force for the inverted fields. The microscopic origin of the inverted fields is, however, not always obvious and
the inherent non-locality of viscosity complicates the identification of correlations, albeit topography and doping profile of the area are known.
Figure~\ref{fig:S15} highlights two extreme cases. In Fig.~\ref{fig:S15}a--c, the encircled area of prominent inverted field (red) 
is related to a topographic fold on the left (height: $\sim 1.4$\,nm ) that is decorated by multiple clusters of dirt with heights of 10-30\,nm. These structures are also visible in the potential map as decreased $E_{\rm F}-E_{\rm D}$, respectively, as a larger hole density (section~\ref{sec:dirt}).
This constitutes an irregular potential barrier likely leading to inhomogeneous flow such that the scenario described above of rather quiet areas surrounded by areas of stronger flow of charge carriers might apply implying a transversal gradient of the vorticity $\omega(x,y)$ leading to inverted electric fields (eq.~(\ref{eq:S27})). The gate dependence of 
$\widehat{E}_{x}^{\rm meas}(x,y)$ of this area is shown in Fig.~\ref{fig5:Gallery_FieldInversion}a--f revealing the consistent appearance of inverted fields around the fold.
Another example where the inverted fields are likely caused by a potential obstacle is shown in Fig.~\ref{fig3:NegFields_ViscousFlow}, main text, where the dipolar field structure is centered at an area of increased electron density. 

In contrast, the inverted field area marked in Fig.~\ref{fig:S15}d--f is not related to a topographic feature or to an exceptional variation of $(E_{\rm F}-E_{\rm D})(x,y)$. In line with the discussion concerning Fig.~\ref{fig4:Relevance_Viscous}, main text, we conjecture that the origin of inverted field in that case is a varying density of short range scatterers that are  not visible by AFM or EFM directly. The inhomogeneity of short range scatterers, that is related to gradients of the local resistivity, probably establishes both, the inhomogeneous flow patterns leading to vorticity $\omega (x,y)$ as well as the emerging viscous flow by local absence of short range scatterers implying gradients of the vorticity.

\section{Outlook}
\label{sec:gallery}

\begin{figure}
	\centering
	\includegraphics[width=165 mm]{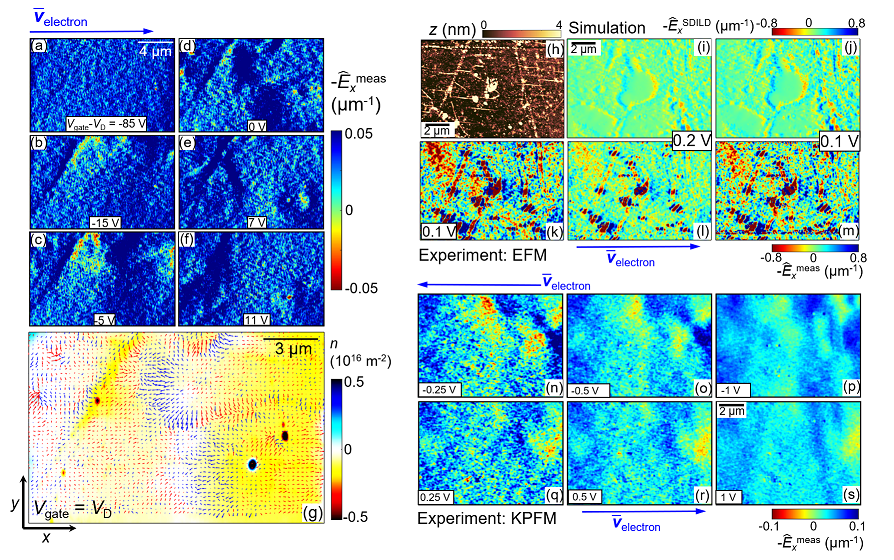}
	\caption{{\bf Ubiquity of inverted $\widehat{E}_{\rm x}^{\rm meas}(x,y)$.} (a)$-$(f) $\widehat{E}_{x}^{\rm meas}(x,y)$ at different $V_{\rm gate}-V_{\rm D}$, $V_{\rm SD}=\,0.5\,V$, KPFM. Corresponding topography in Fig.~\ref{fig:S15}a. Inverted fields mostly appear on the right of a geometric fold with $\sim 1$\,nm height (partly the same as in Fig.~\ref{fig:Relevance_Hydrodynamics_II}a--d).   (g) Same as (d) in different representation: arrows: in-plane electric field vectors ${\mathbf E}^{\rm meas} (x,y) = -\mathbf{\nabla} V_{\rm transport}(x,y)$, blue (red) arrows: $x$ component along (opposing) $V_{\rm SD}$, background color: $n(x,y)$. Source and sink like electric field patterns appear mostly at vanishing charge carrier density.  A background subtraction applied to $V_{\rm CPD}(x,y,V_{\rm SD})$ slightly influences ${\mathbf E}^{\rm meas} (x,y)$ (Fig.~\ref{fig6:ElecField_BG_direction}g, h). (h) Topography of a different graphene area (tapping mode AFM). (i)$-$(j) $\widehat{E}_{ x}^{\rm SDILD}(x,y)$ in the area of h deduced from $V_{\rm CPD}(x,y,V_{\rm SD}=0$\,V) via eqs.~(\ref{Eq_NegField_TrivialEff}) and (\ref{eq:E_SDLD}), $V_{\rm gate}=V_{\rm D}$, $V_{\rm SD}$ marked, EFM.
		(k)$-$(m) $\widehat{E}_{x}^{\rm meas}(x,y)$ in the area of h$-$j, $V_{\rm gate}=V_{\rm D}$, $V_{\rm SD}$ marked, EFM. Comparing i-j with l-m enables distinction between features from SDILD and current induced features. Comparison of k and m showcases reproducibility.
		(n)$-$(s) $\widehat{E}_{ x}^{\rm meas}(x,y)$, $V_{\rm gate}=V_{\rm D}$,  $V_{\rm SD}$ marked, KPFM. The sequence highlights the generally observed contrast reduction with increasing $|V_{\rm SD}|$. 
	}
	\label{fig5:Gallery_FieldInversion}
\end{figure}

Figure~\ref{fig5:Gallery_FieldInversion} showcases additional experimental observations highlighting further opportunities for future studies. Figure~\ref{fig5:Gallery_FieldInversion}a$-$f display $\widehat{E}_{ x}^{\rm meas}(x,y)$ of the area of Fig.~\ref{fig:S15}a--c at different $V_{\rm gate}$.
Inverted $\widehat{E}_{ x}^{\rm meas}(x,y)$ areas appear around $V_{\rm gate}\simeq V_{\rm D}$, but not far away from 
$V_{\rm gate}\simeq V_{\rm D}$ (Fig.~\ref{fig5:Gallery_FieldInversion}a). They appear most prominently  at left of the topographic fold as discussed in section~\ref{sec:origin}.
Figure~\ref{fig5:Gallery_FieldInversion}g shows an arrow plot of the in-plane electric field ${\mathbf E}^{\rm meas} (x,y) = -\mathbf{\nabla} V_{\rm transport}(x,y)$ of the same area at charge neutrality displayed on top of a color plot showcasing $n(x,y)$ (section~\ref{sec4_EField}, \ref{Sec5_DopingDistr_CPD}). This visualizes the correlation between local charge carrier density and current induced curvatures of ${\mathbf E}$ that dominate in areas of low $|n(x,y)|$. Moreover, Fig.~\ref{fig5:Gallery_FieldInversion}h$-$m demonstrate the disentangling of SDILD and current induced electric fields via comparing $\widehat{E}^{\rm meas} (x,y)$ and $\widehat{E}^{\rm SDILD} (x,y)$ of the same area.
Obviously, most of the calculated SDILD patterns are reproduced by the measured
$\widehat{E}^{\rm meas} (x,y)$, such that additional, more extended features can be attributed to current induced features likely due to viscous electron flow. 
Finally, Fig.~\ref{fig5:Gallery_FieldInversion}n$-$s demonstrate how areas of inverted $\widehat{E}_{ x}^{\rm meas}(x,y)$ develop with applied $V_{\rm SD}$. We generally observe that they weaken with increasing $|V_{\rm SD}|$ without a clear understanding yet. 



\begin{mcitethebibliography}{114}
	\providecommand*\natexlab[1]{#1}
	\providecommand*\mciteSetBstSublistMode[1]{}
	\providecommand*\mciteSetBstMaxWidthForm[2]{}
	\providecommand*\mciteBstWouldAddEndPuncttrue
	{\def\EndOfBibitem{\unskip.}}
	\providecommand*\mciteBstWouldAddEndPunctfalse
	{\let\EndOfBibitem\relax}
	\providecommand*\mciteSetBstMidEndSepPunct[3]{}
	\providecommand*\mciteSetBstSublistLabelBeginEnd[3]{}
	\providecommand*\EndOfBibitem{}
	\mciteSetBstSublistMode{f}
	\mciteSetBstMaxWidthForm{subitem}{(\alph{mcitesubitemcount})}
	\mciteSetBstSublistLabelBeginEnd
	{\mcitemaxwidthsubitemform\space}
	{\relax}
	{\relax}
	
	\bibitem[Polini and Geim(2020)Polini, and Geim]{Polini2020}
	Polini,~M.; Geim,~A.~K. Viscous electron fluids. \emph{Phys. Today}
	\textbf{2020}, \emph{73}, 28--34\relax
	\mciteBstWouldAddEndPuncttrue
	\mciteSetBstMidEndSepPunct{\mcitedefaultmidpunct}
	{\mcitedefaultendpunct}{\mcitedefaultseppunct}\relax
	\EndOfBibitem
	\bibitem[Mayzel \latin{et~al.}(2019)Mayzel, Steinberg, and
	Varshney]{Mayzel2019}
	Mayzel,~J.; Steinberg,~V.; Varshney,~A. Stokes flow analogous to viscous
	electron current in graphene. \emph{Nat. Commun.} \textbf{2019}, \emph{10},
	937\relax
	\mciteBstWouldAddEndPuncttrue
	\mciteSetBstMidEndSepPunct{\mcitedefaultmidpunct}
	{\mcitedefaultendpunct}{\mcitedefaultseppunct}\relax
	\EndOfBibitem
	\bibitem[Sch\"{a}fer and Teaney(2009)Sch\"{a}fer, and Teaney]{Schaefer2009}
	Sch\"{a}fer,~T.; Teaney,~D. Nearly perfect fluidity: from cold atomic gases to
	hot quark gluon plasmas. \emph{Rep. Prog. Phys.} \textbf{2009}, \emph{72},
	126001\relax
	\mciteBstWouldAddEndPuncttrue
	\mciteSetBstMidEndSepPunct{\mcitedefaultmidpunct}
	{\mcitedefaultendpunct}{\mcitedefaultseppunct}\relax
	\EndOfBibitem
	\bibitem[Narozhny \latin{et~al.}(2017)Narozhny, Gornyi, Mirlin, and
	Schmalian]{Narozhny2017}
	Narozhny,~B.~N.; Gornyi,~I.~V.; Mirlin,~A.~D.; Schmalian,~J. Hydrodynamic
	Approach to Electronic Transport in Graphene. \emph{Ann. Phys.}
	\textbf{2017}, \emph{529}, 1700043\relax
	\mciteBstWouldAddEndPuncttrue
	\mciteSetBstMidEndSepPunct{\mcitedefaultmidpunct}
	{\mcitedefaultendpunct}{\mcitedefaultseppunct}\relax
	\EndOfBibitem
	\bibitem[Narozhny(2019)]{Narozhny2019}
	Narozhny,~B.~N. Electronic hydrodynamics in graphene. \emph{Ann. Phys.}
	\textbf{2019}, \emph{411}, 167979\relax
	\mciteBstWouldAddEndPuncttrue
	\mciteSetBstMidEndSepPunct{\mcitedefaultmidpunct}
	{\mcitedefaultendpunct}{\mcitedefaultseppunct}\relax
	\EndOfBibitem
	\bibitem[Torre \latin{et~al.}(2015)Torre, Tomadin, Geim, and Polini]{Torre2015}
	Torre,~I.; Tomadin,~A.; Geim,~A.~K.; Polini,~M. Nonlocal transport and the
	hydrodynamic shear viscosity in graphene. \emph{Phys. Rev. B} \textbf{2015},
	\emph{92}, 165433\relax
	\mciteBstWouldAddEndPuncttrue
	\mciteSetBstMidEndSepPunct{\mcitedefaultmidpunct}
	{\mcitedefaultendpunct}{\mcitedefaultseppunct}\relax
	\EndOfBibitem
	\bibitem[Gurzhi(1968)]{Gurzhi1968}
	Gurzhi,~R.~N. Hydrodynamic effects in solids at low temperature. \emph{Sov.
		Phys. Usp.} \textbf{1968}, \emph{11}, 255--270\relax
	\mciteBstWouldAddEndPuncttrue
	\mciteSetBstMidEndSepPunct{\mcitedefaultmidpunct}
	{\mcitedefaultendpunct}{\mcitedefaultseppunct}\relax
	\EndOfBibitem
	\bibitem[Govorov and Heremans(2004)Govorov, and Heremans]{Govorov2004}
	Govorov,~A.~O.; Heremans,~J.~J. Hydrodynamic Effects in Interacting Fermi
	Electron Jets. \emph{Physical Review Letters} \textbf{2004}, \emph{92},
	026803\relax
	\mciteBstWouldAddEndPuncttrue
	\mciteSetBstMidEndSepPunct{\mcitedefaultmidpunct}
	{\mcitedefaultendpunct}{\mcitedefaultseppunct}\relax
	\EndOfBibitem
	\bibitem[Guo \latin{et~al.}(2017)Guo, Ilseven, Falkovich, and Levitov]{Guo2017}
	Guo,~H.; Ilseven,~E.; Falkovich,~G.; Levitov,~L.~S. Higher-than-ballistic
	conduction of viscous electron flows. \emph{Proc. Nat. Acad. Sci.}
	\textbf{2017}, \emph{114}, 3068--3073\relax
	\mciteBstWouldAddEndPuncttrue
	\mciteSetBstMidEndSepPunct{\mcitedefaultmidpunct}
	{\mcitedefaultendpunct}{\mcitedefaultseppunct}\relax
	\EndOfBibitem
	\bibitem[Moessner \latin{et~al.}(2019)Moessner, Morales-Dur{\'{a}}n,
	Sur{\'{o}}wka, and Witkowski]{Moessner2019}
	Moessner,~R.; Morales-Dur{\'{a}}n,~N.; Sur{\'{o}}wka,~P.; Witkowski,~P.
	Boundary-condition and geometry engineering in electronic hydrodynamics.
	\emph{Phys. Rev. B} \textbf{2019}, \emph{100}, 155115\relax
	\mciteBstWouldAddEndPuncttrue
	\mciteSetBstMidEndSepPunct{\mcitedefaultmidpunct}
	{\mcitedefaultendpunct}{\mcitedefaultseppunct}\relax
	\EndOfBibitem
	\bibitem[Holder \latin{et~al.}(2019)Holder, Queiroz, Scaffidi, Silberstein,
	Rozen, Sulpizio, Ella, Ilani, and Stern]{Holder2019}
	Holder,~T.; Queiroz,~R.; Scaffidi,~T.; Silberstein,~N.; Rozen,~A.;
	Sulpizio,~J.~A.; Ella,~L.; Ilani,~S.; Stern,~A. Ballistic and hydrodynamic
	magnetotransport in narrow channels. \emph{Phys. Rev. B} \textbf{2019},
	\emph{100}, 245305\relax
	\mciteBstWouldAddEndPuncttrue
	\mciteSetBstMidEndSepPunct{\mcitedefaultmidpunct}
	{\mcitedefaultendpunct}{\mcitedefaultseppunct}\relax
	\EndOfBibitem
	\bibitem[Mohseni \latin{et~al.}(2005)Mohseni, Shakouri, Ram, and
	Abraham]{Mohseni2005}
	Mohseni,~K.; Shakouri,~A.; Ram,~R.~J.; Abraham,~M.~C. Electron vortices in
	semiconductors devices. \emph{Phys. Fluids} \textbf{2005}, \emph{17},
	100602\relax
	\mciteBstWouldAddEndPuncttrue
	\mciteSetBstMidEndSepPunct{\mcitedefaultmidpunct}
	{\mcitedefaultendpunct}{\mcitedefaultseppunct}\relax
	\EndOfBibitem
	\bibitem[Levitov and Falkovich(2016)Levitov, and Falkovich]{Levitov2016}
	Levitov,~L.; Falkovich,~G. Electron viscosity, current vortices and negative
	nonlocal resistance in graphene. \emph{Nat. Phys.} \textbf{2016}, \emph{12},
	672--676\relax
	\mciteBstWouldAddEndPuncttrue
	\mciteSetBstMidEndSepPunct{\mcitedefaultmidpunct}
	{\mcitedefaultendpunct}{\mcitedefaultseppunct}\relax
	\EndOfBibitem
	\bibitem[Danz and Narozhny(2020)Danz, and Narozhny]{Danz2020}
	Danz,~S.; Narozhny,~B.~N. Vorticity of viscous electronic flow in graphene.
	\emph{2D Materials} \textbf{2020}, \emph{7}, 035001\relax
	\mciteBstWouldAddEndPuncttrue
	\mciteSetBstMidEndSepPunct{\mcitedefaultmidpunct}
	{\mcitedefaultendpunct}{\mcitedefaultseppunct}\relax
	\EndOfBibitem
	\bibitem[Chandra \latin{et~al.}(2019)Chandra, Kataria, Sahdev, and
	Sundararaman]{Chandra2019}
	Chandra,~M.; Kataria,~G.; Sahdev,~D.; Sundararaman,~R. Hydrodynamic and
	ballistic {AC} transport in two-dimensional Fermi liquids. \emph{Phys. Rev.
		B} \textbf{2019}, \emph{99}, 165409\relax
	\mciteBstWouldAddEndPuncttrue
	\mciteSetBstMidEndSepPunct{\mcitedefaultmidpunct}
	{\mcitedefaultendpunct}{\mcitedefaultseppunct}\relax
	\EndOfBibitem
	\bibitem[Lent(1990)]{Lent1990}
	Lent,~C.~S. Ballistic current vortex excitations in electron waveguide
	structures. \emph{Appl. Phys. Lett.} \textbf{1990}, \emph{57},
	1678--1680\relax
	\mciteBstWouldAddEndPuncttrue
	\mciteSetBstMidEndSepPunct{\mcitedefaultmidpunct}
	{\mcitedefaultendpunct}{\mcitedefaultseppunct}\relax
	\EndOfBibitem
	\bibitem[Pellegrino \latin{et~al.}(2016)Pellegrino, Torre, Geim, and
	Polini]{Pellegrino2016}
	Pellegrino,~F. M.~D.; Torre,~I.; Geim,~A.~K.; Polini,~M. Electron hydrodynamics
	dilemma: Whirlpools or no whirlpools. \emph{Phys. Rev. B} \textbf{2016},
	\emph{94}, 155414\relax
	\mciteBstWouldAddEndPuncttrue
	\mciteSetBstMidEndSepPunct{\mcitedefaultmidpunct}
	{\mcitedefaultendpunct}{\mcitedefaultseppunct}\relax
	\EndOfBibitem
	\bibitem[Mendoza \latin{et~al.}(2011)Mendoza, Herrmann, and Succi]{Mendoza2011}
	Mendoza,~M.; Herrmann,~H.~J.; Succi,~S. Preturbulent Regimes in Graphene Flow.
	\emph{Phys. Rev. Lett.} \textbf{2011}, \emph{106}, 156601\relax
	\mciteBstWouldAddEndPuncttrue
	\mciteSetBstMidEndSepPunct{\mcitedefaultmidpunct}
	{\mcitedefaultendpunct}{\mcitedefaultseppunct}\relax
	\EndOfBibitem
	\bibitem[Li \latin{et~al.}(2020)Li, Levchenko, and Andreev]{Li2020}
	Li,~S.; Levchenko,~A.; Andreev,~A.~V. Hydrodynamic electron transport near
	charge neutrality. \emph{Phys. Rev. B} \textbf{2020}, \emph{102},
	075305\relax
	\mciteBstWouldAddEndPuncttrue
	\mciteSetBstMidEndSepPunct{\mcitedefaultmidpunct}
	{\mcitedefaultendpunct}{\mcitedefaultseppunct}\relax
	\EndOfBibitem
	\bibitem[de~Jong and Molenkamp(1995)de~Jong, and Molenkamp]{deJong1995}
	de~Jong,~M. J.~M.; Molenkamp,~L.~W. Hydrodynamic electron flow in high-mobility
	wires. \emph{Phys. Rev. B} \textbf{1995}, \emph{51}, 13389--13402\relax
	\mciteBstWouldAddEndPuncttrue
	\mciteSetBstMidEndSepPunct{\mcitedefaultmidpunct}
	{\mcitedefaultendpunct}{\mcitedefaultseppunct}\relax
	\EndOfBibitem
	\bibitem[Bandurin \latin{et~al.}(2016)Bandurin, Torre, Kumar, Shalom, Tomadin,
	Principi, Auton, Khestanova, Novoselov, Grigorieva, Ponomarenko, Geim, and
	Polini]{Bandurin2016}
	Bandurin,~D.~A.; Torre,~I.; Kumar,~R.~K.; Shalom,~M.~B.; Tomadin,~A.;
	Principi,~A.; Auton,~G.~H.; Khestanova,~E.; Novoselov,~K.~S.;
	Grigorieva,~I.~V.; Ponomarenko,~L.~A.; Geim,~A.~K.; Polini,~M. Negative local
	resistance caused by viscous electron backflow in graphene. \emph{Science}
	\textbf{2016}, \emph{351}, 1055--1058\relax
	\mciteBstWouldAddEndPuncttrue
	\mciteSetBstMidEndSepPunct{\mcitedefaultmidpunct}
	{\mcitedefaultendpunct}{\mcitedefaultseppunct}\relax
	\EndOfBibitem
	\bibitem[Crossno \latin{et~al.}(2016)Crossno, Shi, Wang, Liu, Harzheim, Lucas,
	Sachdev, Kim, Taniguchi, Watanabe, Ohki, and Fong]{Crossno2016}
	Crossno,~J.; Shi,~J.~K.; Wang,~K.; Liu,~X.; Harzheim,~A.; Lucas,~A.;
	Sachdev,~S.; Kim,~P.; Taniguchi,~T.; Watanabe,~K.; Ohki,~T.~A.; Fong,~K.~C.
	Observation of the Dirac fluid and the breakdown of the Wiedemann-Franz law
	in graphene. \emph{Science} \textbf{2016}, \emph{351}, 1058--1061\relax
	\mciteBstWouldAddEndPuncttrue
	\mciteSetBstMidEndSepPunct{\mcitedefaultmidpunct}
	{\mcitedefaultendpunct}{\mcitedefaultseppunct}\relax
	\EndOfBibitem
	\bibitem[Ghahari \latin{et~al.}(2016)Ghahari, Xie, Taniguchi, Watanabe, Foster,
	and Kim]{Ghahari2016}
	Ghahari,~F.; Xie,~H.-Y.; Taniguchi,~T.; Watanabe,~K.; Foster,~M.~S.; Kim,~P.
	Enhanced Thermoelectric Power in Graphene: Violation of the Mott Relation by
	Inelastic Scattering. \emph{Phys. Rev. Lett.} \textbf{2016}, \emph{116},
	136802\relax
	\mciteBstWouldAddEndPuncttrue
	\mciteSetBstMidEndSepPunct{\mcitedefaultmidpunct}
	{\mcitedefaultendpunct}{\mcitedefaultseppunct}\relax
	\EndOfBibitem
	\bibitem[Gallagher \latin{et~al.}(2019)Gallagher, Yang, Lyu, Tian, Kou, Zhang,
	Watanabe, Taniguchi, and Wang]{Gallagher2019}
	Gallagher,~P.; Yang,~C.-S.; Lyu,~T.; Tian,~F.; Kou,~R.; Zhang,~H.;
	Watanabe,~K.; Taniguchi,~T.; Wang,~F. Quantum-critical conductivity of the
	Dirac fluid in graphene. \emph{Science} \textbf{2019}, \emph{364},
	158--162\relax
	\mciteBstWouldAddEndPuncttrue
	\mciteSetBstMidEndSepPunct{\mcitedefaultmidpunct}
	{\mcitedefaultendpunct}{\mcitedefaultseppunct}\relax
	\EndOfBibitem
	\bibitem[Kumar \latin{et~al.}(2017)Kumar, Bandurin, Pellegrino, Cao, Principi,
	Guo, Auton, Shalom, Ponomarenko, Falkovich, Watanabe, Taniguchi, Grigorieva,
	Levitov, Polini, and Geim]{Kumar2017}
	Kumar,~R.~K. \latin{et~al.}  Superballistic flow of viscous electron fluid
	through graphene constrictions. \emph{Nat. Phys.} \textbf{2017}, \emph{13},
	1182--1185\relax
	\mciteBstWouldAddEndPuncttrue
	\mciteSetBstMidEndSepPunct{\mcitedefaultmidpunct}
	{\mcitedefaultendpunct}{\mcitedefaultseppunct}\relax
	\EndOfBibitem
	\bibitem[Berdyugin \latin{et~al.}(2019)Berdyugin, Xu, Pellegrino, Kumar,
	Principi, Torre, Shalom, Taniguchi, Watanabe, Grigorieva, Polini, Geim, and
	Bandurin]{Berdyugin2019}
	Berdyugin,~A.~I.; Xu,~S.~G.; Pellegrino,~F. M.~D.; Kumar,~R.~K.; Principi,~A.;
	Torre,~I.; Shalom,~M.~B.; Taniguchi,~T.; Watanabe,~K.; Grigorieva,~I.~V.;
	Polini,~M.; Geim,~A.~K.; Bandurin,~D.~A. Measuring Hall viscosity of
	graphene's electron fluid. \emph{Science} \textbf{2019}, \emph{364},
	162--165\relax
	\mciteBstWouldAddEndPuncttrue
	\mciteSetBstMidEndSepPunct{\mcitedefaultmidpunct}
	{\mcitedefaultendpunct}{\mcitedefaultseppunct}\relax
	\EndOfBibitem
	\bibitem[Bandurin \latin{et~al.}(2018)Bandurin, Shytov, Levitov, Kumar,
	Berdyugin, Shalom, Grigorieva, Geim, and Falkovich]{Bandurin2018}
	Bandurin,~D.~A.; Shytov,~A.~V.; Levitov,~L.~S.; Kumar,~R.~K.; Berdyugin,~A.~I.;
	Shalom,~M.~B.; Grigorieva,~I.~V.; Geim,~A.~K.; Falkovich,~G. Fluidity onset
	in graphene. \emph{Nat. Commun.} \textbf{2018}, \emph{9}, 4533\relax
	\mciteBstWouldAddEndPuncttrue
	\mciteSetBstMidEndSepPunct{\mcitedefaultmidpunct}
	{\mcitedefaultendpunct}{\mcitedefaultseppunct}\relax
	\EndOfBibitem
	\bibitem[Geurs \latin{et~al.}(2020)Geurs, Kim, Watanabe, Taniguchi, Moon, and
	Smet]{Geurs2020}
	Geurs,~J.; Kim,~Y.; Watanabe,~K.; Taniguchi,~T.; Moon,~P.; Smet,~J.~H.
	Rectification by hydrodynamic flow in an encapsulated graphene Tesla valve.
	\emph{arXiv:} \textbf{2020}, \emph{2008.04862}\relax
	\mciteBstWouldAddEndPuncttrue
	\mciteSetBstMidEndSepPunct{\mcitedefaultmidpunct}
	{\mcitedefaultendpunct}{\mcitedefaultseppunct}\relax
	\EndOfBibitem
	\bibitem[Lucas and Fong(2018)Lucas, and Fong]{Lucas2018}
	Lucas,~A.; Fong,~K.~C. Hydrodynamics of electrons in graphene. \emph{J. Phys.:
		Cond. Matt.} \textbf{2018}, \emph{30}, 053001\relax
	\mciteBstWouldAddEndPuncttrue
	\mciteSetBstMidEndSepPunct{\mcitedefaultmidpunct}
	{\mcitedefaultendpunct}{\mcitedefaultseppunct}\relax
	\EndOfBibitem
	\bibitem[Moll \latin{et~al.}(2016)Moll, Kushwaha, Nandi, Schmidt, and
	Mackenzie]{Moll2016}
	Moll,~P. J.~W.; Kushwaha,~P.; Nandi,~N.; Schmidt,~B.; Mackenzie,~A.~P. Evidence
	for hydrodynamic electron flow in {PdCoO}2. \emph{Science} \textbf{2016},
	\emph{351}, 1061--1064\relax
	\mciteBstWouldAddEndPuncttrue
	\mciteSetBstMidEndSepPunct{\mcitedefaultmidpunct}
	{\mcitedefaultendpunct}{\mcitedefaultseppunct}\relax
	\EndOfBibitem
	\bibitem[Fu \latin{et~al.}(2020)Fu, Guin, Scaffidi, Sun, Saha, Watzman,
	Srivastava, Li, Schnelle, Parkin, Felser, and Gooth]{Fu2020}
	Fu,~C.; Guin,~S.~N.; Scaffidi,~T.; Sun,~Y.; Saha,~R.; Watzman,~S.~J.;
	Srivastava,~A.~K.; Li,~G.; Schnelle,~W.; Parkin,~S. S.~P.; Felser,~C.;
	Gooth,~J. Largely Suppressed Magneto-Thermal Conductivity and Enhanced
	Magneto-Thermoelectric Properties in PtSn$_4$. \emph{Research} \textbf{2020},
	\emph{2020}, 4643507\relax
	\mciteBstWouldAddEndPuncttrue
	\mciteSetBstMidEndSepPunct{\mcitedefaultmidpunct}
	{\mcitedefaultendpunct}{\mcitedefaultseppunct}\relax
	\EndOfBibitem
	\bibitem[Gooth \latin{et~al.}(2018)Gooth, Menges, Kumar, S\"{u}$\upbeta$,
	Shekhar, Sun, Drechsler, Zierold, Felser, and Gotsmann]{Gooth2018}
	Gooth,~J.; Menges,~F.; Kumar,~N.; S\"{u}$\upbeta$,~V.; Shekhar,~C.; Sun,~Y.;
	Drechsler,~U.; Zierold,~R.; Felser,~C.; Gotsmann,~B. Thermal and electrical
	signatures of a hydrodynamic electron fluid in tungsten diphosphide.
	\emph{Nat. Commun.} \textbf{2018}, \emph{9}, 4093\relax
	\mciteBstWouldAddEndPuncttrue
	\mciteSetBstMidEndSepPunct{\mcitedefaultmidpunct}
	{\mcitedefaultendpunct}{\mcitedefaultseppunct}\relax
	\EndOfBibitem
	\bibitem[Block \latin{et~al.}(2020)Block, Principi, Hesp, Cummings, Liebel,
	Watanabe, Taniguchi, Roche, Koppens, van Hulst, and Tielrooij]{Block2020}
	Block,~A.; Principi,~A.; Hesp,~N. C.~H.; Cummings,~A.~W.; Liebel,~M.;
	Watanabe,~K.; Taniguchi,~T.; Roche,~S.; Koppens,~F. H.~L.; van Hulst,~N.~F.;
	Tielrooij,~K.-J. Observation of giant and tunable thermal diffusivity of
	Dirac fluid at room temperature. \emph{arXiv:} \textbf{2020},
	\emph{2008.04189}\relax
	\mciteBstWouldAddEndPuncttrue
	\mciteSetBstMidEndSepPunct{\mcitedefaultmidpunct}
	{\mcitedefaultendpunct}{\mcitedefaultseppunct}\relax
	\EndOfBibitem
	\bibitem[Sulpizio \latin{et~al.}(2019)Sulpizio, Ella, Rozen, Birkbeck, Perello,
	Dutta, Ben-Shalom, Taniguchi, Watanabe, Holder, Queiroz, Principi, Stern,
	Scaffidi, Geim, and Ilani]{Sulpizio2019}
	Sulpizio,~J.~A. \latin{et~al.}  Visualizing Poiseuille flow of hydrodynamic
	electrons. \emph{Nature} \textbf{2019}, \emph{576}, 75--79\relax
	\mciteBstWouldAddEndPuncttrue
	\mciteSetBstMidEndSepPunct{\mcitedefaultmidpunct}
	{\mcitedefaultendpunct}{\mcitedefaultseppunct}\relax
	\EndOfBibitem
	\bibitem[Ku \latin{et~al.}(2020)Ku, Zhou, Li, Shin, Shi, Burch, Anderson,
	Pierce, Xie, Hamo, Vool, Zhang, Casola, Taniguchi, Watanabe, Fogler, Kim,
	Yacoby, and Walsworth]{Ku2020}
	Ku,~M. J.~H. \latin{et~al.}  Imaging viscous flow of the Dirac fluid in
	graphene. \emph{Nature} \textbf{2020}, \emph{583}, 537--541\relax
	\mciteBstWouldAddEndPuncttrue
	\mciteSetBstMidEndSepPunct{\mcitedefaultmidpunct}
	{\mcitedefaultendpunct}{\mcitedefaultseppunct}\relax
	\EndOfBibitem
	\bibitem[Jenkins \latin{et~al.}(2020)Jenkins, Baumann, Zhou, Meynell, Yang,
	Watanabe, Taniguchi, Lucas, Young, and Jayich]{Jenkins2020}
	Jenkins,~A.; Baumann,~S.; Zhou,~H.; Meynell,~S.~A.; Yang,~D.; Watanabe,~K.;
	Taniguchi,~T.; Lucas,~A.; Young,~A.~F.; Jayich,~A. C.~B. Imaging the
	breakdown of ohmic transport in graphene. \emph{arXiv:} \textbf{2020},
	\emph{2002.05065}\relax
	\mciteBstWouldAddEndPuncttrue
	\mciteSetBstMidEndSepPunct{\mcitedefaultmidpunct}
	{\mcitedefaultendpunct}{\mcitedefaultseppunct}\relax
	\EndOfBibitem
	\bibitem[Braem \latin{et~al.}(2018)Braem, Pellegrino, Principi, R\"{o}\"{o}sli,
	Gold, Hennel, Koski, Berl, Dietsche, Wegscheider, Polini, Ihn, and
	Ensslin]{Braem2018}
	Braem,~B.~A.; Pellegrino,~F. M.~D.; Principi,~A.; R\"{o}\"{o}sli,~M.; Gold,~C.;
	Hennel,~S.; Koski,~J.~V.; Berl,~M.; Dietsche,~W.; Wegscheider,~W.;
	Polini,~M.; Ihn,~T.; Ensslin,~K. Scanning gate microscopy in a viscous
	electron fluid. \emph{Phys. Rev. B} \textbf{2018}, \emph{98}, 241304\relax
	\mciteBstWouldAddEndPuncttrue
	\mciteSetBstMidEndSepPunct{\mcitedefaultmidpunct}
	{\mcitedefaultendpunct}{\mcitedefaultseppunct}\relax
	\EndOfBibitem
	\bibitem[Krebs \latin{et~al.}(2021)Krebs, Behn, Li, Smith, Watanabe, Taniguchi,
	Levchenko, and Brar]{krebs2021}
	Krebs,~Z.~J.; Behn,~W.~A.; Li,~S.; Smith,~K.~J.; Watanabe,~K.; Taniguchi,~T.;
	Levchenko,~A.; Brar,~V.~W. Imaging the breaking of electrostatic dams in
	graphene for ballistic and viscous fluids. \emph{arXiv:} \textbf{2021},
	\emph{2106.07212}\relax
	\mciteBstWouldAddEndPuncttrue
	\mciteSetBstMidEndSepPunct{\mcitedefaultmidpunct}
	{\mcitedefaultendpunct}{\mcitedefaultseppunct}\relax
	\EndOfBibitem
	\bibitem[Melitz \latin{et~al.}(2011)Melitz, Shen, Kummel, and Lee]{Melitz2011}
	Melitz,~W.; Shen,~J.; Kummel,~A.~C.; Lee,~S. Kelvin probe force microscopy and
	its application. \emph{Surf. Sci. Rep.} \textbf{2011}, \emph{66}, 1--27\relax
	\mciteBstWouldAddEndPuncttrue
	\mciteSetBstMidEndSepPunct{\mcitedefaultmidpunct}
	{\mcitedefaultendpunct}{\mcitedefaultseppunct}\relax
	\EndOfBibitem
	\bibitem[Xu \latin{et~al.}(2018)Xu, Chen, Li, and Xu]{Xu2018}
	Xu,~J.; Chen,~D.; Li,~W.; Xu,~J. Surface potential extraction from
	electrostatic and Kelvin-probe force microscopy images. \emph{J. Appl. Phys.}
	\textbf{2018}, \emph{123}, 184301\relax
	\mciteBstWouldAddEndPuncttrue
	\mciteSetBstMidEndSepPunct{\mcitedefaultmidpunct}
	{\mcitedefaultendpunct}{\mcitedefaultseppunct}\relax
	\EndOfBibitem
	\bibitem[Falkovich and Levitov(2017)Falkovich, and Levitov]{Falkovich2017}
	Falkovich,~G.; Levitov,~L. Linking Spatial Distributions of Potential and
	Current in Viscous Electronics. \emph{Phys. Rev. Lett.} \textbf{2017},
	\emph{119}, 066601\relax
	\mciteBstWouldAddEndPuncttrue
	\mciteSetBstMidEndSepPunct{\mcitedefaultmidpunct}
	{\mcitedefaultendpunct}{\mcitedefaultseppunct}\relax
	\EndOfBibitem
	\bibitem[Shaygan \latin{et~al.}(2017)Shaygan, Otto, Sagade, Chavarin, Bacher,
	Mertin, and Neumaier]{Shaygan2017}
	Shaygan,~M.; Otto,~M.; Sagade,~A.~A.; Chavarin,~C.~A.; Bacher,~G.; Mertin,~W.;
	Neumaier,~D. Low Resistive Edge Contacts to CVD-Grown Graphene Using a CMOS
	Compatible Metal. \emph{Ann. Phys.} \textbf{2017}, \emph{529}, 1600410\relax
	\mciteBstWouldAddEndPuncttrue
	\mciteSetBstMidEndSepPunct{\mcitedefaultmidpunct}
	{\mcitedefaultendpunct}{\mcitedefaultseppunct}\relax
	\EndOfBibitem
	\bibitem[Panchal \latin{et~al.}(2013)Panchal, Pearce, Yakimova, Tzalenchuk, and
	Kazakova]{Panchal2013}
	Panchal,~V.; Pearce,~R.; Yakimova,~R.; Tzalenchuk,~A.; Kazakova,~O.
	Standardization of surface potential measurements of graphene domains.
	\emph{Sci. Rep.} \textbf{2013}, \emph{3}, 2597\relax
	\mciteBstWouldAddEndPuncttrue
	\mciteSetBstMidEndSepPunct{\mcitedefaultmidpunct}
	{\mcitedefaultendpunct}{\mcitedefaultseppunct}\relax
	\EndOfBibitem
	\bibitem[Martin \latin{et~al.}(2007)Martin, Akerman, Ulbricht, Lohmann, Smet,
	von Klitzing, and Yacoby]{Martin2007}
	Martin,~J.; Akerman,~N.; Ulbricht,~G.; Lohmann,~T.; Smet,~J.~H.; von
	Klitzing,~K.; Yacoby,~A. Observation of electron-hole puddles in graphene
	using a scanning single-electron transistor. \emph{Nat. Phys.} \textbf{2007},
	\emph{4}, 144--148\relax
	\mciteBstWouldAddEndPuncttrue
	\mciteSetBstMidEndSepPunct{\mcitedefaultmidpunct}
	{\mcitedefaultendpunct}{\mcitedefaultseppunct}\relax
	\EndOfBibitem
	\bibitem[Li \latin{et~al.}(2011)Li, Hwang, and Sarma]{Li2011}
	Li,~Q.; Hwang,~E.~H.; Sarma,~S.~D. Temperature-dependent compressibility in
	graphene and two-dimensional systems. \emph{Phys. Rev. B} \textbf{2011},
	\emph{84}, 235407\relax
	\mciteBstWouldAddEndPuncttrue
	\mciteSetBstMidEndSepPunct{\mcitedefaultmidpunct}
	{\mcitedefaultendpunct}{\mcitedefaultseppunct}\relax
	\EndOfBibitem
	\bibitem[Sheehy and Schmalian(2007)Sheehy, and Schmalian]{Sheehy2007}
	Sheehy,~D.~E.; Schmalian,~J. Quantum critical scaling in graphene. \emph{Phys.
		Rev. Lett.} \textbf{2007}, \emph{99}, 226803\relax
	\mciteBstWouldAddEndPuncttrue
	\mciteSetBstMidEndSepPunct{\mcitedefaultmidpunct}
	{\mcitedefaultendpunct}{\mcitedefaultseppunct}\relax
	\EndOfBibitem
	\bibitem[Landauer(1957)]{Landauer1957}
	Landauer,~R. Spatial Variation of Currents and Fields Due to Localized
	Scatterers in Metallic Conduction. \emph{IBM J. Res. Dev.} \textbf{1957},
	\emph{1}, 223--231\relax
	\mciteBstWouldAddEndPuncttrue
	\mciteSetBstMidEndSepPunct{\mcitedefaultmidpunct}
	{\mcitedefaultendpunct}{\mcitedefaultseppunct}\relax
	\EndOfBibitem
	\bibitem[Morr(2017)]{Morr2017}
	Morr,~D.~K. Scanning tunneling potentiometry, charge transport, and Landauer's
	resistivity dipole from the quantum to the classical transport regime.
	\emph{Phys. Rev. B} \textbf{2017}, \emph{95}, 195162\relax
	\mciteBstWouldAddEndPuncttrue
	\mciteSetBstMidEndSepPunct{\mcitedefaultmidpunct}
	{\mcitedefaultendpunct}{\mcitedefaultseppunct}\relax
	\EndOfBibitem
	\bibitem[Giuliani and Vignale(2005)Giuliani, and
	Vignale]{Giuliani_Vignale_2005}
	Giuliani,~G.; Vignale,~G. \emph{Quantum Theory of the Electron Liquid};
	Cambridge University Press, 2005; pp 432--439\relax
	\mciteBstWouldAddEndPuncttrue
	\mciteSetBstMidEndSepPunct{\mcitedefaultmidpunct}
	{\mcitedefaultendpunct}{\mcitedefaultseppunct}\relax
	\EndOfBibitem
	\bibitem[Polini and Vignale(2016)Polini, and
	Vignale]{Polini2016_NoNonsensePhysicist}
	Polini,~M.; Vignale,~G. \emph{No-nonsense Physicist: An Overview of Gabriele
		Giuliani's Work and Life}; Edizioni della Normale, 2016; Vol.~2; pp
	107--124\relax
	\mciteBstWouldAddEndPuncttrue
	\mciteSetBstMidEndSepPunct{\mcitedefaultmidpunct}
	{\mcitedefaultendpunct}{\mcitedefaultseppunct}\relax
	\EndOfBibitem
	\bibitem[Kim \latin{et~al.}(2020)Kim, Xu, Berdyugin, Principi, Slizovskiy, Xin,
	Kumaravadivel, Kuang, Hamer, Kumar, Gorbachev, Watanabe, Taniguchi,
	Grigorieva, Fal'ko, Polini, and Geim]{Kim2020}
	Kim,~M. \latin{et~al.}  Control of electron-electron interaction in graphene by
	proximity screening. \emph{Nat. Commun.} \textbf{2020}, \emph{11}, 2339\relax
	\mciteBstWouldAddEndPuncttrue
	\mciteSetBstMidEndSepPunct{\mcitedefaultmidpunct}
	{\mcitedefaultendpunct}{\mcitedefaultseppunct}\relax
	\EndOfBibitem
	\bibitem[Li and Sarma(2013)Li, and Sarma]{Li2013}
	Li,~Q.; Sarma,~S.~D. Finite temperature inelastic mean free path and
	quasiparticle lifetime in graphene. \emph{Phys. Rev. B} \textbf{2013},
	\emph{87}, 085406\relax
	\mciteBstWouldAddEndPuncttrue
	\mciteSetBstMidEndSepPunct{\mcitedefaultmidpunct}
	{\mcitedefaultendpunct}{\mcitedefaultseppunct}\relax
	\EndOfBibitem
	\bibitem[Sarma \latin{et~al.}(2011)Sarma, Adam, Hwang, and Rossi]{DasSarma2011}
	Sarma,~S.~D.; Adam,~S.; Hwang,~E.~H.; Rossi,~E. Electronic transport in
	two-dimensional graphene. \emph{Rev. Mod. Phys.} \textbf{2011}, \emph{83},
	407--470\relax
	\mciteBstWouldAddEndPuncttrue
	\mciteSetBstMidEndSepPunct{\mcitedefaultmidpunct}
	{\mcitedefaultendpunct}{\mcitedefaultseppunct}\relax
	\EndOfBibitem
	\bibitem[Shon and Ando(1998)Shon, and Ando]{Shon1998}
	Shon,~N.~H.; Ando,~T. Quantum Transport in Two-Dimensional Graphite System.
	\emph{J. Phys. Soc. Jp.} \textbf{1998}, \emph{67}, 2421--2429\relax
	\mciteBstWouldAddEndPuncttrue
	\mciteSetBstMidEndSepPunct{\mcitedefaultmidpunct}
	{\mcitedefaultendpunct}{\mcitedefaultseppunct}\relax
	\EndOfBibitem
	\bibitem[Ohta \latin{et~al.}(2007)Ohta, Bostwick, McChesney, Seyller, Horn, and
	Rotenberg]{Ohta2007}
	Ohta,~T.; Bostwick,~A.; McChesney,~J.~L.; Seyller,~T.; Horn,~K.; Rotenberg,~E.
	Interlayer Interaction and Electronic Screening in Multilayer Graphene
	Investigated with Angle-Resolved Photoemission Spectroscopy. \emph{Phys. Rev.
		Lett.} \textbf{2007}, \emph{98}, 206802\relax
	\mciteBstWouldAddEndPuncttrue
	\mciteSetBstMidEndSepPunct{\mcitedefaultmidpunct}
	{\mcitedefaultendpunct}{\mcitedefaultseppunct}\relax
	\EndOfBibitem
	\bibitem[Plochocka \latin{et~al.}(2008)Plochocka, Faugeras, Orlita, Sadowski,
	Martinez, Potemski, Goerbig, Fuchs, Berger, and de~Heer]{Plochocka2008}
	Plochocka,~P.; Faugeras,~C.; Orlita,~M.; Sadowski,~M.~L.; Martinez,~G.;
	Potemski,~M.; Goerbig,~M.~O.; Fuchs,~J.-N.; Berger,~C.; de~Heer,~W.~A.
	High-Energy Limit of Massless Dirac Fermions in Multilayer Graphene using
	Magneto-Optical Transmission Spectroscopy. \emph{Phys. Rev. Lett.}
	\textbf{2008}, \emph{100}, 087401\relax
	\mciteBstWouldAddEndPuncttrue
	\mciteSetBstMidEndSepPunct{\mcitedefaultmidpunct}
	{\mcitedefaultendpunct}{\mcitedefaultseppunct}\relax
	\EndOfBibitem
	\bibitem[Just \latin{et~al.}(2014)Just, Zimmermann, Kataev, Buechner, Pratzer,
	and Morgenstern]{Just2014}
	Just,~S.; Zimmermann,~S.; Kataev,~V.; Buechner,~B.; Pratzer,~M.;
	Morgenstern,~M. Preferential antiferromagnetic coupling of vacancies in
	graphene on SiO2: Electron spin resonance and scanning tunneling
	spectroscopy. \emph{Phys. Rev. B} \textbf{2014}, \emph{90}, 125449\relax
	\mciteBstWouldAddEndPuncttrue
	\mciteSetBstMidEndSepPunct{\mcitedefaultmidpunct}
	{\mcitedefaultendpunct}{\mcitedefaultseppunct}\relax
	\EndOfBibitem
	\bibitem[Stauber \latin{et~al.}(2007)Stauber, Peres, and Guinea]{Stauber2007}
	Stauber,~T.; Peres,~N. M.~R.; Guinea,~F. Electronic transport in graphene: A
	semiclassical approach including midgap states. \emph{Phys. Rev. B}
	\textbf{2007}, \emph{76}, 205423\relax
	\mciteBstWouldAddEndPuncttrue
	\mciteSetBstMidEndSepPunct{\mcitedefaultmidpunct}
	{\mcitedefaultendpunct}{\mcitedefaultseppunct}\relax
	\EndOfBibitem
	\bibitem[Giannazzo \latin{et~al.}(2011)Giannazzo, Sonde, Rimini, and
	Raineri]{Giannazzo2011}
	Giannazzo,~F.; Sonde,~S.; Rimini,~E.; Raineri,~V. Lateral homogeneity of the
	electronic properties in pristine and ion-irradiated graphene probed by
	scanning capacitance spectroscopy. \emph{Nanosc. Res. Lett.} \textbf{2011},
	\emph{6}, 109\relax
	\mciteBstWouldAddEndPuncttrue
	\mciteSetBstMidEndSepPunct{\mcitedefaultmidpunct}
	{\mcitedefaultendpunct}{\mcitedefaultseppunct}\relax
	\EndOfBibitem
	\bibitem[Lindvall \latin{et~al.}(2012)Lindvall, Kalabukhov, and
	Yurgens]{Lindvall2012}
	Lindvall,~N.; Kalabukhov,~A.; Yurgens,~A. Cleaning graphene using atomic force
	microscope. \emph{J. Appl. Phys.} \textbf{2012}, \emph{111}, 064904\relax
	\mciteBstWouldAddEndPuncttrue
	\mciteSetBstMidEndSepPunct{\mcitedefaultmidpunct}
	{\mcitedefaultendpunct}{\mcitedefaultseppunct}\relax
	\EndOfBibitem
	\bibitem[Goossens \latin{et~al.}(2012)Goossens, Calado, Barreiro, Watanabe,
	Taniguchi, and Vandersypen]{Goossens2012}
	Goossens,~A.~M.; Calado,~V.~E.; Barreiro,~A.; Watanabe,~K.; Taniguchi,~T.;
	Vandersypen,~L. M.~K. Mechanical cleaning of graphene. \emph{Appl. Phys.
		Lett.} \textbf{2012}, \emph{100}, 073110\relax
	\mciteBstWouldAddEndPuncttrue
	\mciteSetBstMidEndSepPunct{\mcitedefaultmidpunct}
	{\mcitedefaultendpunct}{\mcitedefaultseppunct}\relax
	\EndOfBibitem
	\bibitem[Neumaier \latin{et~al.}(2019)Neumaier, Pindl, and Lemme]{Neumaier2019}
	Neumaier,~D.; Pindl,~S.; Lemme,~M.~C. Integrating graphene into semiconductor
	fabrication lines. \emph{Nat. Mater.} \textbf{2019}, \emph{18},
	525--529\relax
	\mciteBstWouldAddEndPuncttrue
	\mciteSetBstMidEndSepPunct{\mcitedefaultmidpunct}
	{\mcitedefaultendpunct}{\mcitedefaultseppunct}\relax
	\EndOfBibitem
	\bibitem[Li \latin{et~al.}(2021)Li, Khodas, and Levchenko]{Li2021}
	Li,~S.; Khodas,~M.; Levchenko,~A. Conformal maps of viscous electron flow in
	the Gurzhi crossover. \emph{arXiv:} \textbf{2021}, \emph{2105.13384}\relax
	\mciteBstWouldAddEndPuncttrue
	\mciteSetBstMidEndSepPunct{\mcitedefaultmidpunct}
	{\mcitedefaultendpunct}{\mcitedefaultseppunct}\relax
	\EndOfBibitem
	\bibitem[Shytov \latin{et~al.}(2018)Shytov, Kong, Falkovich, and
	Levitov]{Shytov2018}
	Shytov,~A.; Kong,~J.~F.; Falkovich,~G.; Levitov,~L. Particle Collisions and
	Negative Nonlocal Response of Ballistic Electrons. \emph{Phys. Rev. Lett.}
	\textbf{2018}, \emph{121}, 176805\relax
	\mciteBstWouldAddEndPuncttrue
	\mciteSetBstMidEndSepPunct{\mcitedefaultmidpunct}
	{\mcitedefaultendpunct}{\mcitedefaultseppunct}\relax
	\EndOfBibitem
	\bibitem[Wang \latin{et~al.}(2019)Wang, Liu, Jiang, and Xie]{Wang2019}
	Wang,~Z.; Liu,~H.; Jiang,~H.; Xie,~X.~C. Numerical study of negative nonlocal
	resistance and backflow current in a ballistic graphene system. \emph{Phys.
		Rev. B} \textbf{2019}, \emph{100}, 155423\relax
	\mciteBstWouldAddEndPuncttrue
	\mciteSetBstMidEndSepPunct{\mcitedefaultmidpunct}
	{\mcitedefaultendpunct}{\mcitedefaultseppunct}\relax
	\EndOfBibitem
	\bibitem[Ella \latin{et~al.}(2019)Ella, Rozen, Birkbeck, Ben-Shalom, Perello,
	Zultak, Taniguchi, Watanabe, Geim, Ilani, and Sulpizio]{Ella2019}
	Ella,~L.; Rozen,~A.; Birkbeck,~J.; Ben-Shalom,~M.; Perello,~D.; Zultak,~J.;
	Taniguchi,~T.; Watanabe,~K.; Geim,~A.~K.; Ilani,~S.; Sulpizio,~J.~A.
	Simultaneous voltage and current density imaging of flowing electrons in two
	dimensions. \emph{Nat. Nanotechnol.} \textbf{2019}, \emph{14}, 480--487\relax
	\mciteBstWouldAddEndPuncttrue
	\mciteSetBstMidEndSepPunct{\mcitedefaultmidpunct}
	{\mcitedefaultendpunct}{\mcitedefaultseppunct}\relax
	\EndOfBibitem
	\bibitem[Sinterhauf \latin{et~al.}(2020)Sinterhauf, Traeger, Pakdehi,
	Sch\"{a}dlich, Willke, Speck, Seyller, Tegenkamp, Pierz, Schumacher, and
	Wenderoth]{Sinterhauf2020}
	Sinterhauf,~A.; Traeger,~G.~A.; Pakdehi,~D.~M.; Sch\"{a}dlich,~P.; Willke,~P.;
	Speck,~F.; Seyller,~T.; Tegenkamp,~C.; Pierz,~K.; Schumacher,~H.~W.;
	Wenderoth,~M. Substrate induced nanoscale resistance variation in epitaxial
	graphene. \emph{Nat. Commun.} \textbf{2020}, \emph{11}, 555\relax
	\mciteBstWouldAddEndPuncttrue
	\mciteSetBstMidEndSepPunct{\mcitedefaultmidpunct}
	{\mcitedefaultendpunct}{\mcitedefaultseppunct}\relax
	\EndOfBibitem
	\bibitem[McCormick \latin{et~al.}(1999)McCormick, Woodside, Huang, Wu, McEuen,
	Duruoz, and Harris]{McCormick1999}
	McCormick,~K.~L.; Woodside,~M.~T.; Huang,~M.; Wu,~M.; McEuen,~P.~L.;
	Duruoz,~C.; Harris,~J.~S. Scanned potential microscopy of edge and bulk
	currents in the quantum Hall regime. \emph{Phys. Rev. B} \textbf{1999},
	\emph{59}, 4654--4657\relax
	\mciteBstWouldAddEndPuncttrue
	\mciteSetBstMidEndSepPunct{\mcitedefaultmidpunct}
	{\mcitedefaultendpunct}{\mcitedefaultseppunct}\relax
	\EndOfBibitem
	\bibitem[Hedberg \latin{et~al.}(2010)Hedberg, Lal, Miyahara, Gr\"{u}tter,
	Gervais, Hilke, Pfeiffer, and West]{Hedberg2010}
	Hedberg,~J.~A.; Lal,~A.; Miyahara,~Y.; Gr\"{u}tter,~P.; Gervais,~G.; Hilke,~M.;
	Pfeiffer,~L.; West,~K.~W. Low temperature electrostatic force microscopy of a
	deep two-dimensional electron gas using a quartz tuning fork. \emph{Appl.
		Phys. Lett.} \textbf{2010}, \emph{97}, 143107\relax
	\mciteBstWouldAddEndPuncttrue
	\mciteSetBstMidEndSepPunct{\mcitedefaultmidpunct}
	{\mcitedefaultendpunct}{\mcitedefaultseppunct}\relax
	\EndOfBibitem
	\bibitem[Li \latin{et~al.}(2009)Li, Zhu, Cai, Borysiak, Han, Chen, Piner,
	Colombo, and Ruoff]{Li2009}
	Li,~X.; Zhu,~Y.; Cai,~W.; Borysiak,~M.; Han,~B.; Chen,~D.; Piner,~R.~D.;
	Colombo,~L.; Ruoff,~R.~S. Transfer of large-area graphene films for
	high-performance transparent conductive electrodes. \emph{Nano Lett.}
	\textbf{2009}, \emph{9}, 4359--4363\relax
	\mciteBstWouldAddEndPuncttrue
	\mciteSetBstMidEndSepPunct{\mcitedefaultmidpunct}
	{\mcitedefaultendpunct}{\mcitedefaultseppunct}\relax
	\EndOfBibitem
	\bibitem[Adam \latin{et~al.}(2007)Adam, Hwang, Galitski, and Sarma]{Adam2007}
	Adam,~S.; Hwang,~E.~H.; Galitski,~V.~M.; Sarma,~S.~D. A self-consistent theory
	for graphene transport. \emph{Proc. Nat. Acad. Sci.} \textbf{2007},
	\emph{104}, 18392–--18397\relax
	\mciteBstWouldAddEndPuncttrue
	\mciteSetBstMidEndSepPunct{\mcitedefaultmidpunct}
	{\mcitedefaultendpunct}{\mcitedefaultseppunct}\relax
	\EndOfBibitem
	\bibitem[El-Barbary \latin{et~al.}(2003)El-Barbary, Telling, Ewels, Heggie, and
	Briddon]{ElBarbary2003}
	El-Barbary,~A.~A.; Telling,~R.~H.; Ewels,~C.~P.; Heggie,~M.~I.; Briddon,~P.~R.
	Structure and energetics of the vacancy in graphite. \emph{Phys. Rev. B}
	\textbf{2003}, \emph{68}, 144107\relax
	\mciteBstWouldAddEndPuncttrue
	\mciteSetBstMidEndSepPunct{\mcitedefaultmidpunct}
	{\mcitedefaultendpunct}{\mcitedefaultseppunct}\relax
	\EndOfBibitem
	\bibitem[Dim()]{DimensionIcon}
	\url{https://www.bruker.com/products/surface-and-dimensional-analysis/atomic-force-microscopes/dimension-icon/overview.html}\relax
	\mciteBstWouldAddEndPuncttrue
	\mciteSetBstMidEndSepPunct{\mcitedefaultmidpunct}
	{\mcitedefaultendpunct}{\mcitedefaultseppunct}\relax
	\EndOfBibitem
	\bibitem[Nečas and Klapetek(2012)Nečas, and Klapetek]{Necas2012}
	Nečas,~D.; Klapetek,~P. Gwyddion: an open-source software for {SPM} data
	analysis. \emph{Central European Journal of Physics} \textbf{2012},
	\emph{10}, 181--188\relax
	\mciteBstWouldAddEndPuncttrue
	\mciteSetBstMidEndSepPunct{\mcitedefaultmidpunct}
	{\mcitedefaultendpunct}{\mcitedefaultseppunct}\relax
	\EndOfBibitem
	\bibitem[Zerweck \latin{et~al.}(2005)Zerweck, Loppacher, Otto, Grafström, and
	Eng]{Zerweck2005}
	Zerweck,~U.; Loppacher,~C.; Otto,~T.; Grafström,~S.; Eng,~L.~M. Accuracy and
	resolution limits of Kelvin probe force microscopy. \emph{Phys. Rev. B}
	\textbf{2005}, \emph{71}, 125424\relax
	\mciteBstWouldAddEndPuncttrue
	\mciteSetBstMidEndSepPunct{\mcitedefaultmidpunct}
	{\mcitedefaultendpunct}{\mcitedefaultseppunct}\relax
	\EndOfBibitem
	\bibitem[SCM()]{SCMPITV2}
	\url{https://www.brukerafmprobes.com/p-3950-scm-pit-v2.aspx}\relax
	\mciteBstWouldAddEndPuncttrue
	\mciteSetBstMidEndSepPunct{\mcitedefaultmidpunct}
	{\mcitedefaultendpunct}{\mcitedefaultseppunct}\relax
	\EndOfBibitem
	\bibitem[S.Hudlet \latin{et~al.}(1998)S.Hudlet, Jean, Guthmann, and
	Berger]{Hudlet1998}
	S.Hudlet,; Jean,~M.~S.; Guthmann,~C.; Berger,~J. Evaluation of the capacitive
	force between an atomic force microscopy tip and a metallic surface.
	\emph{Eur. Phys. J. B} \textbf{1998}, \emph{2}, 5--10\relax
	\mciteBstWouldAddEndPuncttrue
	\mciteSetBstMidEndSepPunct{\mcitedefaultmidpunct}
	{\mcitedefaultendpunct}{\mcitedefaultseppunct}\relax
	\EndOfBibitem
	\bibitem[Strassburg \latin{et~al.}(2005)Strassburg, Boag, and
	Rosenwaks]{Strassburg2005}
	Strassburg,~E.; Boag,~A.; Rosenwaks,~Y. Reconstruction of electrostatic force
	microscopy images. \emph{Rev. Sci. Instr.} \textbf{2005}, \emph{76},
	083705\relax
	\mciteBstWouldAddEndPuncttrue
	\mciteSetBstMidEndSepPunct{\mcitedefaultmidpunct}
	{\mcitedefaultendpunct}{\mcitedefaultseppunct}\relax
	\EndOfBibitem
	\bibitem[Li \latin{et~al.}(2012)Li, Mao, Lan, and Liu]{Li2012}
	Li,~G.; Mao,~B.; Lan,~F.; Liu,~L. Practical aspects of single-pass scan Kelvin
	probe force microscopy. \emph{Rev. Sci. Instrum} \textbf{2012}, \emph{83},
	113701\relax
	\mciteBstWouldAddEndPuncttrue
	\mciteSetBstMidEndSepPunct{\mcitedefaultmidpunct}
	{\mcitedefaultendpunct}{\mcitedefaultseppunct}\relax
	\EndOfBibitem
	\bibitem[Yu \latin{et~al.}(2009)Yu, Zhao, Ryu, Brus, Kim, and Kim]{Yu2009}
	Yu,~Y.-J.; Zhao,~Y.; Ryu,~S.; Brus,~L.~E.; Kim,~K.~S.; Kim,~P. Tuning the
	graphene work function by electric field effect. \emph{Nano Lett.}
	\textbf{2009}, \emph{9}, 3430--3434\relax
	\mciteBstWouldAddEndPuncttrue
	\mciteSetBstMidEndSepPunct{\mcitedefaultmidpunct}
	{\mcitedefaultendpunct}{\mcitedefaultseppunct}\relax
	\EndOfBibitem
	\bibitem[Willke \latin{et~al.}(2016)Willke, M\"{o}hle, Sinterhauf, Kotzott, Yu,
	Wodtke, and Wenderoth]{Wilke2016}
	Willke,~P.; M\"{o}hle,~C.; Sinterhauf,~A.; Kotzott,~T.; Yu,~H.~K.; Wodtke,~A.;
	Wenderoth,~M. Local transport measurements in graphene on SiO$_2$ using
	Kelvin probe force microscopy. \emph{Carbon} \textbf{2016}, \emph{102},
	470--476\relax
	\mciteBstWouldAddEndPuncttrue
	\mciteSetBstMidEndSepPunct{\mcitedefaultmidpunct}
	{\mcitedefaultendpunct}{\mcitedefaultseppunct}\relax
	\EndOfBibitem
	\bibitem[Girard(2001)]{Girard2001}
	Girard,~P. Electrostatic force microscopy: Principles and some applications to
	semiconductors. \emph{Nanotechnology} \textbf{2001}, \emph{12},
	485--490\relax
	\mciteBstWouldAddEndPuncttrue
	\mciteSetBstMidEndSepPunct{\mcitedefaultmidpunct}
	{\mcitedefaultendpunct}{\mcitedefaultseppunct}\relax
	\EndOfBibitem
	\bibitem[Altvater \latin{et~al.}(2019)Altvater, Wu, Zhang, Zhu, Li, Watanabe,
	Taniguchi, and Andrei]{Altvater2019}
	Altvater,~M.~A.; Wu,~S.; Zhang,~Z.; Zhu,~T.; Li,~G.; Watanabe,~K.;
	Taniguchi,~T.; Andrei,~E.~Y. Electrostatic imaging of encapsulated graphene.
	\emph{2D Mat.} \textbf{2019}, \emph{6}, 045034\relax
	\mciteBstWouldAddEndPuncttrue
	\mciteSetBstMidEndSepPunct{\mcitedefaultmidpunct}
	{\mcitedefaultendpunct}{\mcitedefaultseppunct}\relax
	\EndOfBibitem
	\bibitem[Giessibl(2003)]{Giessibl2003}
	Giessibl,~F.~J. Advances in atomic force microscopy. \emph{Rev. Mod. Phys.}
	\textbf{2003}, \emph{75}, 136802\relax
	\mciteBstWouldAddEndPuncttrue
	\mciteSetBstMidEndSepPunct{\mcitedefaultmidpunct}
	{\mcitedefaultendpunct}{\mcitedefaultseppunct}\relax
	\EndOfBibitem
	\bibitem[Samaddar \latin{et~al.}(2016)Samaddar, Coraux, Martin, Grévin,
	Courtois, and Winkelmann]{Samaddar2016}
	Samaddar,~S.; Coraux,~J.; Martin,~S.~C.; Grévin,~B.; Courtois,~H.;
	Winkelmann,~C.~B. Equal variations of the Fermi level and work function in
	graphene at the nanoscale. \emph{Nanoscale} \textbf{2016}, \emph{8},
	15162--15166\relax
	\mciteBstWouldAddEndPuncttrue
	\mciteSetBstMidEndSepPunct{\mcitedefaultmidpunct}
	{\mcitedefaultendpunct}{\mcitedefaultseppunct}\relax
	\EndOfBibitem
	\bibitem[Dombrowski \latin{et~al.}(1999)Dombrowski, Steinebach, Wittneven,
	Morgenstern, and Wiesendanger]{Dombrowski1999}
	Dombrowski,~R.; Steinebach,~C.; Wittneven,~C.; Morgenstern,~M.;
	Wiesendanger,~R. Tip-induced band bending by scanning tunneling spectroscopy
	of the states of the tip-induced quantum dot on {InAs}(110). \emph{Phys. Rev.
		B} \textbf{1999}, \emph{59}, 8043--8048\relax
	\mciteBstWouldAddEndPuncttrue
	\mciteSetBstMidEndSepPunct{\mcitedefaultmidpunct}
	{\mcitedefaultendpunct}{\mcitedefaultseppunct}\relax
	\EndOfBibitem
	\bibitem[Loppacher \latin{et~al.}(2004)Loppacher, Zerweck, and
	Eng]{Loppacher2004}
	Loppacher,~C.; Zerweck,~U.; Eng,~L.~M. Kelvin probe force microscopy of alkali
	chloride thin films on Au(111). \emph{Nanotechnology} \textbf{2004},
	\emph{15}, S9--S13\relax
	\mciteBstWouldAddEndPuncttrue
	\mciteSetBstMidEndSepPunct{\mcitedefaultmidpunct}
	{\mcitedefaultendpunct}{\mcitedefaultseppunct}\relax
	\EndOfBibitem
	\bibitem[Pivetta \latin{et~al.}(2005)Pivetta, Patthey, Stengel, Baldereschi,
	and Schneider]{Pivetta2005}
	Pivetta,~M.; Patthey,~F.; Stengel,~M.; Baldereschi,~A.; Schneider,~W.-D. Local
	work function Moir{\'{e}} pattern on ultrathin ionic films: NaCl on Ag(100).
	\emph{Phys. Rev. B} \textbf{2005}, \emph{72}, 115404\relax
	\mciteBstWouldAddEndPuncttrue
	\mciteSetBstMidEndSepPunct{\mcitedefaultmidpunct}
	{\mcitedefaultendpunct}{\mcitedefaultseppunct}\relax
	\EndOfBibitem
	\bibitem[Ploigt \latin{et~al.}(2007)Ploigt, Brun, Pivetta, Patthey, and
	Schneider]{Ploigt2007}
	Ploigt,~H.-C.; Brun,~C.; Pivetta,~M.; Patthey,~F.; Schneider,~W.-D. Local work
	function changes determined by field emission resonances: NaCl/Ag(100).
	\emph{Phys. Rev. B} \textbf{2007}, \emph{76}, 195404\relax
	\mciteBstWouldAddEndPuncttrue
	\mciteSetBstMidEndSepPunct{\mcitedefaultmidpunct}
	{\mcitedefaultendpunct}{\mcitedefaultseppunct}\relax
	\EndOfBibitem
	\bibitem[Prada \latin{et~al.}(2008)Prada, Martinez, and Pacchioni]{Prada2008}
	Prada,~S.; Martinez,~U.; Pacchioni,~G. Work function changes induced by
	deposition of ultrathin dielectric films on metals: A theoretical analysis.
	\emph{Phys. Rev. B} \textbf{2008}, \emph{78}, 235423\relax
	\mciteBstWouldAddEndPuncttrue
	\mciteSetBstMidEndSepPunct{\mcitedefaultmidpunct}
	{\mcitedefaultendpunct}{\mcitedefaultseppunct}\relax
	\EndOfBibitem
	\bibitem[Teyssedre \latin{et~al.}(2021)Teyssedre, Zheng, Boudou, and
	Laurent]{Teyssedre2021}
	Teyssedre,~G.; Zheng,~F.; Boudou,~L.; Laurent,~C. Charge trap spectroscopy in
	polymer dielectrics: a critical review. \emph{J. Phys. D: Appl. Phys.}
	\textbf{2021}, \emph{54}, 263001\relax
	\mciteBstWouldAddEndPuncttrue
	\mciteSetBstMidEndSepPunct{\mcitedefaultmidpunct}
	{\mcitedefaultendpunct}{\mcitedefaultseppunct}\relax
	\EndOfBibitem
	\bibitem[Melios \latin{et~al.}(2016)Melios, Centeno, Zurutuza, Panchal, Giusca,
	Spencer, Silva, and Kazakova]{Melios2016}
	Melios,~C.; Centeno,~A.; Zurutuza,~A.; Panchal,~V.; Giusca,~C.~E.; Spencer,~S.;
	Silva,~S. R.~P.; Kazakova,~O. Effects of humidity on the electronic
	properties of graphene prepared by chemical vapour deposition. \emph{Carbon}
	\textbf{2016}, \emph{103}, 273--280\relax
	\mciteBstWouldAddEndPuncttrue
	\mciteSetBstMidEndSepPunct{\mcitedefaultmidpunct}
	{\mcitedefaultendpunct}{\mcitedefaultseppunct}\relax
	\EndOfBibitem
	\bibitem[Wang \latin{et~al.}(2011)Wang, Wang, Zhang, Li, Fang, and
	Qiu]{Wang2011}
	Wang,~R.; Wang,~S.; Zhang,~D.; Li,~Z.; Fang,~Y.; Qiu,~X. Control of Carrier
	Type and Density in Exfoliated Graphene by Interface Engineering. \emph{ACS
		Nano} \textbf{2011}, \emph{5}, 418 -- 412\relax
	\mciteBstWouldAddEndPuncttrue
	\mciteSetBstMidEndSepPunct{\mcitedefaultmidpunct}
	{\mcitedefaultendpunct}{\mcitedefaultseppunct}\relax
	\EndOfBibitem
	\bibitem[Behn \latin{et~al.}(2021)Behn, Krebs, Smith, Watanabe, Taniguchi, and
	Brar]{Behm2021}
	Behn,~W.~A.; Krebs,~Z.~J.; Smith,~K.~J.; Watanabe,~K.; Taniguchi,~T.;
	Brar,~V.~W. Measuring and Tuning the Potential Landscape of Electrostatically
	Defined Quantum Dots in Graphene. \emph{Nano Lett.} \textbf{2021}, \emph{21},
	5013--5020\relax
	\mciteBstWouldAddEndPuncttrue
	\mciteSetBstMidEndSepPunct{\mcitedefaultmidpunct}
	{\mcitedefaultendpunct}{\mcitedefaultseppunct}\relax
	\EndOfBibitem
	\bibitem[Schweizer \latin{et~al.}(2020)Schweizer, Dolle, Dasler, Abellan,
	Hauke, Hirsch, and Spiecker]{Schweizer2020}
	Schweizer,~P.; Dolle,~C.; Dasler,~D.; Abellan,~G.; Hauke,~F.; Hirsch,~A.;
	Spiecker,~E. Mechanical cleaning of graphene using in situ electron
	microscopy. \emph{Nat. Commun.} \textbf{2020}, \emph{11}, 1743\relax
	\mciteBstWouldAddEndPuncttrue
	\mciteSetBstMidEndSepPunct{\mcitedefaultmidpunct}
	{\mcitedefaultendpunct}{\mcitedefaultseppunct}\relax
	\EndOfBibitem
	\bibitem[Xia \latin{et~al.}(2009)Xia, Chen, Li, and Tao]{Xia2009}
	Xia,~J.; Chen,~F.; Li,~J.; Tao,~N. Measurement of the quantum capacitance of
	graphene. \emph{Nat. Nanotechnol.} \textbf{2009}, \emph{4}, 505--509\relax
	\mciteBstWouldAddEndPuncttrue
	\mciteSetBstMidEndSepPunct{\mcitedefaultmidpunct}
	{\mcitedefaultendpunct}{\mcitedefaultseppunct}\relax
	\EndOfBibitem
	\bibitem[Elias \latin{et~al.}(2011)Elias, Gorbachev, Mayorov, Morozov, Zhukov,
	Blake, Ponomarenko, Grigorieva, Novoselov, Guinea, and Geim]{Elias2011}
	Elias,~D.~C.; Gorbachev,~R.~V.; Mayorov,~A.~S.; Morozov,~S.~V.; Zhukov,~A.~A.;
	Blake,~P.; Ponomarenko,~L.~A.; Grigorieva,~I.~V.; Novoselov,~K.~S.;
	Guinea,~F.; Geim,~A.~K. Dirac cones reshaped by interaction effects in
	suspended graphene. \emph{Nat. Phys.} \textbf{2011}, \emph{7}, 701--704\relax
	\mciteBstWouldAddEndPuncttrue
	\mciteSetBstMidEndSepPunct{\mcitedefaultmidpunct}
	{\mcitedefaultendpunct}{\mcitedefaultseppunct}\relax
	\EndOfBibitem
	\bibitem[Li \latin{et~al.}(2009)Li, Luican, and Andrei]{Li2009b}
	Li,~G.; Luican,~A.; Andrei,~E.~Y. Scanning Tunneling Spectroscopy of Graphene
	on Graphite. \emph{Phys. Rev. Lett,} \textbf{2009}, \emph{102}, 176804\relax
	\mciteBstWouldAddEndPuncttrue
	\mciteSetBstMidEndSepPunct{\mcitedefaultmidpunct}
	{\mcitedefaultendpunct}{\mcitedefaultseppunct}\relax
	\EndOfBibitem
	\bibitem[Siegel \latin{et~al.}(2011)Siegel, Park, Hwang, Deslippe, Fedorov,
	Louie, and Lanzara]{Siegel2011}
	Siegel,~D.~A.; Park,~C.-H.; Hwang,~C.; Deslippe,~J.; Fedorov,~A.~V.;
	Louie,~S.~G.; Lanzara,~A. Many-body interactions in quasi-freestanding
	graphene. \emph{Proc. Natl. Acad. Sci.} \textbf{2011}, \emph{108},
	11365--11369\relax
	\mciteBstWouldAddEndPuncttrue
	\mciteSetBstMidEndSepPunct{\mcitedefaultmidpunct}
	{\mcitedefaultendpunct}{\mcitedefaultseppunct}\relax
	\EndOfBibitem
	\bibitem[Eisenstein \latin{et~al.}(1992)Eisenstein, Pfeiffer, and
	West]{Eisenstein1992}
	Eisenstein,~J.~P.; Pfeiffer,~L.~N.; West,~K.~W. Negative compressibility of
	interacting two-dimensional electron and quasiparticle gases. \emph{Phys.
		Rev. Lett.} \textbf{1992}, \emph{68}, 674--677\relax
	\mciteBstWouldAddEndPuncttrue
	\mciteSetBstMidEndSepPunct{\mcitedefaultmidpunct}
	{\mcitedefaultendpunct}{\mcitedefaultseppunct}\relax
	\EndOfBibitem
	\bibitem[Larentis \latin{et~al.}(2014)Larentis, Tolsma, Fallahazad, Dillen,
	Kim, MacDonald, and Tutuc]{Larentis2014}
	Larentis,~S.; Tolsma,~J.~R.; Fallahazad,~B.; Dillen,~D.~C.; Kim,~K.;
	MacDonald,~A.~H.; Tutuc,~E. Band Offset and Negative Compressibility in
	Graphene-{MoS}2 Heterostructures. \emph{Nano Lett.} \textbf{2014}, \emph{14},
	2039--2045\relax
	\mciteBstWouldAddEndPuncttrue
	\mciteSetBstMidEndSepPunct{\mcitedefaultmidpunct}
	{\mcitedefaultendpunct}{\mcitedefaultseppunct}\relax
	\EndOfBibitem
	\bibitem[Li \latin{et~al.}(2013)Li, Chen, Wang, He, Wu, Cai, Zhang, Wang, Han,
	Lortz, Zhang, Sheng, and Wang]{Li2013b}
	Li,~W.; Chen,~X.; Wang,~L.; He,~Y.; Wu,~Z.; Cai,~Y.; Zhang,~M.; Wang,~Y.;
	Han,~Y.; Lortz,~R.~W.; Zhang,~Z.-Q.; Sheng,~P.; Wang,~N. Density of States
	and Its Local Fluctuations Determined by Capacitance of Strongly Disordered
	Graphene. \emph{Sci. Rep.} \textbf{2013}, \emph{3}, 1772\relax
	\mciteBstWouldAddEndPuncttrue
	\mciteSetBstMidEndSepPunct{\mcitedefaultmidpunct}
	{\mcitedefaultendpunct}{\mcitedefaultseppunct}\relax
	\EndOfBibitem
	\bibitem[Hu \latin{et~al.}(2008)Hu, Hwang, and Sarma]{Hu2008}
	Hu,~B. Y.-K.; Hwang,~E.~H.; Sarma,~S.~D. Density of states of disordered
	graphene. \emph{Phys. Rev. B} \textbf{2008}, \emph{78}, 165411\relax
	\mciteBstWouldAddEndPuncttrue
	\mciteSetBstMidEndSepPunct{\mcitedefaultmidpunct}
	{\mcitedefaultendpunct}{\mcitedefaultseppunct}\relax
	\EndOfBibitem
	\bibitem[Hwang \latin{et~al.}(2007)Hwang, Hu, and Sarma]{Hwang2007}
	Hwang,~E.~H.; Hu,~B. Y.-K.; Sarma,~S.~D. Inelastic carrier lifetime in
	graphene. \emph{Phys. Rev. B} \textbf{2007}, \emph{76}, 115434\relax
	\mciteBstWouldAddEndPuncttrue
	\mciteSetBstMidEndSepPunct{\mcitedefaultmidpunct}
	{\mcitedefaultendpunct}{\mcitedefaultseppunct}\relax
	\EndOfBibitem
	\bibitem[Sarma \latin{et~al.}(2007)Sarma, Hwang, and Tse]{DasSarma2007}
	Sarma,~S.~D.; Hwang,~E.~H.; Tse,~W.-K. Many-body interaction effects in doped
	and undoped graphene: Fermi liquid versus non-Fermi liquid. \emph{Phys. Rev.
		B} \textbf{2007}, \emph{75}, 121406(R)\relax
	\mciteBstWouldAddEndPuncttrue
	\mciteSetBstMidEndSepPunct{\mcitedefaultmidpunct}
	{\mcitedefaultendpunct}{\mcitedefaultseppunct}\relax
	\EndOfBibitem
	\bibitem[Chen \latin{et~al.}(2008)Chen, Jang, Xiao, Ishigami, and
	Fuhrer]{Chen2008}
	Chen,~J.-H.; Jang,~C.; Xiao,~S.; Ishigami,~M.; Fuhrer,~M.~S. Intrinsic and
	extrinsic performance limits of graphene devices on ${\rm SiO_2}$. \emph{Nat.
		Nanotechnol.} \textbf{2008}, \emph{3}, 206--209\relax
	\mciteBstWouldAddEndPuncttrue
	\mciteSetBstMidEndSepPunct{\mcitedefaultmidpunct}
	{\mcitedefaultendpunct}{\mcitedefaultseppunct}\relax
	\EndOfBibitem
	\bibitem[Fratini and Guinea(2008)Fratini, and Guinea]{Fratini2008}
	Fratini,~S.; Guinea,~F. Substrate-limited electron dynamics in graphene.
	\emph{Phys. Rev. B} \textbf{2008}, \emph{77}, 195415\relax
	\mciteBstWouldAddEndPuncttrue
	\mciteSetBstMidEndSepPunct{\mcitedefaultmidpunct}
	{\mcitedefaultendpunct}{\mcitedefaultseppunct}\relax
	\EndOfBibitem
	\bibitem[Zhu \latin{et~al.}(2010)Zhu, Neumayer, Perebeinos, and
	Avouris]{Zhu2010}
	Zhu,~W.; Neumayer,~D.; Perebeinos,~V.; Avouris,~P. Silicon nitride gate
	dielectrics and band gap engineering in graphene layers. \emph{Nano Lett.}
	\textbf{2010}, \emph{10}, 3572--3576\relax
	\mciteBstWouldAddEndPuncttrue
	\mciteSetBstMidEndSepPunct{\mcitedefaultmidpunct}
	{\mcitedefaultendpunct}{\mcitedefaultseppunct}\relax
	\EndOfBibitem
	\bibitem[Kiselev and Schmalian(2019)Kiselev, and Schmalian]{Kiselev2019b}
	Kiselev,~E.~I.; Schmalian,~J. Boundary conditions of viscous electron flow.
	\emph{Phys. Rev. B} \textbf{2019}, \emph{99}, 035430\relax
	\mciteBstWouldAddEndPuncttrue
	\mciteSetBstMidEndSepPunct{\mcitedefaultmidpunct}
	{\mcitedefaultendpunct}{\mcitedefaultseppunct}\relax
	\EndOfBibitem
	\bibitem[Keser \latin{et~al.}(2021)Keser, Wang, Klochan, Ho, Tkachenko,
	Tkachenko, Culcer, Adam, Farrer, Ritchie, Sushkov, and Hamilton]{keser2021}
	Keser,~A.~C.; Wang,~D.~Q.; Klochan,~O.; Ho,~D. Y.~H.; Tkachenko,~O.~A.;
	Tkachenko,~V.~A.; Culcer,~D.; Adam,~S.; Farrer,~I.; Ritchie,~D.~A.;
	Sushkov,~O.~P.; Hamilton,~A.~R. Geometric control of universal hydrodynamic
	flow in a two dimensional electron fluid. \emph{arXiv:} \textbf{2021},
	2103.09463\relax
	\mciteBstWouldAddEndPuncttrue
	\mciteSetBstMidEndSepPunct{\mcitedefaultmidpunct}
	{\mcitedefaultendpunct}{\mcitedefaultseppunct}\relax
	\EndOfBibitem
	\bibitem[Ariel and Natan(2013)Ariel, and Natan]{Ariel2013}
	Ariel,~V.; Natan,~A. Electron effective mass in graphene. 2013 International
	Conference on Electromagnetics in Advanced Applications ({ICEAA}). 2013\relax
	\mciteBstWouldAddEndPuncttrue
	\mciteSetBstMidEndSepPunct{\mcitedefaultmidpunct}
	{\mcitedefaultendpunct}{\mcitedefaultseppunct}\relax
	\EndOfBibitem
	\bibitem[L\"{u}pke \latin{et~al.}(2017)L\"{u}pke, Eschbach, Heider, Lanius,
	Sch\"{u}ffelgen, Rosenbach, von~den Driesch, Cherepanov, Mussler, Plucinski,
	Gr\"{u}tzmacher, Schneider, and Voigtl\"{a}nder]{Luepke2017}
	L\"{u}pke,~F.; Eschbach,~M.; Heider,~T.; Lanius,~M.; Sch\"{u}ffelgen,~P.;
	Rosenbach,~D.; von~den Driesch,~N.; Cherepanov,~V.; Mussler,~G.;
	Plucinski,~L.; Gr\"{u}tzmacher,~D.; Schneider,~C.~M.; Voigtl\"{a}nder,~B.
	Electrical resistance of individual defects at a topological insulator
	surface. \emph{Nat. Commun.} \textbf{2017}, \emph{8}, 15704\relax
	\mciteBstWouldAddEndPuncttrue
	\mciteSetBstMidEndSepPunct{\mcitedefaultmidpunct}
	{\mcitedefaultendpunct}{\mcitedefaultseppunct}\relax
	\EndOfBibitem
	\bibitem[Hui \latin{et~al.}(2020)Hui, Lederer, Oganesyan, and Kim]{Hui2020}
	Hui,~A.; Lederer,~S.; Oganesyan,~V.; Kim,~E.-A. Quantum aspects of hydrodynamic
	transport from weak electron-impurity scattering. \emph{Phys. Rev. B}
	\textbf{2020}, \emph{101}, 121107\relax
	\mciteBstWouldAddEndPuncttrue
	\mciteSetBstMidEndSepPunct{\mcitedefaultmidpunct}
	{\mcitedefaultendpunct}{\mcitedefaultseppunct}\relax
	\EndOfBibitem
\end{mcitethebibliography}
\providecommand{\latin}[1]{#1}
\makeatletter
\providecommand{\doi}
{\begingroup\let\do\@makeother\dospecials
	\catcode`\{=1 \catcode`\}=2 \doi@aux}
\providecommand{\doi@aux}[1]{\endgroup\texttt{#1}}
\makeatother
\providecommand*\mcitethebibliography{\thebibliography}
\csname @ifundefined\endcsname{endmcitethebibliography}
{\let\endmcitethebibliography\endthebibliography}{}

\end{document}